\documentclass[12pt]{article}
\usepackage{epsfig}
\usepackage{amsmath}
\usepackage{hhline}
\usepackage{amssymb}
\usepackage{times}
\usepackage{cite}
\usepackage{time}
\usepackage{psfrag}
\usepackage{caption2}
\usepackage[left]{lineno}

\newlength{\dinwidth}
\newlength{\dinmargin}
\begin{document}
\newcommand{\picb}{pb$^{-1}$ }
\newcommand{\Qsq}{\ensuremath{Q^2} }
\newcommand{\Msq}{\ensuremath{M_{\psi}^2} }
\newcommand{\GeVsq}{\ensuremath{\,\mathrm{GeV}^2} }
\newcommand{\GeV}{\ensuremath{\,\mathrm{GeV}} }
\newcommand{\nb}{\ensuremath{\,\mathrm{nb}} }
\newcommand{\Wgp}{\ensuremath{W_{\gamma p}} }
\newcommand{\Wgsp}{\ensuremath{W_{\gamma p}}}
\newcommand{\gp}{\ensuremath{\gamma p} }
\newcommand{\gsp}{\ensuremath{\gamma p} }
\newcommand{\Jpsi}{\boldmath\ensuremath{J/\psi} }
\newcommand{\jpsi}{\ensuremath{J/\psi} }
\newcommand{\mumu}{\ensuremath{\mu^+\mu^-} }
\newcommand{\mee}{\ensuremath{M_{ee}}}
\newcommand{\psip}{\ensuremath{\psi(2S)} }
\newcommand{\cc}{\ensuremath{c\bar{c}} }
\newcommand{\qq}{\ensuremath{q\bar{q}} }
\renewcommand{\deg}{\ensuremath{^{\circ}} }
\newcommand{\stat}{\ensuremath{(\mathrm{stat.})}}
\newcommand{\sys}{\ensuremath{(\mathrm{syst.})}}
\newcommand{\qexp}{\ensuremath{2.486\pm0.080\stat\pm0.068\sys}}
\newcommand{\wexpphp}{\ensuremath{0.75\pm0.03\pm0.03}}
\newcommand{\aophp}{\ensuremath{1.224\pm0.010\pm0.012}}
\newcommand{\bophp}{\ensuremath{4.630\pm0.060^{+0.043}_{-0.163}}}
\newcommand{\apphp}{\ensuremath{0.164\pm0.028\pm0.030}}
\newcommand{\aodis}{\ensuremath{1.183\pm0.054\pm0.030}}
\newcommand{\bodis}{\ensuremath{3.86\pm0.13\pm0.31}}
\newcommand{\apdis}{\ensuremath{0.019\pm0.139\pm0.076}}
\newcommand{\Pom}{I\!P}
\newcommand{\VM}{\ensuremath{\rm VM}}

\def\Journal#1#2#3#4{{#1} {\bf #2} (#3) #4}
\def\NCA{\em Nuovo Cimento}
\def\NIM{\em Nucl. Instrum. Methods}
\def\NIMA{{\em Nucl. Instrum. Methods} {\bf A}}
\def\NPB{{\em Nucl. Phys.}   {\bf B}}
\def\PLB{{\em Phys. Lett.}   {\bf B}}
\def\PRL{\em Phys. Rev. Lett.}
\def\PRD{{\em Phys. Rev.}    {\bf D}}
\def\ZPC{{\em Z. Phys.}      {\bf C}}
\def\EJC{{\em Eur. Phys. J.} {\bf C}}
\def\CPC{\em Comp. Phys. Commun.}

\begin{titlepage}

\begin{flushleft}

DESY 05-161 \hfill ISSN 0418-9833 \\
October 2005
\end{flushleft}

\vspace{2cm}

\begin{center}
\begin{Large}
  
  {\bf Elastic \Jpsi Production at HERA}

\vspace{2cm}

H1 Collaboration

\end{Large}
\end{center}

\vspace{2cm}

\begin{abstract}
  \noindent
  Cross sections for elastic production of \jpsi mesons in photoproduction
  and electroproduction are measured in electron proton collisions at HERA
  using an integrated luminosity of 55~pb$^{-1}$. 
  Results are presented for photon virtualities \Qsq up to
  $80\GeVsq$. The dependence on the photon-proton centre of mass energy \Wgp
  is analysed in the range $40\le\Wgp\le305\GeV$ in photoproduction and
  $40\le\Wgp\le160\GeV$ in electroproduction. The \Wgp dependences of the
  cross sections do not change significantly with \Qsq and can be described
  by models based on perturbative QCD. Within such models, the data show a 
  high sensitivity to the gluon  density of the proton in the domain of low
  Bjorken $x$ and low \Qsq.
  Differential cross sections d$\sigma/\mbox{d}t$, 
  where $t$ is the squared four-momentum transfer at the proton vertex,
  are measured in the range $|t|<1.2\GeVsq$ as functions of \Wgp and $Q^2$.
  Effective Pomeron trajectories are determined for
  photoproduction and electroproduction.
  The \jpsi\ production and decay angular distributions are consistent
  with $s$-channel helicity conservation.
  The ratio of the cross sections for longitudinally and transversely 
  polarised photons is
  measured as a function of \Qsq and is found to be described by perturbative
  QCD based models. 
\end{abstract}

\vspace{1.5cm}

\begin{center}
Submitted to \EJC
\end{center}

\end{titlepage}



\begin{flushleft}

A.~Aktas$^{10}$,               
V.~Andreev$^{26}$,             
T.~Anthonis$^{4}$,             
B.~Antunovic$^{27}$,             
S.~Aplin$^{10}$,               
A.~Asmone$^{34}$,              
A.~Astvatsatourov$^{4}$,       
A.~Babaev$^{25}$,              
S.~Backovic$^{31}$,            
J.~B\"ahr$^{39}$
% ...

\bigskip{\it
 $ ^{1}$ I. Physikalisches Institut der RWTH, Aachen, Germany$^{ a}$ \\
% ...
$ ^{4}$ Inter-University Institute for High Energies ULB-VUB, Brussels;
          Universiteit Antwerpen, Antwerpen; Belgium$^{ c}$ \\
% ...
$ ^{10}$ DESY, Hamburg, Germany \\
% ...
$ ^{25}$ Institute for Theoretical and Experimental Physics,
          Moscow, Russia$^{ l}$ \\
 $ ^{26}$ Lebedev Physical Institute, Moscow, Russia$^{ e}$ \\
 $ ^{27}$ Max-Planck-Institut f\"ur Physik, M\"unchen, Germany \\
% ...
$ ^{31}$ Faculty of Science, University of Montenegro,
          Podgorica, Serbia and Montenegro$^{ e}$ \\
% ...
$ ^{34}$ Dipartimento di Fisica Universit\`a di Roma Tre
          and INFN Roma~3, Roma, Italy \\
% ...
$ ^{39}$ DESY, Zeuthen, Germany \\
% ...
$ ^a$ Supported by the Bundesministerium f\"ur Bildung und Forschung, FRG,
      under contract numbers 05 H1 1GUA /1, 05 H1 1PAA /1, 05 H1 1PAB /9,
      05 H1 1PEA /6, 05 H1 1VHA /7 and 05 H1 1VHB /5 \\
% ...
$ ^c$ Supported by FNRS-FWO-Vlaanderen, IISN-IIKW and IWT
      and  by Interuniversity
Attraction Poles Programme,
      Belgian Science Policy \\
 $ ^d$ Partially Supported by the Polish State Committee for Scientific
      Research, SPUB/DESY/P003/DZ 118/2003/2005 \\
 $ ^e$ Supported by the Deutsche Forschungsgemeinschaft \\
% ...
$ ^l$ Partly Supported by Russian Foundation
      for Basic Research,  grants  03-02-17291
      and  04-02-16445 \\
}

\end{flushleft}

\newpage
\section{Introduction}

Quantum Chromodynamics (QCD), the field theory of quark and gluon
 interactions, is expected to describe the strong force
between hadrons.
QCD is a successful theory in the limit of short distances,
corresponding to small values of the strong coupling constant $\alpha_s$, where
perturbative methods can be applied (perturbative QCD, pQCD).
The bulk of the scattering cross section of hadrons however, is 
dominated by long-range forces (``soft interactions''), where a satisfactory
understanding of QCD still remains a challenge. A large fraction of these soft
interactions is mediated by vacuum quantum number exchange and is termed
``diffractive''. In hadronic interactions, diffraction is well described by
Regge theory, where it is due to the $t$-channel exchange of a leading
trajectory with vacuum quantum numbers, called the ``Pomeron'' trajectory. In
the high energy limit, Pomeron exchange dominates over all other
contributions to the scattering amplitude and leads to an almost
energy-independent total cross section. Elastic photoproduction of
vector mesons, $\gp\to \VM$ $p$, 
is a particular example for a diffractive process. Measurements of the cross sections 
for the elastic production of light vector mesons ($\rho, 
\omega,$ and $\phi$) in low \Qsq electron-proton
collisions at HERA as function of the photon-proton centre of mass
energy \Wgp~\cite{Aid:1996ee,Derrick:1995vq} have verified the
expected universal Regge behaviour.
   
The cross section for elastic photoproduction of \jpsi mesons, $\gp\to\jpsi
\, p$, on the contrary,  rises steeply with
\Wgp~\cite{Aid:1996dn,Adloff:2000vm,Breitweg:1997rg,Chekanov:2002xi}, 
incompatible with a universal Pomeron. Due to the 
large mass of the \jpsi meson, which provides a ``hard'' scale (equivalent to a short range of the
forces involved), the elastic photoproduction of \jpsi
mesons is expected to be described by pQCD. In
electroproduction the photon virtuality \Qsq can provide a second hard
scale in addition to the \jpsi mass, allowing the interplay between 
these two scales to be studied. 

\vspace*{-0.5cm}
\begin{figure}[h!]
  \begin{center}
    \setlength{\unitlength}{1pt}

    \begin{picture}(1,1)(0,0)
      \thicklines
      \put(-45.0,85.0){{\bf a)}}
      \put(-45.0,55.0){{\small \bf \boldmath \Wgp}}
       \put(38.0,45.0){{\small \Pom}}
      \put(170.0,85.0){{\bf b)}}
    \end{picture}~\hspace{-2cm}~
    \includegraphics[bb=0 0 350 300,scale=0.50,keepaspectratio]
     {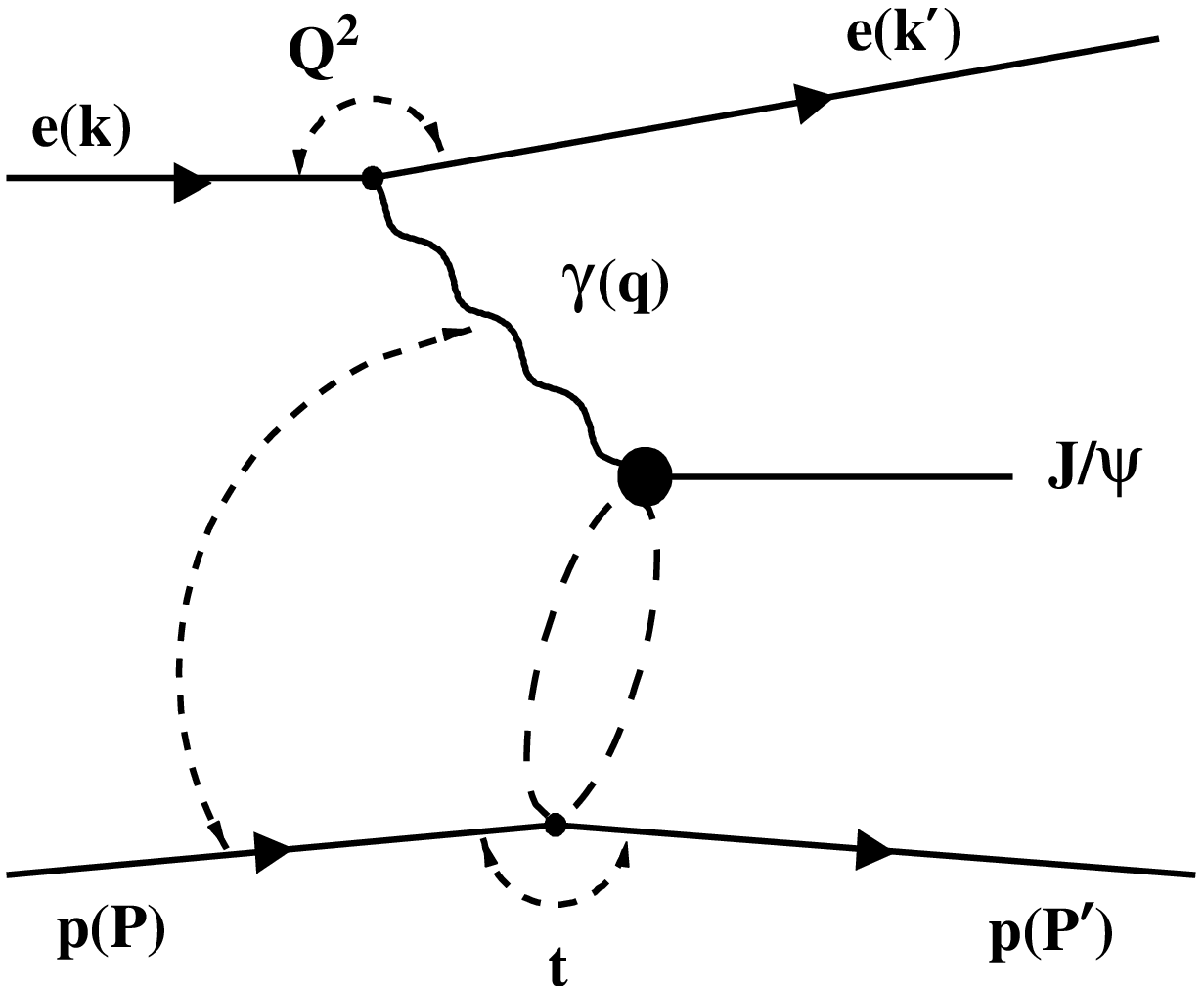}~\hspace{1.5cm}~
  \includegraphics[bb=50 -10 350 300,scale=0.55,keepaspectratio]
     {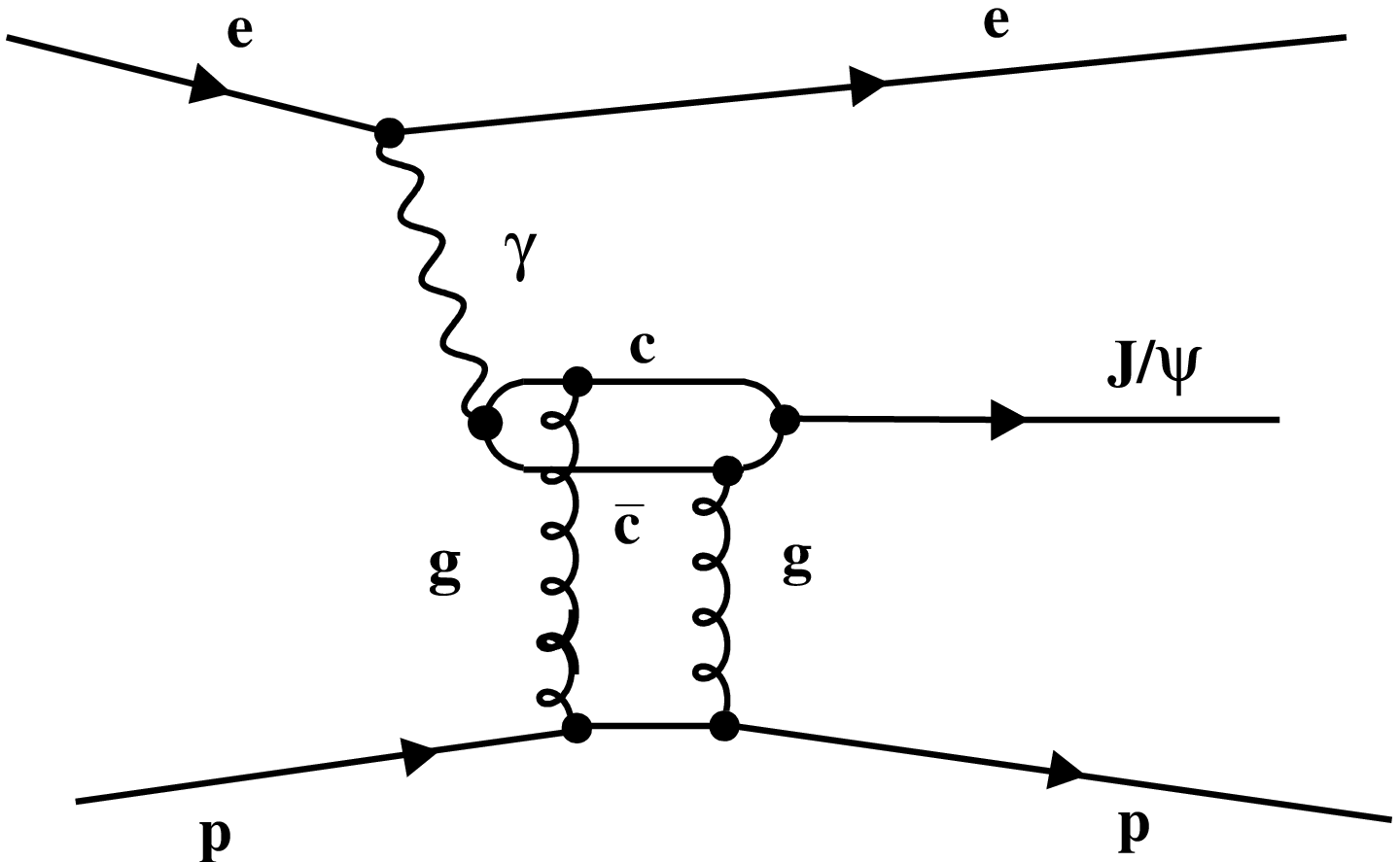}\\
\caption{Elastic \jpsi production, a)~in an approach based on Pomeron (\Pom) exchange and 
b)~in a pQCD approach via two gluon exchange. 
  The kinematic variables are indicated in a).} 
    \label{feyn}
  \end{center}
\end{figure}
The elastic production of \jpsi mesons is illustrated in figure \ref{feyn}.
In QCD at lowest order the process 
is mediated by a colour-singlet state of two gluons (figure \ref{feyn}b) and the cross
section is related to the square of the gluon density in the proton.
The gluon's momentum fraction
$x$ is kinematically related to \Wgp: the steep rise in the gluon density
towards low values of $x$ thus explains the steep rise of the cross section with
increasing \Wgp observed in the data. 
 Beyond this approximation correlations between the gluons have to 
be taken into account and the cross section for elastic \jpsi production involves
   the generalised gluon density (see for example~\cite{Diehl:2003ny} for a review).

The dependence of the elastic \jpsi cross sections on the squared four-momentum
transfer $t$ at the proton vertex shows a fast fall with increasing $|t|$. 
This dependence can be parameterised as an exponential function $e^{bt}$ at low values of $|t|$,  
although other shapes have also been proposed (for example \cite{Frankfurt:2002ka}).
In Regge theory with a $t$ dependent Pomeron trajectory, the $t$ dependence of the cross
section varies with $\Wgp$, the slope parameter $b$ increasing logarithmically
with $\Wgp$ (``shrinkage'' of the diffractive peak). 
In QCD-based models, on the other hand, the dependence of $b$ on \Wgp is 
expected to be weak~\cite{Frankfurt:2000ez,Brodsky:1998kn}. In addition to the elastic process, in
which the proton remains intact, diffractive \jpsi production can lead to
proton dissociation, $\gp\to\jpsi \,Y$, in which a low mass baryonic state $Y$
is produced. This process is expected to be important at large values of $|t|$.  

In the past, diffractive \jpsi cross sections have been
measured using photon and electron beams in fixed target experiments
up to centre of mass energies of about 20~GeV. 
At HERA the kinematic range is extended up to photon-proton
centre of mass energies of $\Wgp\sim 290\,$GeV in
photoproduction, and in electroproduction up to photon
virtualities of $\Qsq \lesssim 100
\GeVsq$~\cite{Aid:1996ee,Aid:1996dn,Adloff:1999zs,Adloff:2000vm,Adloff:2002re,Aktas:2003zi,Breitweg:1997rg,Breitweg:1998nh,Chekanov:2002rm,Chekanov:2002xi,Chekanov:2004mw}.
In this paper new data are presented on the $\Qsq$, \Wgp and $t$ dependence of the
cross section for elastic \jpsi production. The data correspond to a factor of
three more integrated luminosity than our previous
publication for photoproduction~\cite{Adloff:2000vm} and a factor of two more for
electroproduction~\cite{Adloff:1999zs}.
The kinematic range is extended
to values of $\Wgp$ up to $305\GeV$ in photoproduction, while in
electroproduction the range covered is $40<\Wgp<160\GeV$. Furthermore, the 
angular distributions for production and decay of the \jpsi mesons are
determined in order to extract the cross sections of longitudinally and
transversely polarised photons and to 
test the hypothesis of $s$-channel helicity conservation (SCHC),
which predicts that the helicity of the \jpsi meson in the final state is the
same as that of the initial (virtual) photon.

\section{Models for Elastic \boldmath{\Jpsi} Production}
Within the Regge framework  (see 
for example~\cite{Collins:1977} for a review) the cross section for diffractive photoproduction of vector
mesons at low values of $|t|$ approximately 
follows a power law $\sigma_{\gp}\propto \Wgp^{\delta}$
with $\delta \simeq 0.2$~\cite{DoLan:1995}. 
The power is related to the Pomeron trajectory $\delta\sim 4\,(\alpha_{\Pom}(t)-1)$
where $\alpha_{\Pom}(t) = \alpha_{\Pom}(0) + \alpha_{\Pom}^{\prime} t$.
The existing measurements for \jpsi mesons, however, indicate a much steeper
dependence on \Wgp $(\delta \simeq 0.8)$ than is predicted by the universal
(``soft'') Pomeron. There are also indications that the slope
$\alpha_{\Pom}^{\prime}$, responsible for the shrinkage of the diffractive
peak of elastic \jpsi photoproduction, is smaller
~\cite{Adloff:2000vm} than the value of 0.25 GeV$^{-2}$ expected from 
the soft Pomeron trajectory.
To overcome this difficulty, Donnachie and Landshoff have suggested an
additional ``hard'' Pomeron trajectory~\cite{DoLan:1998} for processes which 
involve a hard scale, such as the vector meson mass or a large
momentum transfer $Q^2$. With this conjecture
the concept of a single universal Pomeron trajectory becomes obsolete for
hard scattering processes. It has become customary, however, to introduce an
effective Pomeron trajectory $\alpha(t) = \alpha_0 + \alpha^\prime t$, where
the intercept $\alpha_0$ can be calculated within certain 
QCD models (see for example~\cite{Bartels:2000ze} for a review). 
  
In photoproduction of \jpsi mesons the mass $M_{\jpsi}$ may serve as a hard scale and in
electroproduction both $M_{\jpsi}$ and $Q^2$. A third hard scale
may be provided by a sufficiently large momentum transfer $|t|$
at the proton vertex~\cite{Aktas:2003zi,Chekanov:2002rm}.   
In the presence of a hard scale QCD factorisation methods 
(e.g. collinear factorisation, $k_T$
factorisation) may be 
applied. Factorisation allows the separation of the scattering amplitude into a
perturbative hard scattering coefficient function and non-perturbative
quantities, such as the input gluon density for the proton and the vector 
meson wave function.   

Early pQCD predictions, for example~\cite{Ryskin:1992ui}, assume that the two
exchanged gluons have the same longitudinal momentum fraction $x$ with
respect to the proton, where
$x\simeq(Q^2+M_\psi^2)/(Q^2+\Wgp^2)$, and that each of the quarks making up the \jpsi 
meson carries half of the photon momentum. Such models correspond to a leading
approximation in $\log {1/x}$, and at high energies the cross
section depends on the square of the gluon density within the proton.  
More recently generalised or ``skewed'' parton distributions have been 
considered, where the two gluons have different fractional momenta
~\cite{Martin:1997wy,Martin:1999wb,Teubner:1999pm,Ivanov:2004vd,Ivanov:2004ax}. 

The data presented here are compared with a pQCD model by
Martin, Ryskin and Teubner (MRT~\cite{Martin:1999wb}) which is based on
$k_T$ factorisation and uses a parton-hadron duality ansatz avoiding the
large uncertainties from the poorly known \jpsi meson wave function.
In this model, effects beyond the leading logarithmic approximation in $\log
{Q^2}$ are included at the amplitude level, requiring an integration over the transverse
momenta of the two gluons and hence the use of unintegrated gluon
distributions. In the MRT calculations these distributions are 
derived~\cite{Teubner:1999pm} from the conventional integrated
parton distributions\footnote{
In this model the unintegrated gluon distribution is determined from
the derivative of the standard gluon distribution with respect to $\log Q^2$, i.e.
essentially from the second derivative of the proton structure function 
$F_2(x,Q^2)$ with respect to $\log Q^2$, which in the kinematic region
of the \Jpsi analysis, at low $x$ and $Q^2$, is not well measured at HERA yet.},
 as extracted from inclusive deep-inelastic scattering. The skewing
effects are estimated independently by applying a 
factor to the amplitude~\cite{Shuvaev:1999ce}. Since these calculations only apply to the imaginary
part of the scattering amplitude, dispersion relations are used to estimate
the effects of the real part. In the parton-hadron duality  approach the
correct spin-parity
state ($J^P = 1^{--}$) of the $c\bar{c}$ pair is projected out by using the 
appropriate rotation matrices in the integrals over the resonance mass
region. Since the choice of the mass range in the integration is arbitrary 
to some extent,
the normalisation of the cross sections is predicted with limited
accuracy. The overall normalisation contains additional uncertainties due to 
missing higher order corrections. The approximations used are
however believed to have little influence on the \Wgp and \Qsq
dependences of the cross sections~\cite{teubnerdis}.  
Predictions are provided both in the photoproduction and electroproduction
regimes.  

The calculations by Frankfurt, McDermott and Strikman
(FMS~\cite{Frankfurt:2000ez}) are based on the dipole approach. Here the exchanged
photon turns into a  $q\bar{q}$ pair long before the interaction with the
proton. A leading logarithmic approximation
for the interaction of this $q\bar{q}$ pair, described as a small transverse-size
 dipole, is used. For the interaction with the proton two-gluon exchange is assumed. 
In addition the effect of a running quark mass, a \Wgp dependent
slope of the exponential $t$ distribution, and generalised gluon distributions
are considered in this calculation. Similarly to the MRT calculations the
model does not provide an
accurate normalisation of the cross section. Predictions are only available for 
photoproduction.

\section{Data Analysis}
The data were recorded with the H1 detector in the years 1999 and 2000 when HERA
was operated with electrons or predominantly with positrons\footnote{Hereafter the term
  `positron' is used for all lepton beam particles, whereas `electron' is
  used for both electrons and positrons from \jpsi decays.} of $27.5\GeV$ and
protons of $920\GeV$. 
 The \jpsi mesons are detected via their
decays into $\mu^+\mu^-$ or $e^+e^-$ pairs. They are selected from data
corresponding to an integrated luminosity of 55~pb$^{-1}$.

\subsection{The H1 Detector}
\vspace*{-1.5ex} The H1 detector is described in detail in~\cite{Abt:1996hi}.
Charged particles are detected in the central and forward\footnote{The
  positive $z$-axis is defined by the proton beam direction. The polar angle
  $\theta$ is measured with respect to the $z$ axis and $\theta<90\deg$ is
  called the `forward' direction.} tracking detectors (CTD and FTD), which consist
of drift and proportional chambers that provide a polar angle coverage between
$7\deg$ and $165\deg$. Tracks at large $\theta$ are detected in the backward
silicon tracker (BST~\cite{Eick:1996gv}, $165\deg<\theta<175\deg$). 
The central liquid argon (LAr)~\cite{Abt:1996hi} and backward lead
scintillator (SpaCal) 
calorimeters~\cite{Nicholls:1995di,Appuhn:1996na} cover the polar
angle regions $4\deg<\theta<153\deg$ and $153\deg<\theta<177.5\deg$, respectively.
For $\Qsq\gtrsim 2\,\GeVsq$ the scattered positron is detected in the SpaCal, while the decay electrons 
from the \jpsi\ meson are identified in the LAr and SpaCal calorimeters.
Muons are
identified as minimum ionising particles in the LAr calorimeter or in the
instrumented iron return yoke of the solenoidal magnet which surrounds the
central detector (central muon detector, CMD, $4\deg<\theta <171\deg$).

Dissociated proton states $Y$ with masses $M_Y\gtrsim 1.6\GeV$ may, 
after a secondary interaction, be measured in a set of detectors in the  
forward direction. These are 
the proton remnant tagger (PRT), an array of scintillators covering
$0.06\deg<\theta<0.17\deg$, the drift chambers of the forward muon
detector (FMD)~\cite{Biddulph:1993bz} closest to the beam interaction region
in the angular range $3\deg<\theta<17\deg$ and the forward region
 of the LAr calorimeter ($\theta<10\deg$).

H1 uses a multi-stage trigger system. At level 1 signals from the
CTD, SpaCal, and CMD are used to obtain the present data sets. At level 2 
information from these detectors and the LAr calorimeter is used in 
neural network algorithms~\cite{Kohne:1997ph}.

The luminosity is determined from the rate of Bethe Heitler events.

\subsection{Kinematics}
\vspace*{-1.5ex}
The kinematics of the process $ep\to ep\jpsi$ are described by the following
variables: the square of the $ep$ centre of mass energy $s=(p+k)^2$; the
negative four-momentum transfer squared at the lepton vertex
$Q^2=-q^2=-(k-k')^2$; the four-momentum transfer squared at the proton vertex
$t=(p-p')^2$ and the inelasticity $y=(p \cdot q)/(p \cdot k)$. The
four-momenta $k$, $k'$, $p$, $p'$ and $q$ refer to the incident and scattered
positron, the incoming and outgoing proton (or dissociated system $Y$) and 
the exchanged photon, respectively. The centre of mass energy of the
photon-proton system $\Wgp$  
is given by $\Wgsp^2 = (p+q)^2 = ys-\Qsq$ neglecting the proton mass.

In electroproduction the event kinematics are reconstructed using the double
angle method~\cite{Bentvelsen:1992fu},

\vspace*{-3ex}
\begin{eqnarray*}
  y &=& \frac{\sin{\theta_{e}}(1-\cos{\theta_\psi})}{\sin{\theta_\psi}
  +\sin{\theta_{e}}-\sin(\theta_{e}+\theta_\psi)}\\
  \Qsq &=& 4E_e^2
  \frac{\sin{\theta_\psi}(1+\cos{\theta_{e}})}{\sin{\theta_\psi}
  +\sin{\theta_{e}}-\sin(\theta_{e}+\theta_\psi)}.
\end{eqnarray*}
Here $E_e$ is the energy of the incident positron and $\theta_\psi$ and
$\theta_{e}$ are the polar angles of the \jpsi meson and the scattered
positron, respectively. The variable $t$ is calculated as 
$t\simeq -(\vec{p}_{t,\psi}+\vec{p}_{t,e})^2$, where 
$\vec{p}_{t,\psi}$ is the transverse momentum of the \jpsi meson
candidate and $\vec{p}_{t,e}$ that of the outgoing positron. 

In photoproduction, where the positron is not observed in the central
detector, $y$ is reconstructed via $y=(E-p_z)_{\psi}/(2E_e)$
\cite{Jacquet:1979}, where $E$ and $p_z$ denote the energy and the
longitudinal component of the momentum of the \jpsi meson. 
The variable $t$ is approximated as 
$t\simeq -p^2_{t,\psi}$ (see also the section on d$\sigma/\mbox{d}t$ below).

\subsection{Monte Carlo Simulation}
\vspace*{-1.5ex}
Monte Carlo simulations are used to calculate acceptances and the 
efficiencies for triggering, track reconstruction, event selection and lepton
identification.

The elastic \jpsi signal events are generated using the program
DIFFVM~\cite{diffvm} which is based on the Vector Dominance Model
and permits separate variation of the dependence on \Wgsp, $t$ and \Qsq.
The parameters are iteratively adjusted to those of the present measurements.
 DIFFVM is also used to
generate \jpsi production with proton dissociation. A mass dependence of
d$\sigma/\mbox{d}M_Y^2 \propto f(M_Y^2)\,M_Y^{-\beta}$ is implemented, where
$f(M_Y^2)=1$ for $M_Y^2>3.6\GeVsq$. At lower $M_Y^2$ the function $f(M_Y^2)$
takes into account the production of excited nucleon states.
The decay angular distributions of the \jpsi meson are
simulated assuming $s$-channel helicity conservation. For electroproduction,
radiative corrections are included using the generator
HERACLES~\cite{Kwiatkowski:1990es}, where contributions up to order
$\alpha^3_{QED}$ are taken into account. 

The non-resonant background is estimated using the generators LPAIR
\cite{Baranov:1991yq}, which simulates the process $\gamma \gamma \rightarrow
\ell^+\ell^-$ and COMPTON~\cite{Carli:1991yn} for the QED Compton process
$ep\to e\gp$. Cross checks with the generator GRAPE \cite{grape}
did not show significant deviations from the results of LPAIR
in the region of the present analysis. 

For all processes, detector effects are simulated in detail with 
the GEANT program~\cite{Brun:1987ma}. The detector
response including trigger efficiencies is tuned using independent data.
Remaining differences are included in the systematic errors. The simulated
events are passed through the same reconstruction software as the data.

\subsection{Event Selection}\label{secsel}
\vspace*{-1.5ex}
Elastic \jpsi events are selected by requiring two muons or two electrons 
and, in the case of electroproduction, a scattered positron candidate. 
For photoproduction the absence of any such candidate is required. As described
in table~\ref{tab:sel} four data sets are defined covering different
regions of \Qsq and $\Wgp$ and corresponding to  
different signatures of the \jpsi decay leptons.

\begin{table}[b!]
  \begin{center}
    \begin{tabular}{|l|c|p{2.em}|c|c|}
      \hline
      Data set & I & \centering II & III & IV\\
      \hline
      Kinematic region & \multicolumn{1}{|c|}{Electroproduction} & \multicolumn{3}{|c|}{Photoproduction}\\
      \hline\hline 
 & & \multicolumn{3}{|c|}{}\\[-2.3ex] 
   $\Qsq$ range $[\GeVsq]$ & $2-80$ & \multicolumn{3}{|c|}{$<1$}\\ 
\hline  
       & & \multicolumn{3}{|c|}{}\\   [-2.3ex]
      $\langle\Qsq\rangle$ $[\GeVsq]$ & $8.9$ & \multicolumn{3}{|c|}{$0.05$}\\ 
\hline   
      \Wgp $[\GeV]$&\multicolumn{2}{|c|}{$40 - 160$}&$135 - 235$&$205 -305$\\  
\hline
     &\multicolumn{4}{|c|}{}   \\[-2.3ex]
    $|t|$ $[\GeVsq]$ &\multicolumn{4}{|c|}{$< 1.2$}\\

      \hline
      Decay channel & \multicolumn{2}{|c|}{$\jpsi\to\mu^+\mu^-$} &
      \multicolumn{2}{|c|}{$\jpsi\to e^+e^-$} \\
      \hline
      Lepton
      signature&\multicolumn{2}{|c|}{Track-Track}&Track-Cluster&Cluster-Cluster
      \\
      \hline
      $\!\!\begin{array}[c]{l}
        \mbox{Lepton polar}\\ \mbox{angle region }[\deg] 
      \end{array}$
      & \multicolumn{2}{|c|}{$20 - 160$} &
      $\!\!\begin{array}[c]{l}
        \theta_1: 80 - 155\\
        \theta_2: 160 - 177
      \end{array}$
      &
      $\!\!\begin{array}[c]{l}
        \theta_1: 160 - 174\\
        \theta_2: 160 - 175.5
      \end{array}$ \\
      \hline
      $\!\!\begin{array}[c]{l}
        \mbox{Lepton energy}\\ { }[\GeV] 
      \end{array}$
      & \multicolumn{2}{|c|}{$p_t>0.8$} &
      $\!\!\begin{array}[c]{l}
        p_{t,1}>0.7, p_1>0.8\\
        E_2>4.2
      \end{array}$
      &
      $\!\!\begin{array}[c]{l}
        E_{1,2}>4.2\\
        \mbox{max}(E_1,E_2)>6
      \end{array}$\\
           \hline
    Elastic selection  &\multicolumn{4}{|c|}{No signal in forward detectors}\\
         \hline
 & \multicolumn{2}{|c|}{} & \multicolumn{2}{|c|}{}\\[-2.3ex]
       $\int L\,dt$ $[{\rm pb}^{-1}]$&\multicolumn{2}{|c|}{$54.79$}& $30.26$ &
      $26.90$ \\
      \hline
    \end{tabular}
    \caption{Summary of the most important event selection criteria for the four different data
      sets together with the corresponding integrated luminosities.}
    \label{tab:sel}
  \end{center}
\end{table}

For data set I (electroproduction) the scattered positron must be detected
with an energy of at least $12\GeV$ in the SpaCal and the reconstructed value
of \Qsq must be within $2<\Qsq<80\GeVsq$. To
suppress photoproduction background and to reduce the fraction of events with
initial state QED radiation, events are rejected if {$\sum(E-p_z)<45\GeV$},
where the sum runs over all final state particles including the scattered
positron. Neglecting radiative effects this variable is expected to be twice
the incident positron energy due to longitudinal momentum conservation.  

For the selection of photoproduction events (data sets II--IV) the
absence of any candidate for the scattered positron is required,
restricting the accepted range of negative four-momentum transfer squared
$Q^2$ to below about 1 GeV$^2$, with $\langle\Qsq\rangle = 0.05\, \GeVsq$.

Data sets I and II ($40< \Wgsp < 160\GeV$) contain $\jpsi\to\mu^+\mu^-$ events.
Exactly two oppositely charged particles must
be present in the CTD, with transverse momenta (with respect to the beamline)
$p_t>0.8\GeV$. A reconstructed vertex within $\pm40\,$cm of the $z$
coordinate of the nominal beam interaction point is required. At least one particle 
must be identified as a muon in the
central calorimeter or in the CMD.  For data set II background from cosmic
ray muons is rejected using an acollinearity cut as well as timing
information from the CTD. Further details of this analysis may be found
in~\cite{Fleischmann:2004uj}.

Data sets III and IV are selected to cover photoproduction at high values of 
$W_{\gamma p}$, which are 
related to large polar angles of the \jpsi decay leptons. The \jpsi decay
into $e^+e^-$ pairs is used. Data set III ($135<\Wgp<235\GeV$) requires one 
decay electron to be measured in the CTD coming from within $\pm40\ $ cm of the
nominal beam interaction point and one in the backward calorimeter 
SpaCal. The selected polar angle regions are given in table \ref{tab:sel}. The
electron measured in the CTD must have a momentum $p_1>0.8\GeV$ and a
transverse momentum $p_{t,1}>0.7\GeV$ and must be identified by a matching
electromagnetic energy deposition in the central calorimeter. The other
electron is selected by requiring a cluster in the SpaCal
with an energy $E_2>4.2\GeV$. Data set IV ($185<\Wgp<305\GeV$)
requires both electrons to be detected as energy clusters in the SpaCal 
with energies
$E_{1,2}>4.2\GeV$ and the more energetic cluster to be above $6\GeV$.  
At least one electron must be in the acceptance region of the BST and
every electron in the BST acceptance region must be validated by a BST track
from the nominal interaction point.
This requirement rejects most of the non-resonant background from Compton
scattering. In both data
sets III and IV the energy in the SpaCal outside the selected electron
cluster(s) must be negligible. Further details of this analysis may be found
in~\cite{Janauschek:2004}.

In order to suppress background from proton dissociative or inelastic \jpsi
production, no additional tracks are allowed in the CTD or FTD and the selected events
are required to have no significant signals in the forward detectors (PRT, FMD
and LAr). The fraction of proton dissociation is further suppressed by
limiting $t$ to the range $|t|<1.2\GeVsq$, where elastic processes are
dominant. These requirements reject most of the proton dissociative background. 
The remaining fraction 
is 14\% on average, ranging from 8\% at $|t|\approx 0$ to  35\% at $|t|\approx
1.2\,\GeVsq$. It is corrected for using the MC simulation, which is
tuned to give a good description of the forward detectors. 
A further correction is applied to account for \psip
decays into \jpsi and neutral mesons. This correction is estimated to be 4\% for
data sets I and II and approximately 2\% in sets III and IV, where the
neutral decays are partly rejected by the cut on the energy in the SpaCal.

Triggers based on muon and track signatures from the decay leptons are used
for data sets I and II. For data set I a trigger signal is also derived from 
the scattered positron.
The triggers for data sets III and IV are based on signals due to the 
\jpsi decay electrons from the SpaCal and the CTD (set III) . 
In addition the triggers for data sets II-IV use second level triggers based 
on neural network algorithms.

\begin{figure}[p]
   \begin{center}
     \setlength{\unitlength}{1pt}
     \begin{picture}(1,1)(0,0)
       \thicklines
      \put(45.0,215.0){Data set I}  
     \end{picture}~
     \includegraphics[scale=0.95,bb=294 510 526 748,keepaspectratio]{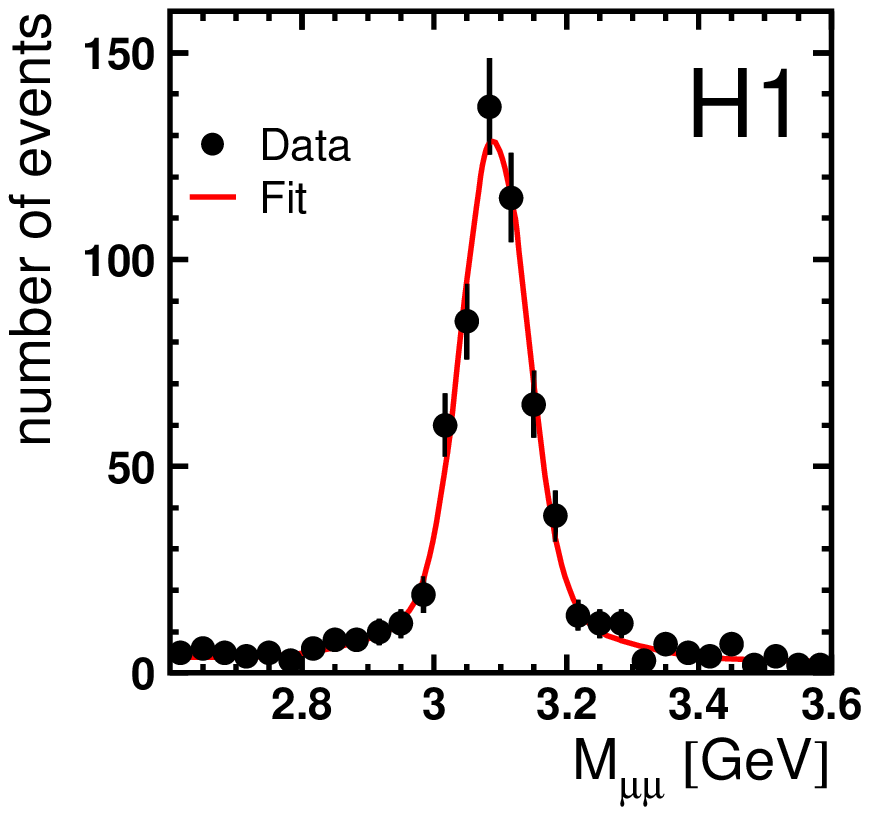}~
     \hspace*{0.2cm}
     \setlength{\unitlength}{1pt}
     \begin{picture}(1,1)(0,0)
       \thicklines
       \put(50.0,215.0){Data set II}
     \end{picture}~
     \includegraphics[scale=0.95,bb=294 510 526 748,keepaspectratio]{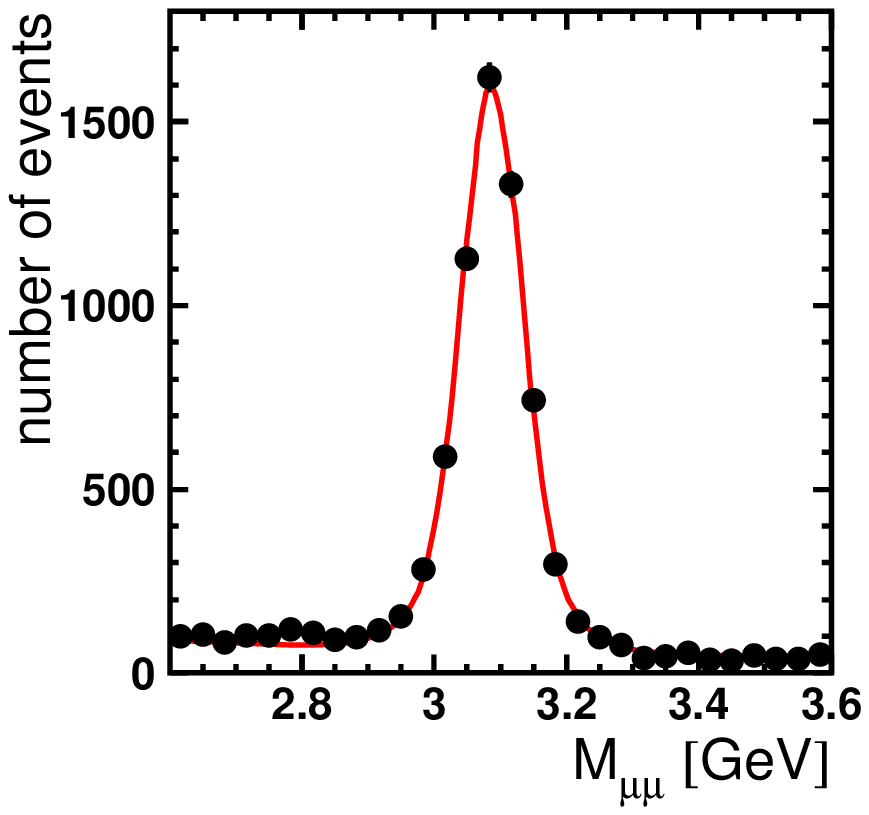}\\[1ex]
     \setlength{\unitlength}{1pt}
     \begin{picture}(1,1)(0,0)
       \thicklines
       \put(50.0,215.0){Data set III}
     \end{picture}~
     \includegraphics[scale=0.97,bb=294 510 526 748,width=0.485\textwidth,keepaspectratio]{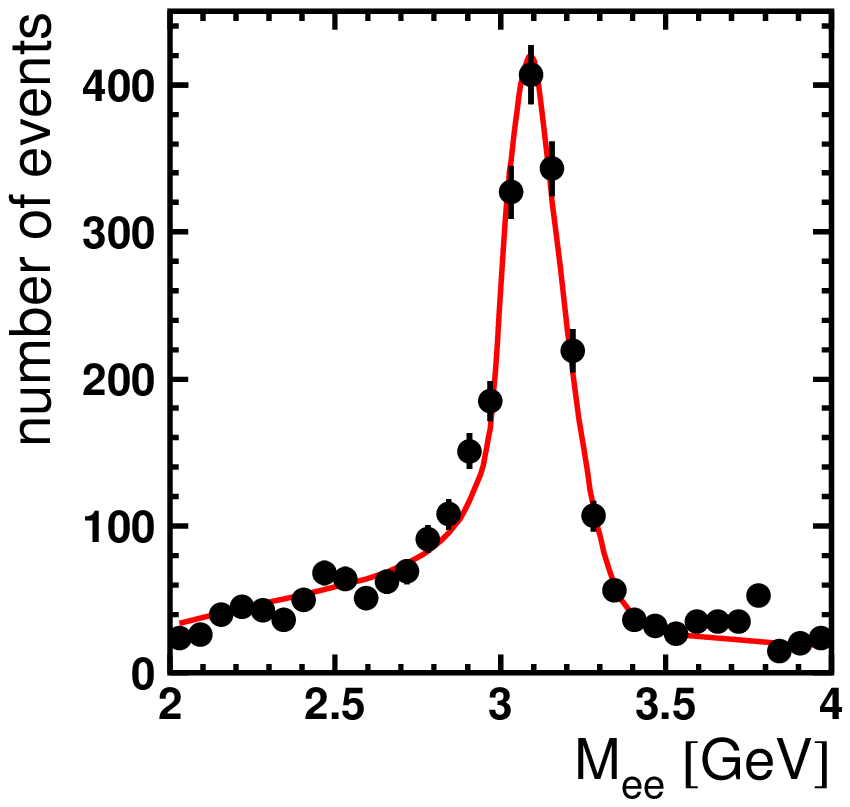}~ 
     \hspace*{0.2cm}
         \setlength{\unitlength}{1pt}
     \begin{picture}(1,1)(0,0)
       \thicklines
       \put(50.0,215.0){Data set IV}
     \end{picture}~
     \includegraphics[scale=0.97,bb=294 510 526 748,width=0.485\textwidth,keepaspectratio]{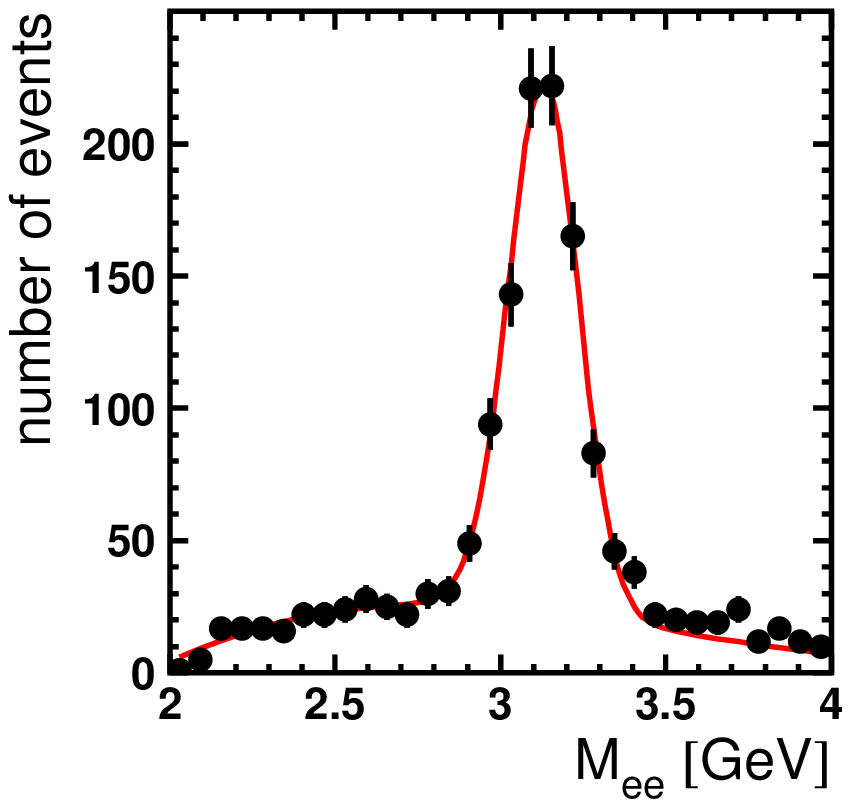}
     \caption{The dilepton invariant mass distributions (data and fits) in the four
       kinematic regions defined in table~\ref{tab:sel}.} 
     \label{fig:peaks}
   \end{center}
 \end{figure}

Figure~\ref{fig:peaks} shows the two-lepton invariant mass distributions for
the four data sets. The shapes of the \jpsi signal peaks reflect the usage of different 
detectors with different resolutions and a different response to electrons, muons and photons.
The signal of data set III shows a tail towards low masses due to radiative energy 
losses of the electron reconstructed in the tracking detector. In all data
sets the non resonant background below the \jpsi signal peaks
is dominated by $\gamma\gamma\to\ell^+\ell^-$, where one photon originates
from each of the positron and the proton. 
At high \Wgp a potential source of
background is Compton scattering $ep\to e\gp$ where the final state electron
and photon can form an invariant mass of the same order as the \jpsi mass. It is 
efficiently suppressed by the BST track requirements explained above.
In addition the BST tracks lead to an improved mass resolution in data set IV.

The number of \jpsi events is determined in each analysis bin by a fit of the
sum of a signal and a background function to the dilepton mass distribution. 
For data sets I and II ($\jpsi\to\mu^+\mu^-$) the signal shape is a Gaussian
function, and the background is fitted using a power law distribution.  For the
signal in data set III ($\jpsi\to e^+e^-$) the radiative tail
is taken into account by fitting a modified Gaussian distribution,
$f(\mee)\propto
\frac{1}{\sigma^\prime}\exp{(-(\mee-\mu)^2/(2\,\sigma^\prime)})$ 
where $\sigma^\prime=(\sigma+r\,(|\mee-\mu|-\mee+\mu))^2$. Here $\mu$ and
$\sigma$ denote the peak position and the standard deviation and 
$r$ parameterises the contribution of the radiative tail. In
data set IV ($\jpsi\to e^+e^-$) a single Gaussian function is adequate to describe the signal. In 
data sets III and IV the shape of the background is found to depend strongly
on $\Wgp$. The shapes are
reasonably well described by the predicted shapes of the Monte Carlo 
simulations LPAIR and COMPTON, which are therefore used in the fit.

\begin{figure}[p]
   \setlength{\unitlength}{1cm}
   \begin{picture}(1,1)(0,0)
     \thicklines
     \put(5.8,-0.6){\small Data set I}
     \put(5.8,-4.5){\small Data set II}
     \put(5.8,-8.3){\small Data set III} 
     \put(5.8,-12.3){\small Data set IV}
   \end{picture}

    \includegraphics[bb=53 216 303 454,width=0.24\textwidth,keepaspectratio]{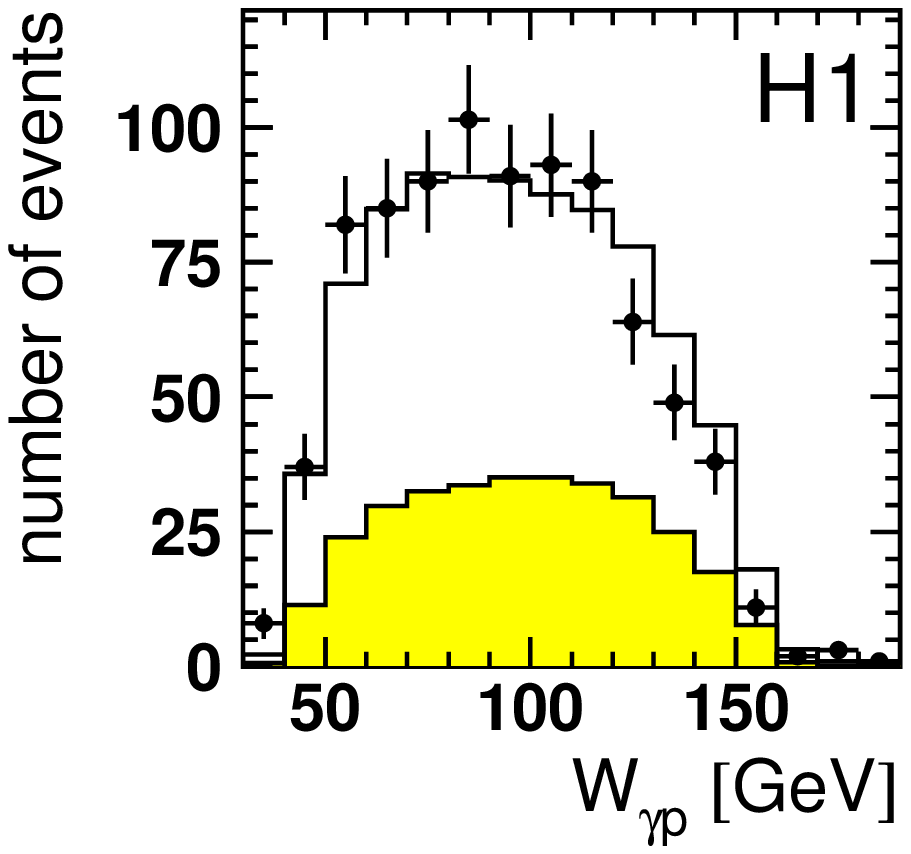}~
    \includegraphics[bb=50 216 303 454,width=0.24\textwidth,keepaspectratio]{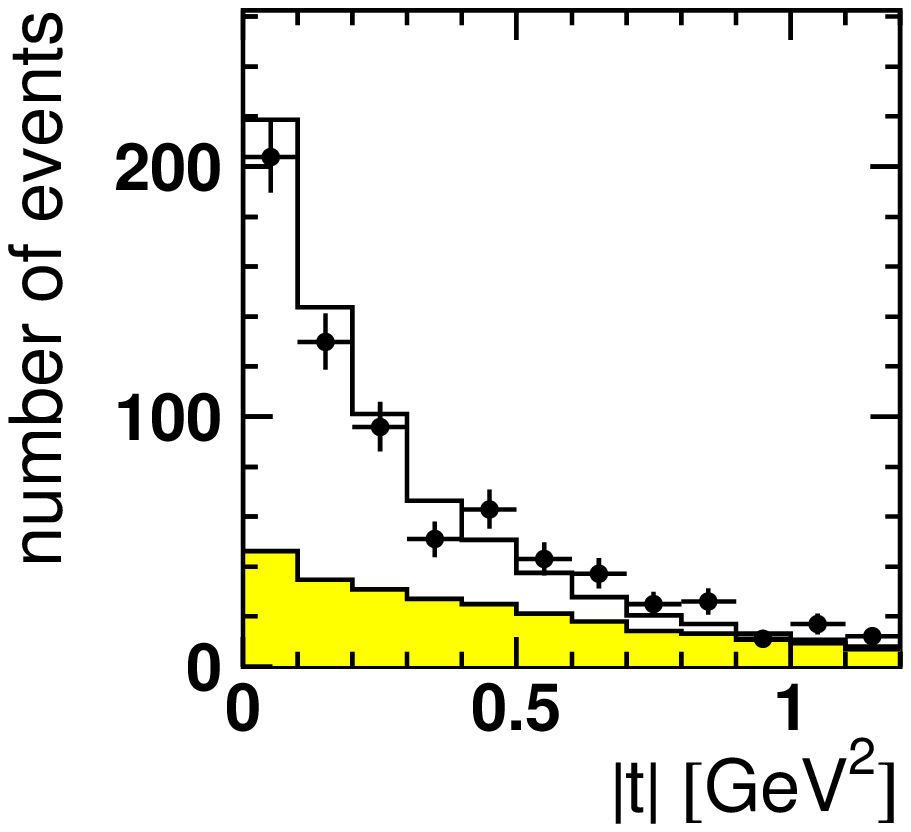}~
    \includegraphics[bb=50 216 303 454,width=0.24\textwidth,keepaspectratio]{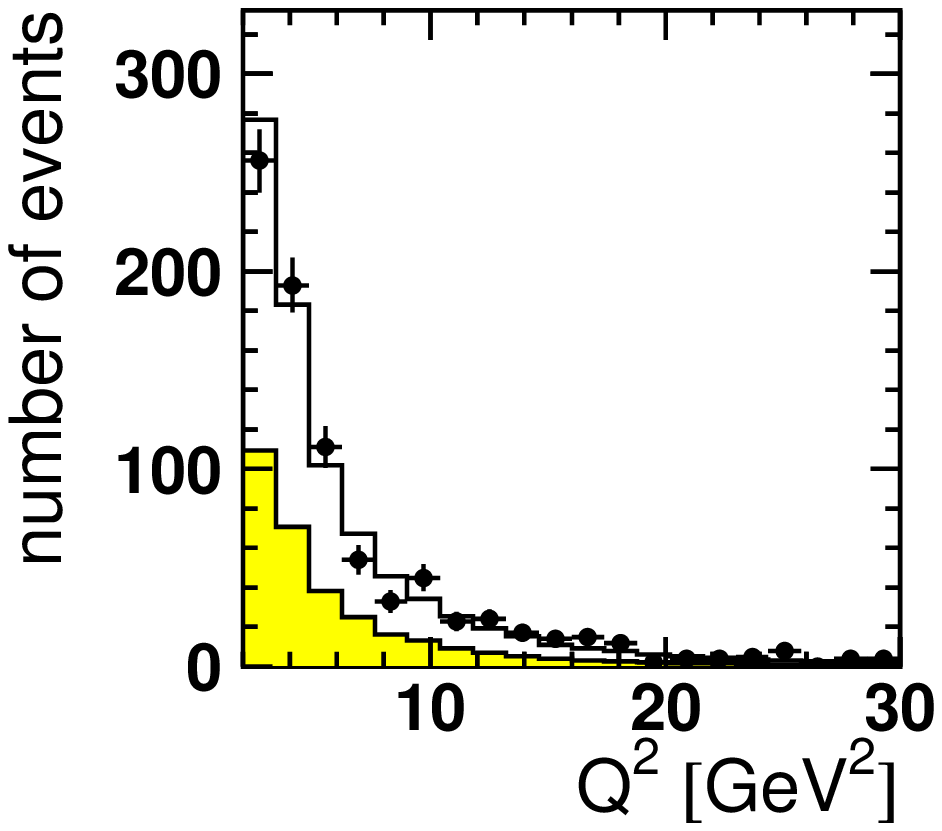}~
    \includegraphics[bb=50 216 303 454,width=0.24\textwidth,keepaspectratio]{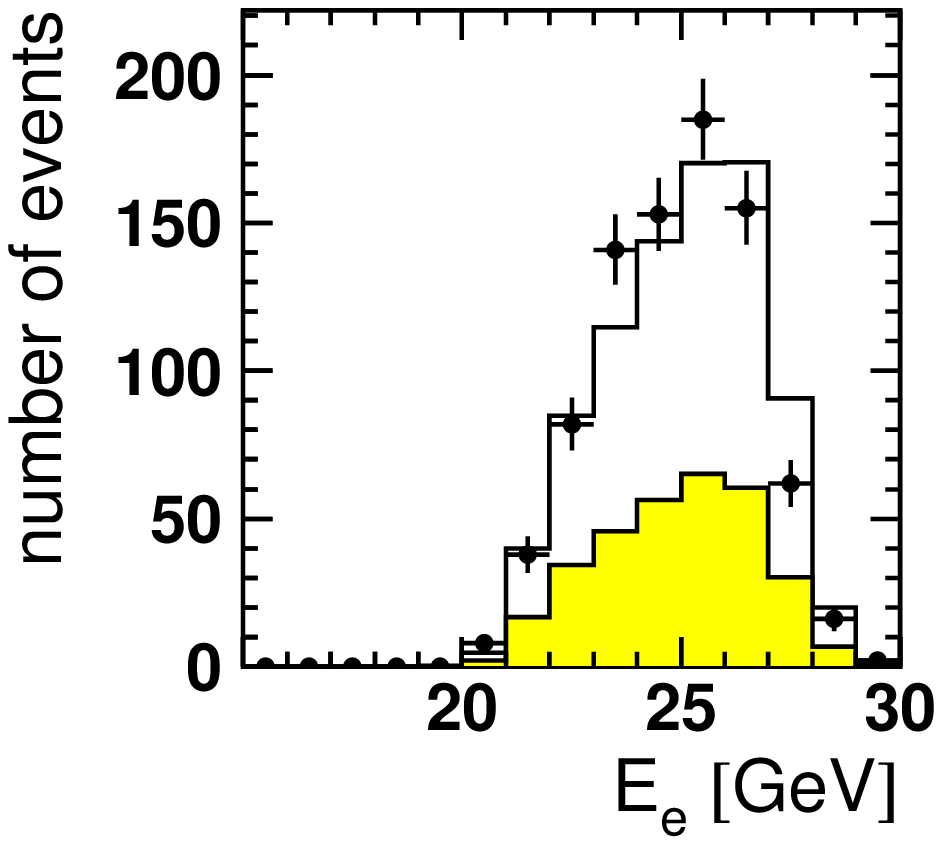}\\[1ex]
    \includegraphics[bb=55 216 303 454,width=0.24\textwidth,keepaspectratio]{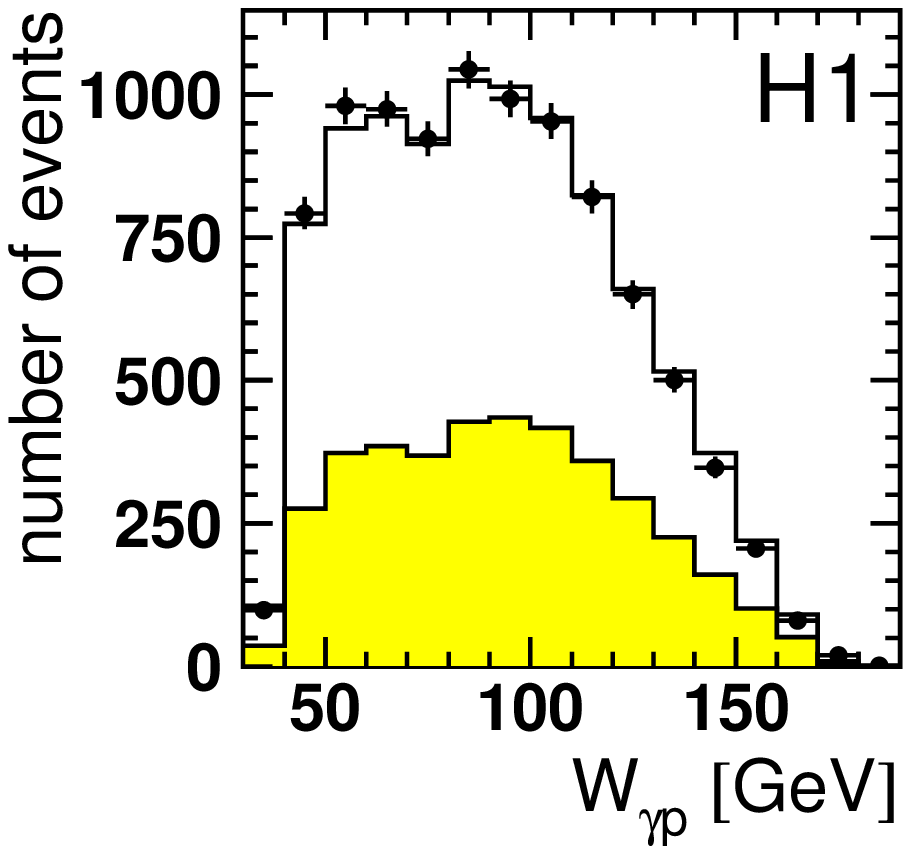}~
    \includegraphics[bb=55 216 303 454,width=0.24\textwidth,keepaspectratio]{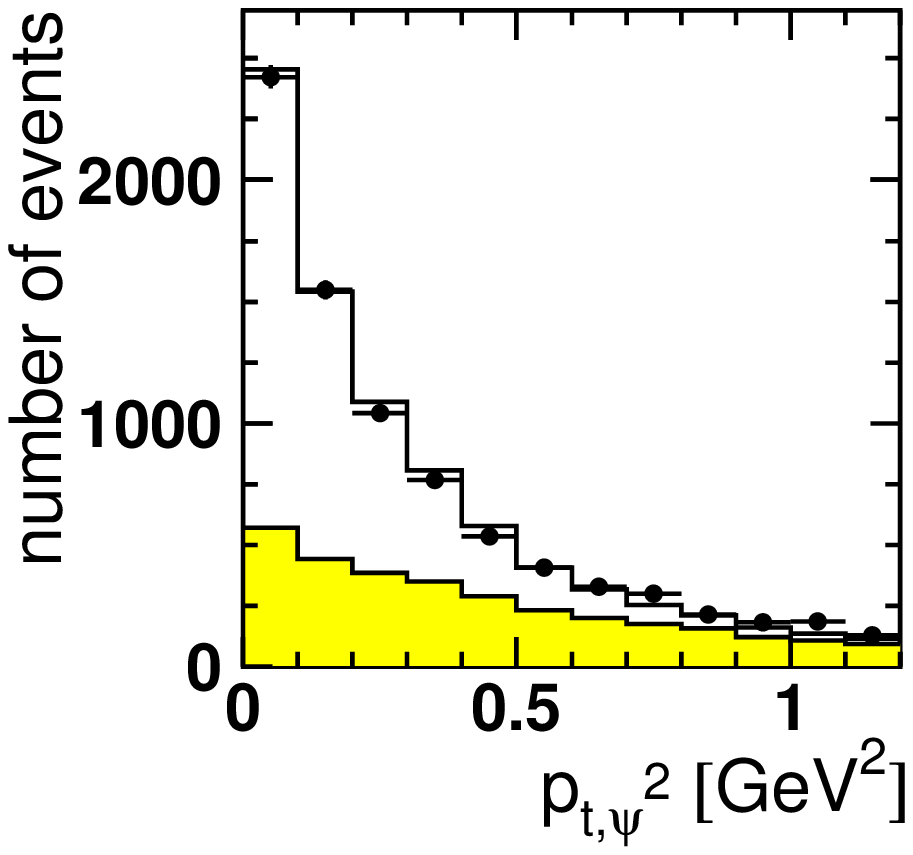}~
    \includegraphics[bb=55 216 303 454,width=0.24\textwidth,keepaspectratio]{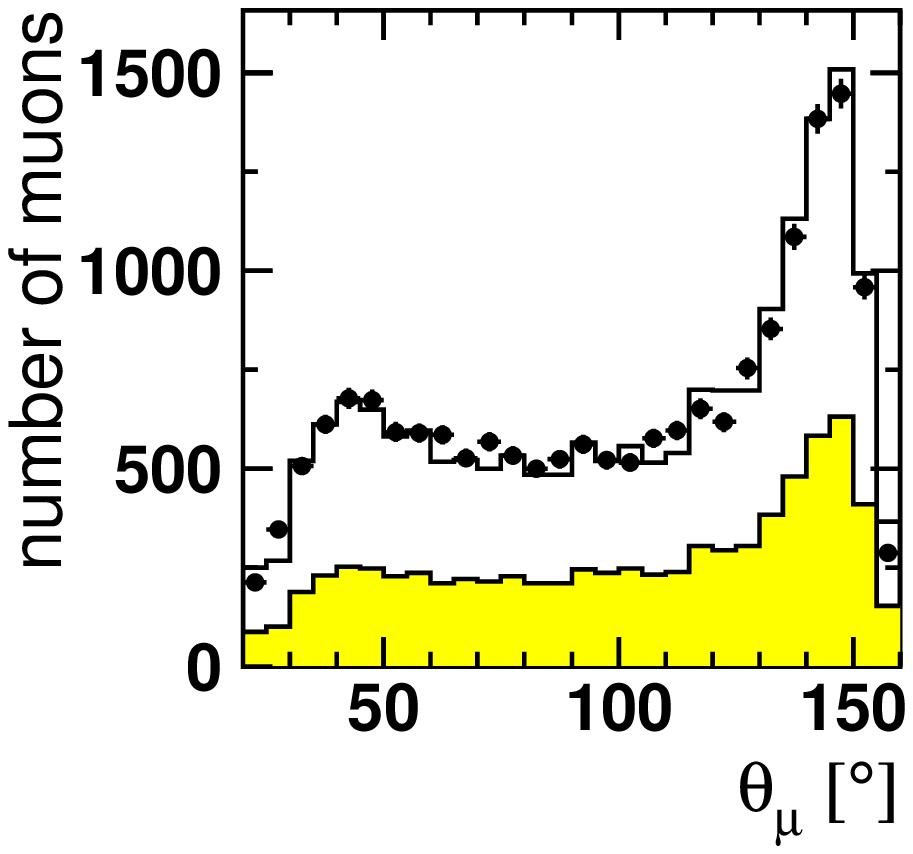}
 \includegraphics[bb=80 250 280 400,width=0.24\textwidth]{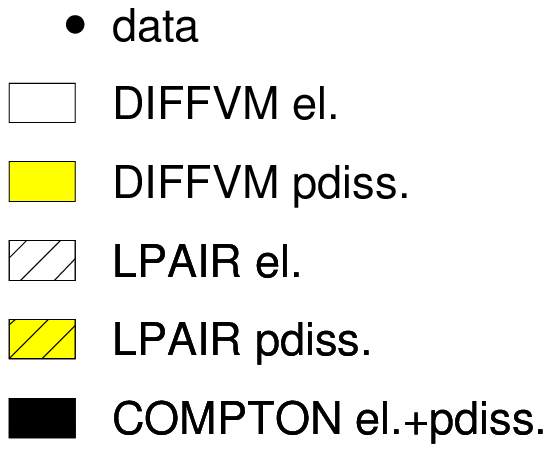}\\[3ex]
    \includegraphics[bb=55 214 295 430,width=0.235\textwidth,keepaspectratio]{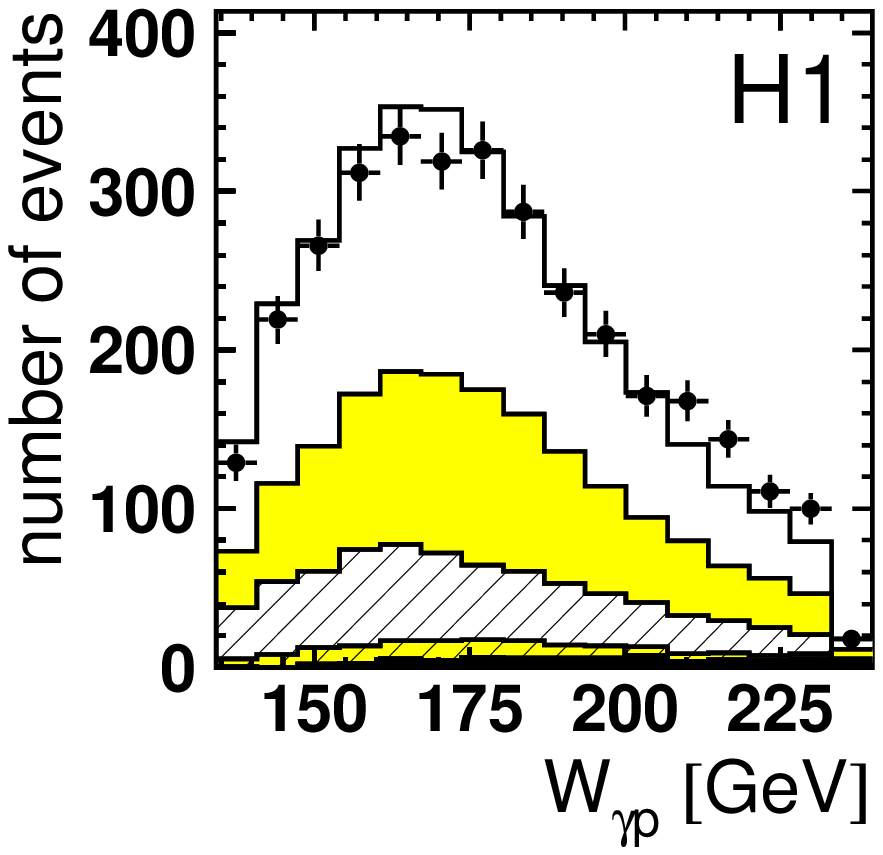}~
    \includegraphics[bb=55 214 295 430,width=0.235\textwidth,keepaspectratio]{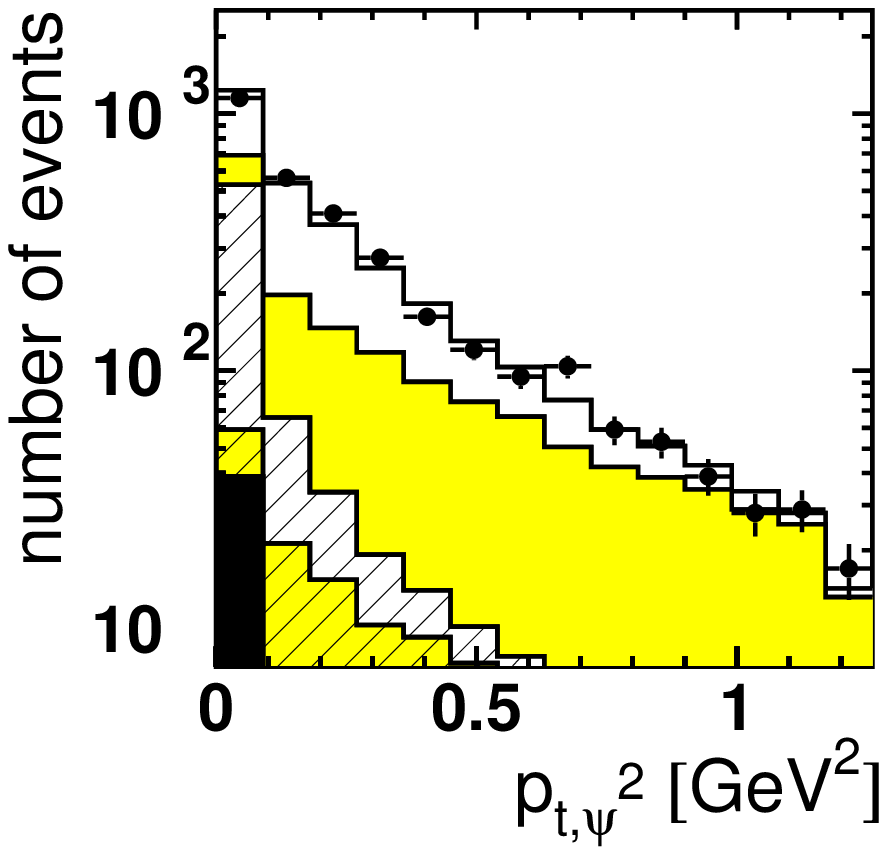}~
    \includegraphics[bb=55 214 295 430,width=0.235\textwidth,keepaspectratio]{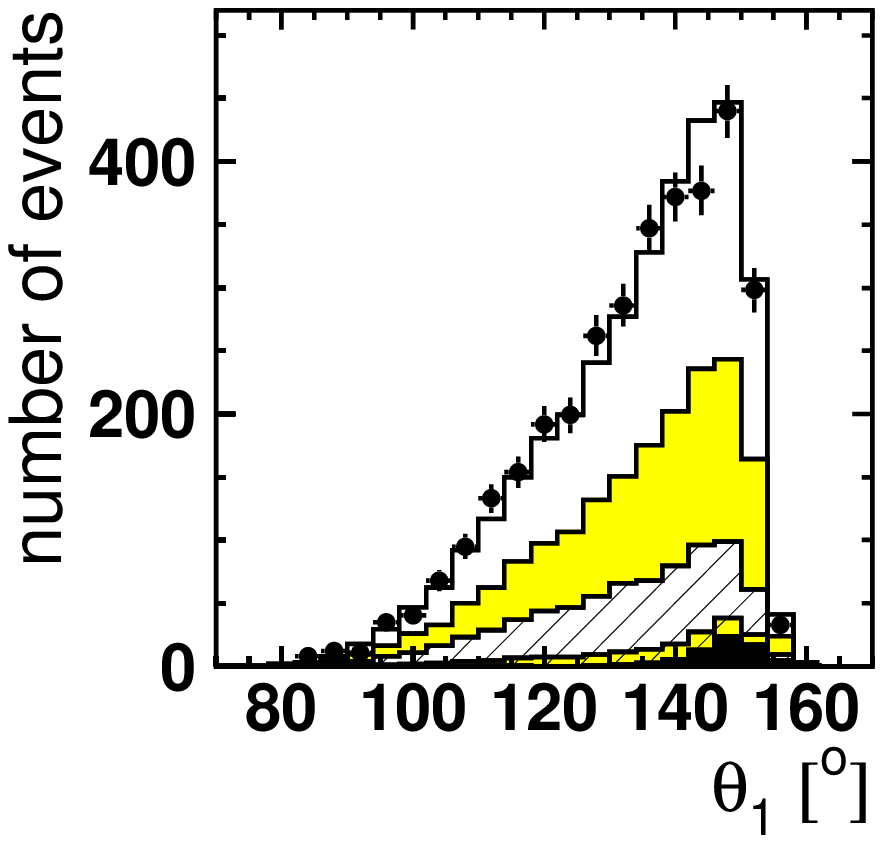}~
\includegraphics[bb=55 214 295 430,width=0.235\textwidth,keepaspectratio]{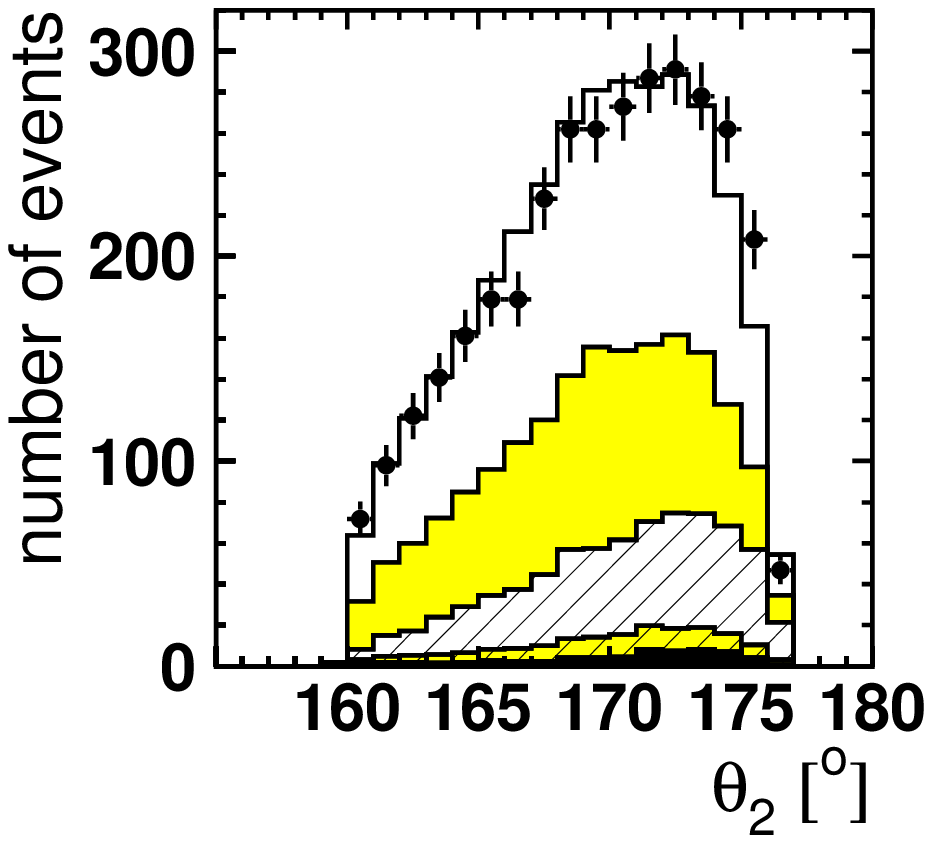}\\[3ex]
    \includegraphics[bb=45 214 295 430,width=0.235\textwidth,keepaspectratio]{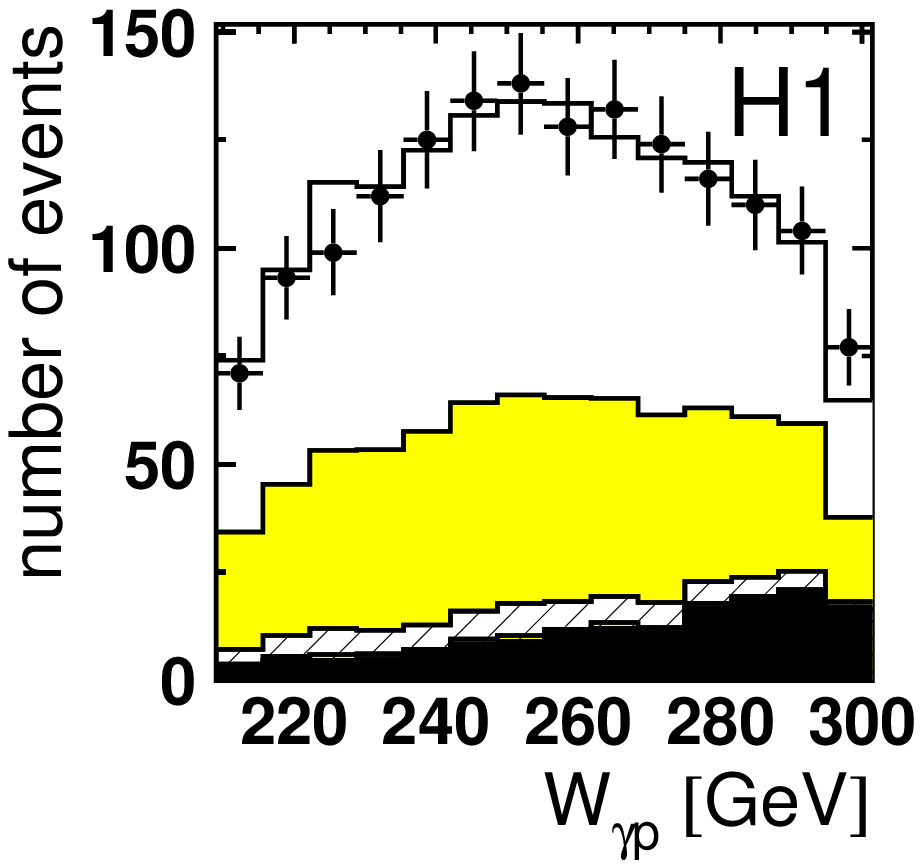}~
    \includegraphics[bb=45 214 295 430,width=0.235\textwidth,keepaspectratio]{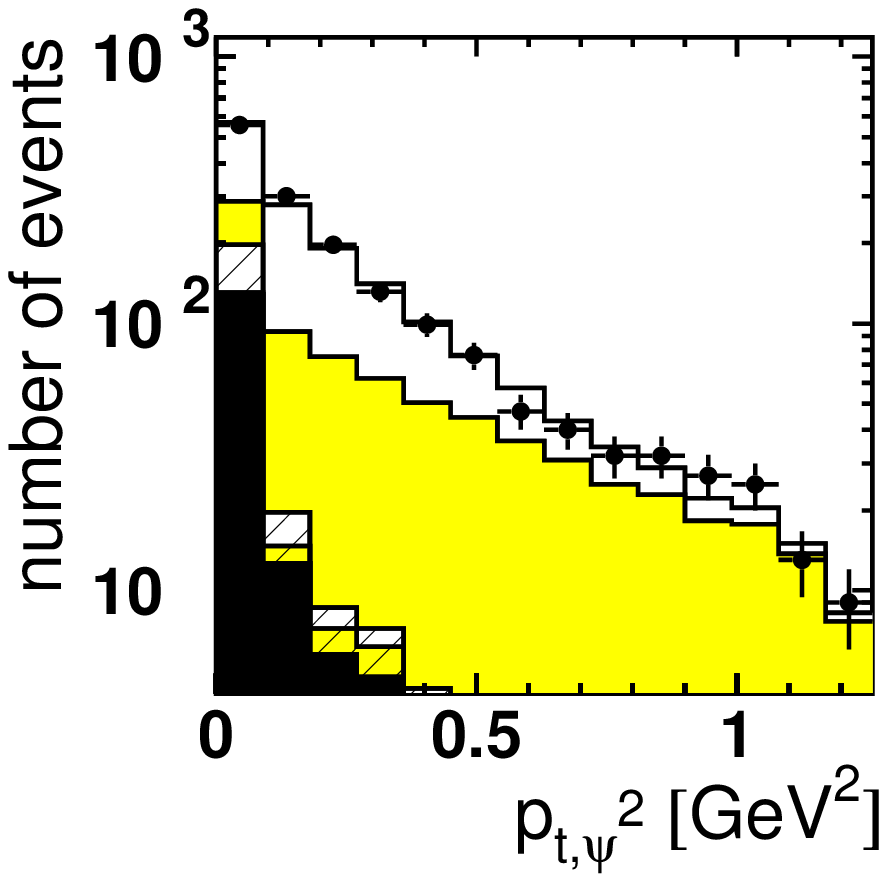}~
    \includegraphics[bb=45 214 295 430,width=0.235\textwidth,keepaspectratio]{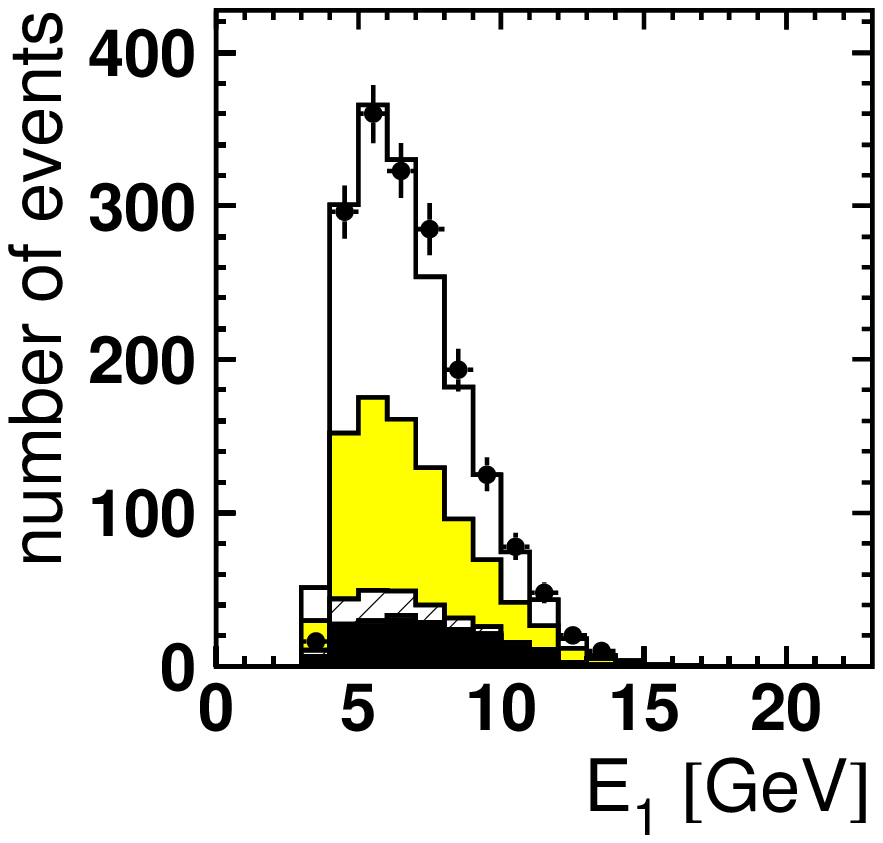}~
    \includegraphics[bb=45 214 295 430,width=0.235\textwidth,keepaspectratio]{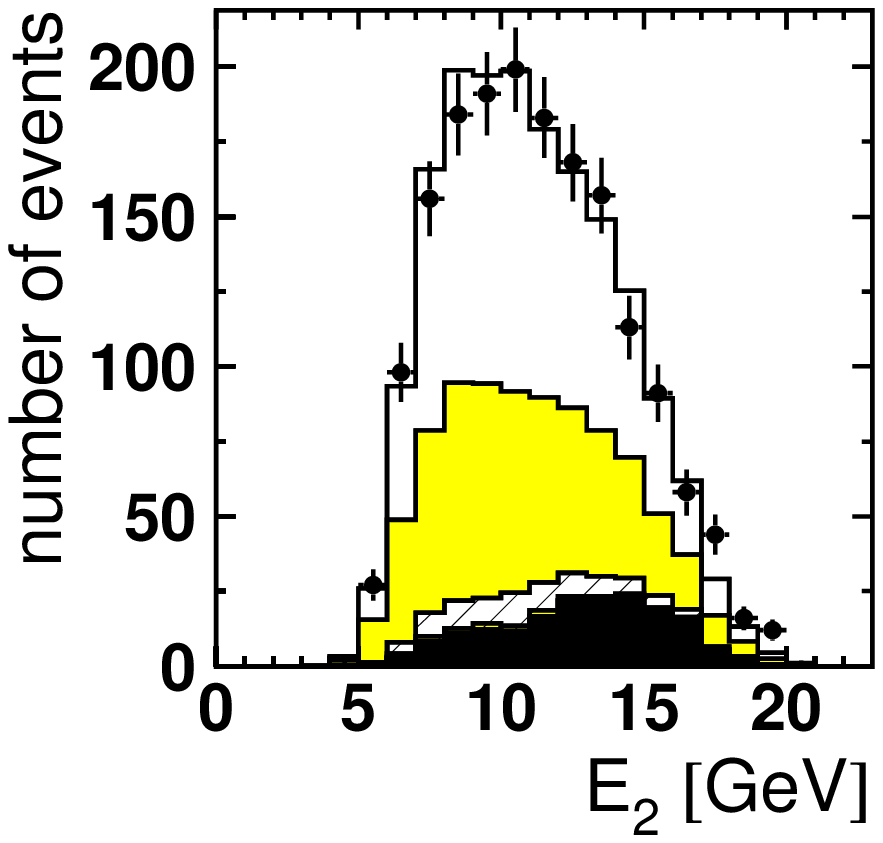}
 \begin{center}
   \caption{Observed event distributions 
      for the four data sets (points) defined in table~\ref{tab:sel}, omitting the
      forward detector cuts against proton dissociative events.   
      The first two rows correspond to the selected 
      $\jpsi\to\mu^+\mu^-$ candidates where the small non-resonant background has
      been subtracted. The data are shown with the elastic (signal) simulations (DIFFVM el.,
      white area) and proton dissociation MC (DIFFVM pdiss., shaded
      area).  Rows three and four correspond to $\jpsi\to e^+e^-$
      candidates, where
      the non-resonant background is not subtracted. Here, in addition to the
      elastic and proton dissociative  
      \jpsi\ simulations the contributions from $\gamma \gamma \rightarrow
      e^+e^-$ (LPAIR) and Compton scattering (COMPTON)
      are shown. The normalisations are obtained from a fit
      of the overall mass peak of each data set. The variables \Wgp, $Q^2$, $t$
      and $p^2_{t,\psi}$ are defined in the text.
      $\theta_\mu$ refers to the decay muons of data set II. In row three
      $\theta_{1}$ and $\theta_{2}$ refer to the decay electrons which are selected in
      different polar angular regions. In row four $E_{1}$ and $E_2$ refer to the
      energies of the decay electrons.}
    \label{fig:control}
  \end{center}
\end{figure}

Data and MC simulation are compared in figure~\ref{fig:control}. Each row
corresponds to one of the four data sets. The selected events from a mass
window around the nominal \jpsi mass ($\pm 0.2\GeV$ in data sets I and II,
$\pm0.3\GeV$ in data set IV and $2.6< M_{ee}<3.4\GeV$ for data set III) are shown
before applying the cuts on the forward detectors.  For data sets I
and II 
the non-resonant background, which is small in this kinematic region, has been
subtracted and the data are described by
a combination of DIFFVM for elastic and proton dissociative \jpsi production.
For data sets III and IV the selected events are shown but without subtracting
the non resonant background. The data are seen to be reasonably well described by
a sum of simulations for elastic and proton dissociative \jpsi production (DIFFVM), $\gamma\gamma\to
e^+e^-$ (LPAIR) and $ep\to e\gp$ (COMPTON).

\subsection{Systematic Uncertainties}
\vspace*{-1.5ex}
The systematic uncertainties on the cross sections are dominated by detector effects
which are not perfectly modelled in the Monte Carlo simulation.
Most uncertainties are obtained by comparisons of data with simulation after
tuning the detector simulation with independent data sets. The uncertainties on the measured 
cross sections are then estimated by variations of the simulation. In the following
the main sources of the uncertainties are summarised and typical values are 
given for the uncertainty on the total cross section.

\begin{itemize}
\item The uncertainty due to the track reconstruction efficiency in the CTD
  is $1\%$ per track.  The track information from the BST has two
  sources of uncertainty: coherent signal losses (3.0\%) and track
  reconstruction efficiency (1.5\%).
\item The uncertainty on the lepton identification efficiency leads to a cross
  section uncertainty of $1.5\%$ for
  muons and 2\% for electrons measured in the CTD. 
The uncertainty on the energy measurement of the decay electrons 
in the backward calorimeter is estimated to vary
  linearly from 2.7\% at 3~GeV to 0.5\% at 27.5~GeV from an analysis of Compton 
scattering~\cite{Aktas:2004ek}. The resulting
  uncertainties on the cross sections vary from 1\% to 7\%, depending on
  $\Wgsp$.  A small additional uncertainty for data set IV arises due to an
  uncertainty of 0.3 mrad in the reconstruction of the polar angle of the
  decay electrons in the BST, leading to a \Wgp dependent cross section uncertainty of
  $1-3$\%.
\item The uncertainties of the trigger efficiencies 
are determined to be
  $1.6\%$, $5\%$, $6.5\%$ and $5\%$ for data sets I to IV, respectively.
\item The separation of elastic events from proton dissociation
  leads to a systematic uncertainty of $4-6\%$ due to the modelling of the
  response of the forward detectors, with a small dependence on
  \Wgp and $|t|$. The error due to the simulation of the 
   dependence of the cross section on $M_Y$ was found to be negligible by comparison. 
 \item The uncertainty in the modelling of the $z$ position of the interaction region affects 
   the  \Wgp dependence of the cross section and is found to be $1\%$ on
   average for data sets I and II, $0.5-2.6$\% for III and $2.0$\% for IV.
 \item Varying the methods of determination of the number of signal events
        (e.g. by using a counting method instead of fits, or by changing the
        shapes of the background functions), results
   in a $1\%$ uncertainty for data sets I and II ($\mu^+\mu^-$).  For data sets III
   and IV (electrons) an uncertainty between 3\% and 6\% is estimated, which
   is due to the uncertainties in the signal and background shapes.
\item For the electroproduction sample, an additional uncertainty of $4\%$ is
  estimated which covers uncertainties in the reconstruction of the
  energy and angle of the scattered positron.
\item Other sources of systematic uncertainties are the luminosity
  measurement (1.5\%), the $J/\psi$ branching ratio (1.7\%) and the \psip
  background (0.5\% for sets I and II, 1.5\% on average for III and IV).
\end{itemize}

The systematic uncertainties are calculated in each analysis bin and
the total uncertainty is obtained by adding all individual
contributions in quadrature. The average values for the total systematic
uncertainties on the cross sections are $8\%$, $9\%$, $10\%$ and $11\%$ for the data sets I to
IV, respectively. The correlated part of the error, which affects all bins
equally,  is estimated to be approximately 5\% and is 
not included in subsequent fits unless mentioned otherwise.

\section{Results}
Cross sections are calculated for the individual data sets I--IV using 
the number $N$ of selected events after correcting for
non resonant, proton dissociative and \psip backgrounds as described in the
previous section. 
The efficiencies $A$ for the event selection are in general determined from the MC
simulation. In the equivalent photon approximation the $\gamma p$ cross 
section is given by:

\vspace*{-3ex}
\begin{equation}
  \label{eq:xsec}
\sigma(\gamma p\rightarrow \jpsi p) = \frac{N}
    {A\,\cdot B\!R\cdot\mathcal{L}\cdot \Phi_{\gamma}}. 
\end{equation}

Here $\Phi_{\gamma}$~\cite{Frixione:1993yw} denotes the photon flux in the 
$\Qsq$ and \Wgp range considered, $\mathcal{L}$ the integrated luminosity
and $B\!R$ the branching ratio for the decay of the
\jpsi mesons\footnote{Branching fractions $(5.88\pm0.10)\%$ and 
$(5.93\pm0.10)\%$~\cite{Eidelman:2004wy} are used for $\jpsi\rightarrow 
\mu^+\mu^-$ and $e^+e^-$, respectively.}.

Note that this cross section corresponds to  $\sigma_{\gsp} = \sigma^T_{\gsp} +
\varepsilon\sigma^L_{\gsp}$, where $\sigma^T_{\gsp}$ and $\sigma^L_{\gsp}$ are
the cross sections for transversely and longitudinally polarised photons, respectively, and
$\varepsilon$ is the polarisation parameter of the virtual photon\footnote{The
  present results will be compared with results from the ZEUS collaboration~\cite{Chekanov:2004mw}, where
  $\sigma^T_{\gsp} + \sigma^L_{\gsp}$ is extracted. In the present kinematic region
  the difference is however small compared with the measurement errors.}. The parameter 
$\varepsilon$ depends only on the kinematics, $\varepsilon=(1-y)/(1-y+\frac{1}{2}y^2)$.
In the kinematic range of the present analysis $\varepsilon$ is generally
above $0.95$ with $\langle\varepsilon\rangle=0.993$.
Cross sections are
given at `bin centres', $\langle\Wgp\rangle$, $\langle\Qsq\rangle$ and 
$\langle t\rangle$,
which are determined taking into account the measured \Wgp, $\Qsq$ and $t$ dependences.

\subsection{\boldmath\Qsq Dependence}
\begin{figure}[p]
  \begin{center}
    \setlength{\unitlength}{1pt}
    \begin{picture}(1,1)(0,0)
    \end{picture}
    \includegraphics[bb=150 250 520 740,scale=0.99,keepaspectratio]{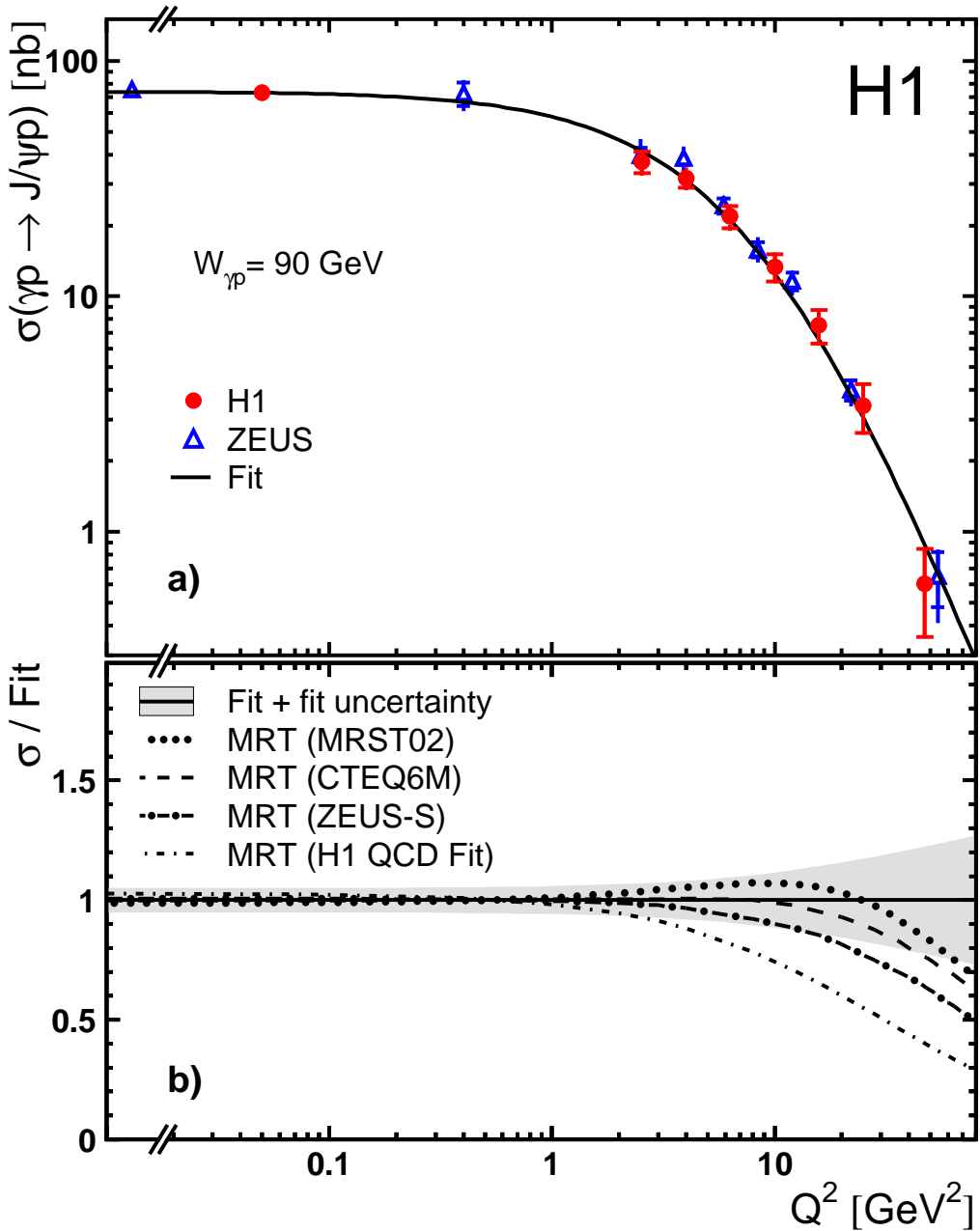}
    \caption{a) Total cross section for elastic \jpsi production as a
      function of $Q^2$ in the range $|t|<1.2\GeVsq$ at $\Wgp=90\GeV$. The
      inner error bars show the statistical errors, while the outer error bars
      show the statistical and systematic uncertainties added in quadrature.
      The solid line is a fit to the H1 data of the form
      $\sigma_{\gsp}\propto (M_\psi^2+Q^2)^{-n}$. Data from the ZEUS
      experiment~\cite{Chekanov:2002xi,Chekanov:2004mw} are also shown. 
 b) The ratio of the MRT calculations~\cite{Martin:1999wb} to the fit from a).
      The MRT QCD predictions are based on different gluon
      distributions~\cite{Pumplin:2002vw,Martin:2001es,Adloff:2003uh,
      Chekanov:2002pv}. 
      The curves are individually normalised to the measurements across the
      complete \Qsq range yielding factors between 1.5 and 2.8. The shaded 
      band represents the uncertainty of the fit result.
      }
    \label{fig:qxsec}
  \end{center}
\end{figure}

\vspace*{-1.5ex}
The cross sections for elastic \jpsi production as a
function of \Qsq at $\Wgp = 90\,\GeV$ are listed in 
table~\ref{tab:qxsec} and shown in figure~\ref{fig:qxsec}a. The photoproduction
point is obtained from the fit described in the next section. 

A phenomenological fit of the form
$\sigma_{\gsp}\propto(M_\psi^2+\Qsq)^{-n}$ to the H1 data yields a value of
$n=\qexp$. This result confirms, with smaller errors, the \Qsq dependence
observed previously by H1~\cite{Adloff:1999zs}. The quality of the fit is good
($\chi^2$/ndf$=0.5$). 
Recent results from the ZEUS
collaboration~\cite{Chekanov:2002xi,Chekanov:2004mw} are also shown in
figure~\ref{fig:qxsec}a, which agree well with the present data in the entire
range of $Q^2$.

In figure~\ref{fig:qxsec}b the pQCD calculations `MRT' of Martin et 
al.~\cite{Martin:1999wb} 
are compared with the fit result quoted above. 
Results with four different gluon distributions
(CTEQ6M~\cite{Pumplin:2002vw}, MRST02~\cite{Martin:2001es},
 H1QCDFIT~\cite{Adloff:2003uh} and ZEUS-S~\cite{Chekanov:2002pv}) derived from 
global fits to current inclusive $F_2$ measurements and other data are shown. 
A normalisation factor is determined individually for each
prediction by comparing with the data across the complete \Qsq\ range. 
The different factors, which are mainly given by the photoproduction measurement,
 are between $1.5$ and $2.8$.
The theoretical predictions of the shape of the \Qsq\ dependence are
consistent with the fit to the data within the experimental uncertainties, 
which are shown as 
a grey band in figure~\ref{fig:qxsec}b.

\subsection{\boldmath\Wgp Dependence}\label{wgpdep}
\begin{figure}[p]
  \begin{center}  
\includegraphics[scale=0.85,bb=62 492 521 748,keepaspectratio]
                                   {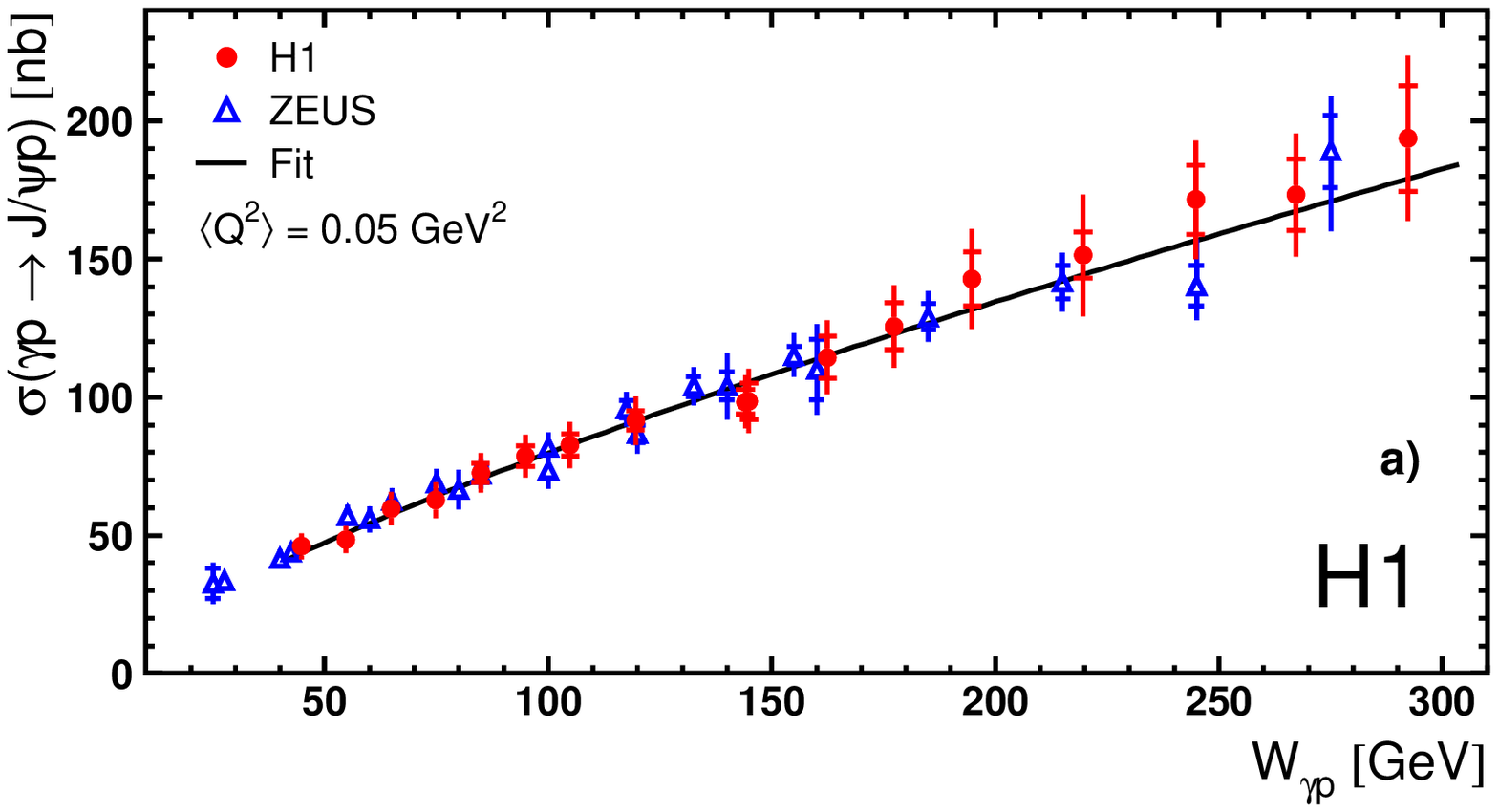}\\[1ex]
\includegraphics[scale=0.85,bb=62 492 521 748,keepaspectratio]
                                      {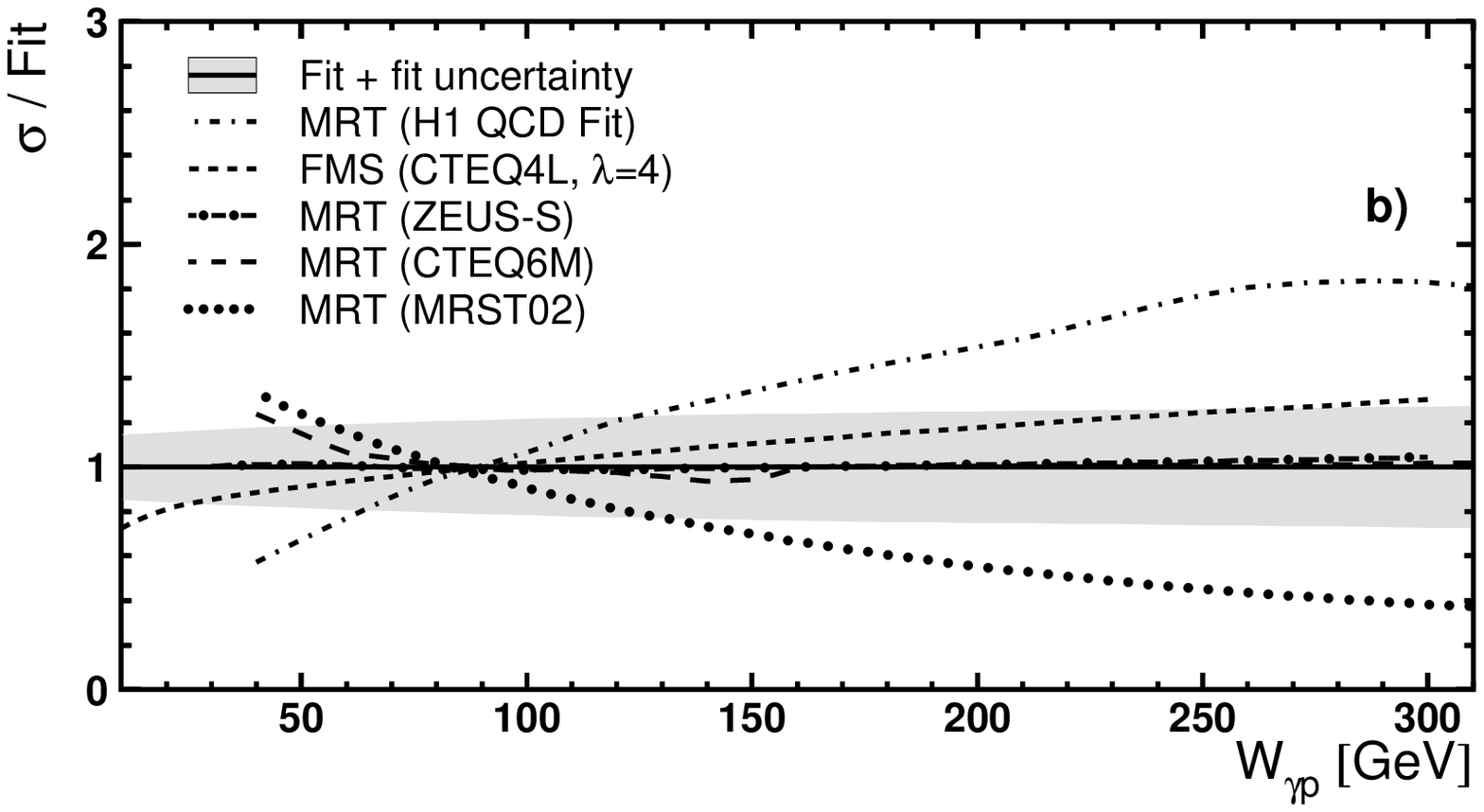}
\caption{a) Total cross sections for elastic \jpsi production as a
     function of \Wgp in the range $|t|<1.2\GeVsq$ in photoproduction.
     The inner error bars show the statistical errors, while the outer error bars
     show the statistical and systematic uncertainties added in quadrature.
     The solid line shows a fit to the H1 data of the form
     $\sigma\propto\Wgp^{\delta}$. 
     Results from the ZEUS experiment~\cite{Chekanov:2002xi} in a similar
     kinematic range are also shown. b) The ratio of theoretical predictions to the fit to 
            the H1 data in a). The shaded band represents the uncertainties of
     the fit result.
     Predictions from MRT QCD calculations~\cite{Martin:1999wb} and a dipole
     model (FMS, ~\cite{Frankfurt:2000ez}) based on different gluon
     distributions~\cite{Pumplin:2002vw,Martin:2001es,Adloff:2003uh,Chekanov:2002pv,
      Lai:1996mg}  are shown. For the MRT curves the normalisation factors determined from the 
      \Qsq distributions are used. The FMS prediction is normalised to the fit
     result at $\Wgp=90\GeV$.}
    \label{fig:wxsec}
  \end{center}
\end{figure}

\begin{figure}[p]
  \begin{center}
    \includegraphics[scale=0.99,bb=294 510 521 736,width=0.5\textwidth]
                                   {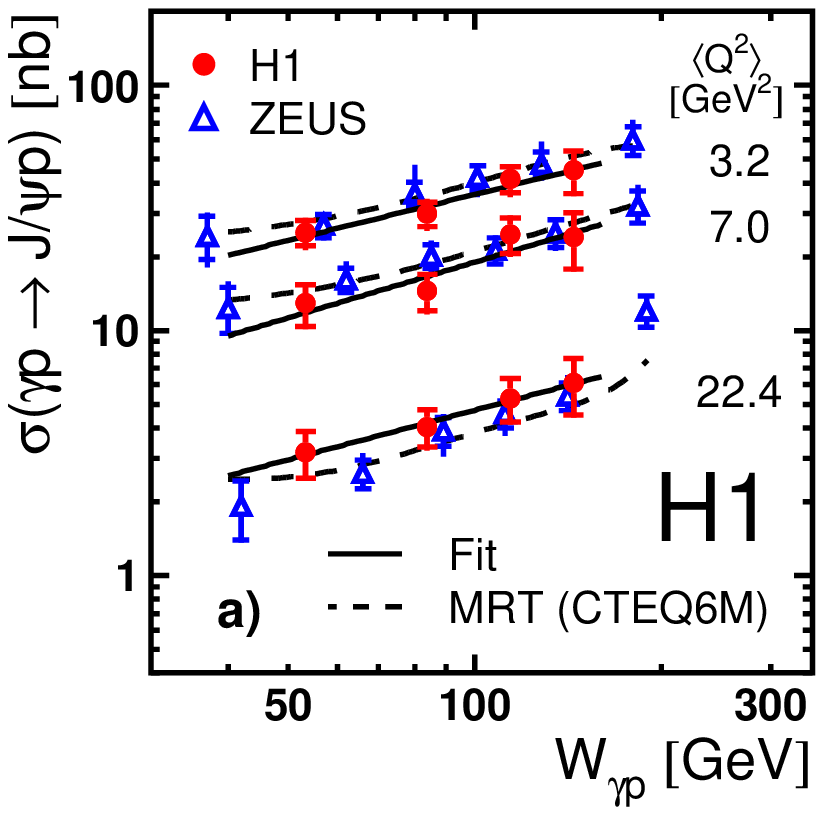}~ 
    \includegraphics[scale=0.99,bb=294 510 521 736,keepaspectratio]
         {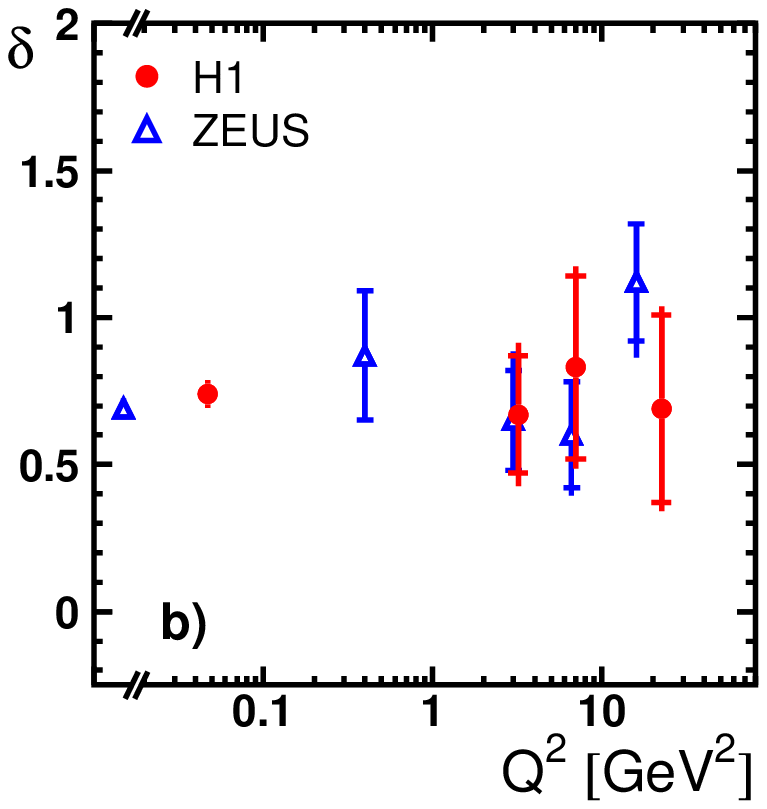}
   \caption{a) Total cross sections for elastic \jpsi production as a
     function of \Wgp in the range $|t|<1.2\GeVsq$ in electroproduction 
     in three bins of $Q^2$. $\langle\Qsq\rangle$
     indicates the bin centre value in the \Qsq range considered. The solid
     lines show fits to the H1 data of the form
     $\sigma\propto\Wgp^{\delta}$. The dashed curves show the MRT QCD
     prediction 
     based on the gluon distribution CTEQ6M~\cite{Pumplin:2002vw} 
     with the normalisation factors from the fit to the \Qsq distribution.
     Results from the ZEUS experiment~\cite{Chekanov:2004mw} in a similar
     kinematic range are also shown. They have been scaled to the
     given $\langle\Qsq\rangle$ values using the \Qsq dependence measured by
     ZEUS.
     b)~The fit parameter $\delta$ as a function of
     $Q^2$. The inner error bars show the statistical error, while the outer
     error bars show the statistical and systematic uncertainties added in
     quadrature.}
    \label{fig:wxsec_dis}
  \end{center}
\end{figure}

\vspace*{-1.5ex}
The \gp cross section for elastic \jpsi production is
presented as a function of \Wgp in figures~\ref{fig:wxsec}a and ~\ref{fig:wxsec_dis}a 
and in tables~\ref{tab:wxsec_php} and \ref{tab:wxsec_dis} for photoproduction and
electroproduction, respectively. 


In figure~\ref{fig:wxsec}a the photoproduction data are shown with the result
of a fit of the form $\sigma_{\gp}\propto\Wgp^\delta$.
Separate relative normalisation factors for the three data sets 
are additional fit parameters which take into account the correlated systematic
uncertainties. The fit yields a value of $\delta=\wexpphp$. The first error is
obtained using only the statistical uncertainties in the fit while the second one
reflects the systematic uncertainties.
The fit result is in agreement with our previous 
result~\cite{Adloff:2000vm}. Similar data from
the ZEUS collaboration~\cite{Chekanov:2002xi} (also shown in
figure~\ref{fig:wxsec}a) agree well with the present data.
 
A comparison with theoretical predictions is shown in figure~\ref{fig:wxsec}b, 
where the ratio of theory to the fit result is shown. The uncertainty of the
fit result is indicated by the grey band. 
The MRT predictions are normalised using the factors obtained from the 
$Q^2$ distributions. 
The same four gluon distributions are used to calculate the
respective unintegrated  skewed gluon distributions which are required by MRT. 
The \Wgp dependence is observed to be quite sensitive to the shape of
the gluon distribution\footnote{For a detailed discussion of the sensitivity
and the uncertainties of the model assumptions see \cite{teubnerdis}.}. 
While the results based on the gluon
distributions CTEQ6M and ZEUS-S describe the shape of
the data well, the gluon distribution from the H1 fit to inclusive data leads to a
steeper \Wgp dependence and the one from MRST02 to a flatter \Wgp dependence
than is observed. The dipole model
result~FMS~\cite{Frankfurt:2000ez} based on the CTEQ4L~\cite{Lai:1996mg}
gluon density is somewhat too steep.  Note, however, that these observations
are based on the central values of the respective gluon distributions and do
not take into account their uncertainties.
The kinematic range used in the MRT calculations extends to lower values
of Bjorken $x$ and \Qsq than was available in the inclusive data used for 
the determination of the gluon densities and an extrapolation to very low
values of \Qsq is performed.  

In figure~\ref{fig:wxsec_dis}a the electroproduction cross section is shown
in three bins of \Qsq ($2<\Qsq<5\GeVsq$, $5<\Qsq<10\GeVsq$ and
$10<\Qsq<80\GeVsq$). Data from the ZEUS experiment~\cite{Chekanov:2004mw}, 
which are shifted to the present bin centres using the \Qsq dependence
measured by ZEUS, are in agreement. In figure~\ref{fig:wxsec_dis}a the results 
from MRT based on the gluon density CTEQ6M using the same normalisation factor 
as above are also shown and give a reasonable description of the data. 

The \Wgp dependence is found to be similar to that
obtained in photoproduction. When parameterised in the form $\Wgp^\delta$,
the fits to the H1 electroproduction data yield $\delta$ values which are compatible 
with photoproduction within the rather large experimental errors (see table \ref{tab:bq}).
The fitted values for $\delta$ describing the \Wgp dependence of elastic
\jpsi production from this analysis and from
\cite{Chekanov:2002xi,Chekanov:2004mw} are displayed in
figure~\ref{fig:wxsec_dis}b as a function of $\Qsq$. Within the present
experimental accuracy no dependence on \Qsq is observed.

\subsection{Differential Cross Sections d\boldmath$\sigma/\mbox{d}t$}
\begin{figure}[p]
  \begin{center} 
\setlength{\unitlength}{1cm}
     \begin{picture}(1,1)(0,0)
       \thicklines
  \end{picture}

\includegraphics[scale=0.99,bb=60 408  530 740,keepaspectratio]{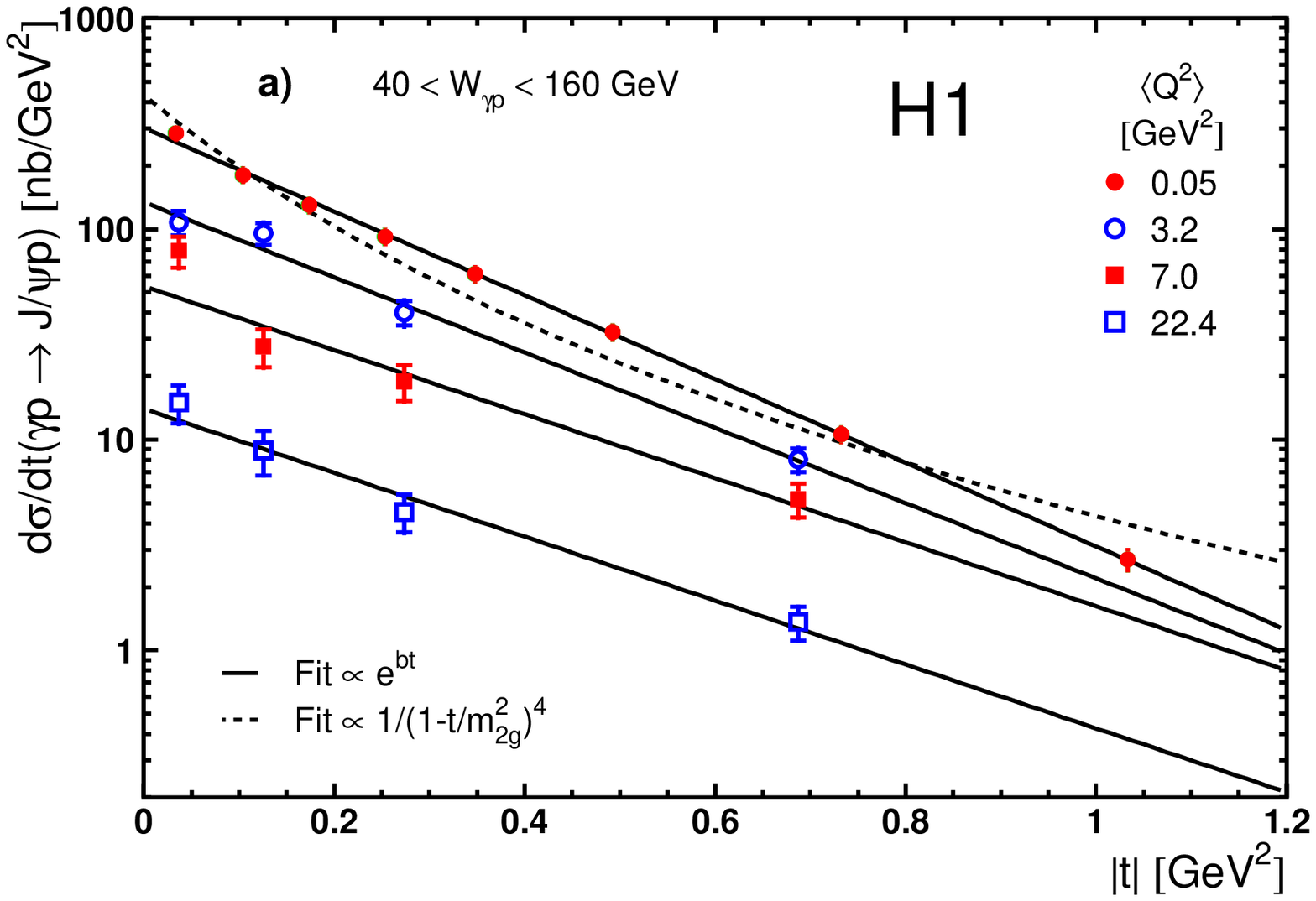}\\[1ex]
    \includegraphics[scale=0.75,keepaspectratio]{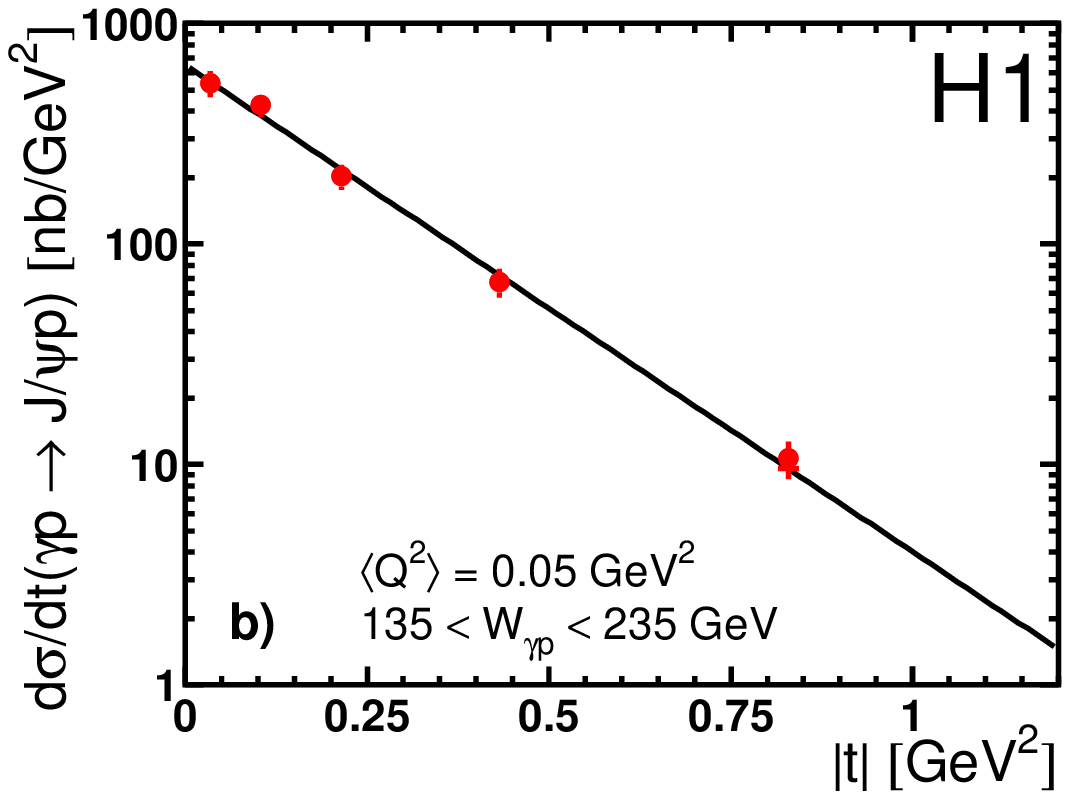}~
    \includegraphics[scale=0.75,keepaspectratio]{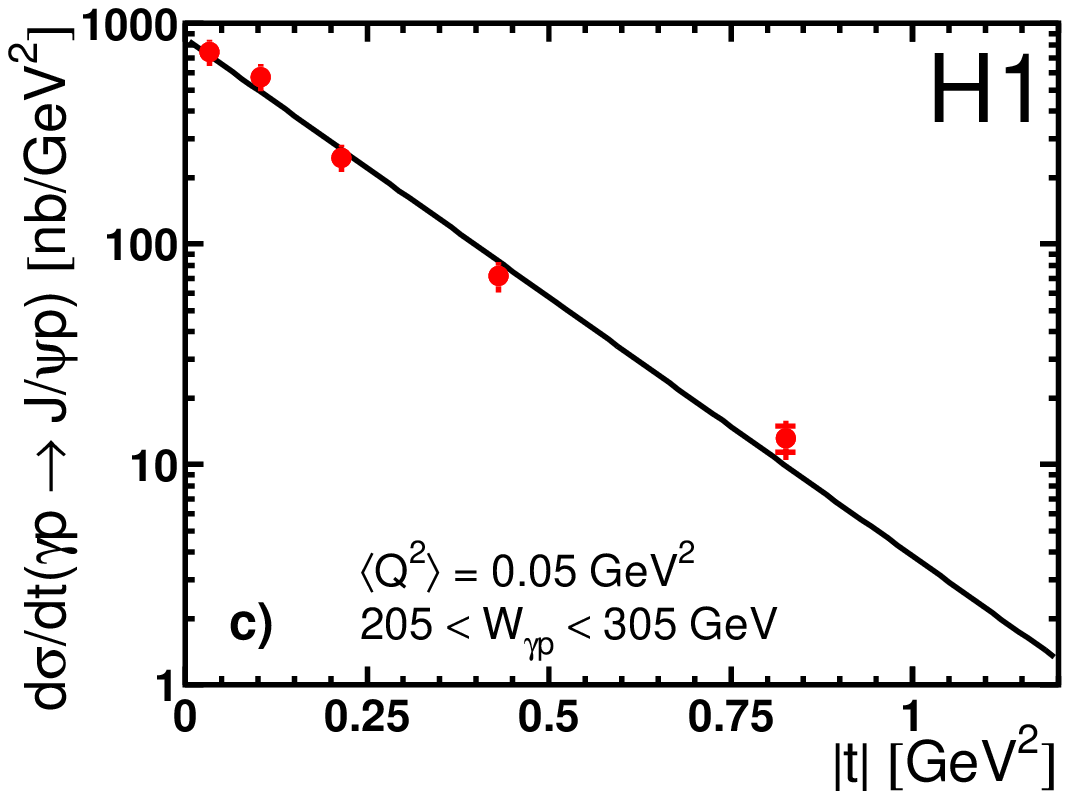}
\caption{Differential cross section d$\sigma/\mbox{d}t$ for elastic \jpsi
  production as a function of $|t|$ a) in four bins of $Q^2$ in the range
  $40<\Wgsp<160\GeV$. $\langle\Qsq\rangle$ indicates the  bin centre
  value in the \Qsq range considered. The inner error bars show the
  statistical error, while the outer error bars show the statistical and
  systematic uncertainties added in quadrature. The solid lines show fits to
  the data of the form d$\sigma/\mbox{d}t\propto e^{bt}$. The dashed curve 
  shows the result of a fit proposed by Frankfurt and 
  Strikman~\cite{Frankfurt:2002ka}. Figures b) and c) show the 
  photoproduction measurements in the ranges $135 <\Wgp< 235$ GeV and
  $205 <\Wgp<305$ GeV.}
    \label{fig:txsec}
  \end{center}
\end{figure}

\vspace*{-1.5ex}
The $t$ dependence of the elastic cross section for \jpsi
meson production is studied in the range $40<\Wgsp<160\GeV$ for different
$Q^2$ bins. The differential cross sections d$\sigma/\mbox{d}t$ as derived from data sets I and II
are listed in table~\ref{tab:txsec} and shown in figure~\ref{fig:txsec}a with
fits of the form d$\sigma/\mbox{d}t\propto e^{bt}$. 
The resulting $b$ values (table~\ref{tab:bq}) for electroproduction 
are systematically lower than the value for photoproduction 
but are compatible within the errors. 

In the context of
developing the calculations using generalised parton densities,
Frankfurt and Strikman~\cite{Frankfurt:2002ka} have proposed an alternative
$t$ dependence. It is based on a dipole function with a $t$ dependent two-gluon form
factor, leading to d$\sigma/\mbox{d}t\propto(1-t/m_{\rm 2g}^2)^{-4}$.  
In a fit to the photoproduction data the two-gluon invariant mass 
$m_{\rm 2g}$ is left as a free parameter.  A value of $m_{\rm
  2g}=(0.679\pm0.006\pm0.011)\GeV$  
is obtained with $\chi^2/$ndf $=5.5$
compared to $\chi^2/$ndf $=0.25$ for the exponential function. The dipole 
form is thus strongly disfavoured by the data.

For photoproduction, the measurement of the $t$ dependence has been extended
to significantly higher \Wgp than in our previous
publication~\cite{Adloff:2000vm} using  data sets III ($135 <\Wgp< 235$ GeV)
and  IV ($205 <\Wgp<305$ GeV). 
Due to the reconstruction of the \jpsi electrons via calorimeter signals 
the resolution in $p_{t,\psi}^2$, which is used to approximate $t$, is worse
than in the track based measurements. The 
differential cross sections d$\sigma/\mbox{d}t$ are obtained using an unfolding 
procedure~\cite{nim:DAgostini}. The results (last two lines in 
table~\ref{tab:wtxsecPhP}) are shown in 
figures~\ref{fig:txsec}b and c with exponential fits, which
describe the data well. The resulting $b$ values are listed in table
\ref{tab:bw} and are discussed further in the following section.

\subsection{Effective Pomeron Trajectories}
\begin{figure}[p]
  \begin{center}
    \includegraphics[scale=0.99,keepaspectratio]{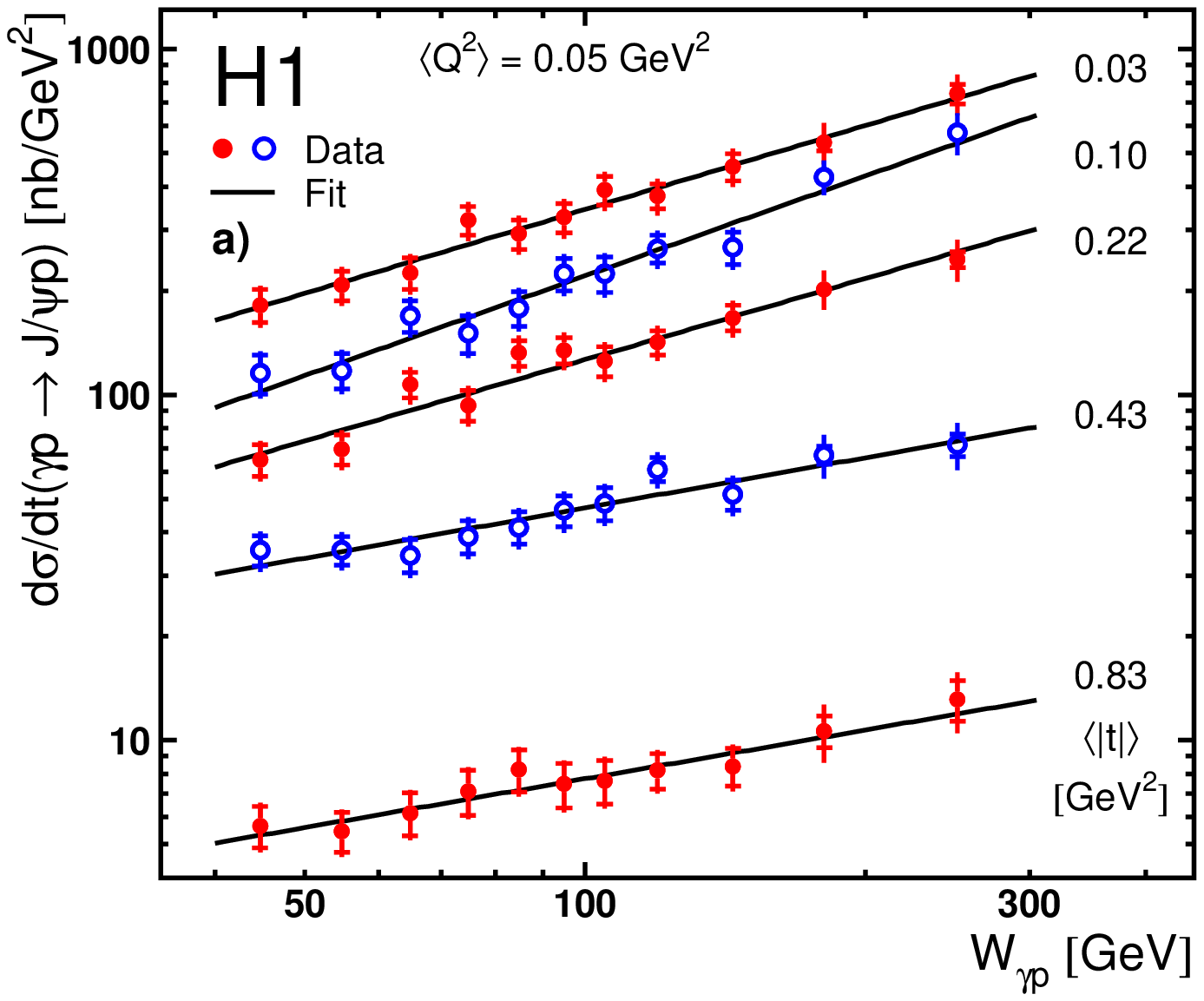}\\[1ex]
    \includegraphics[scale=0.99,keepaspectratio]{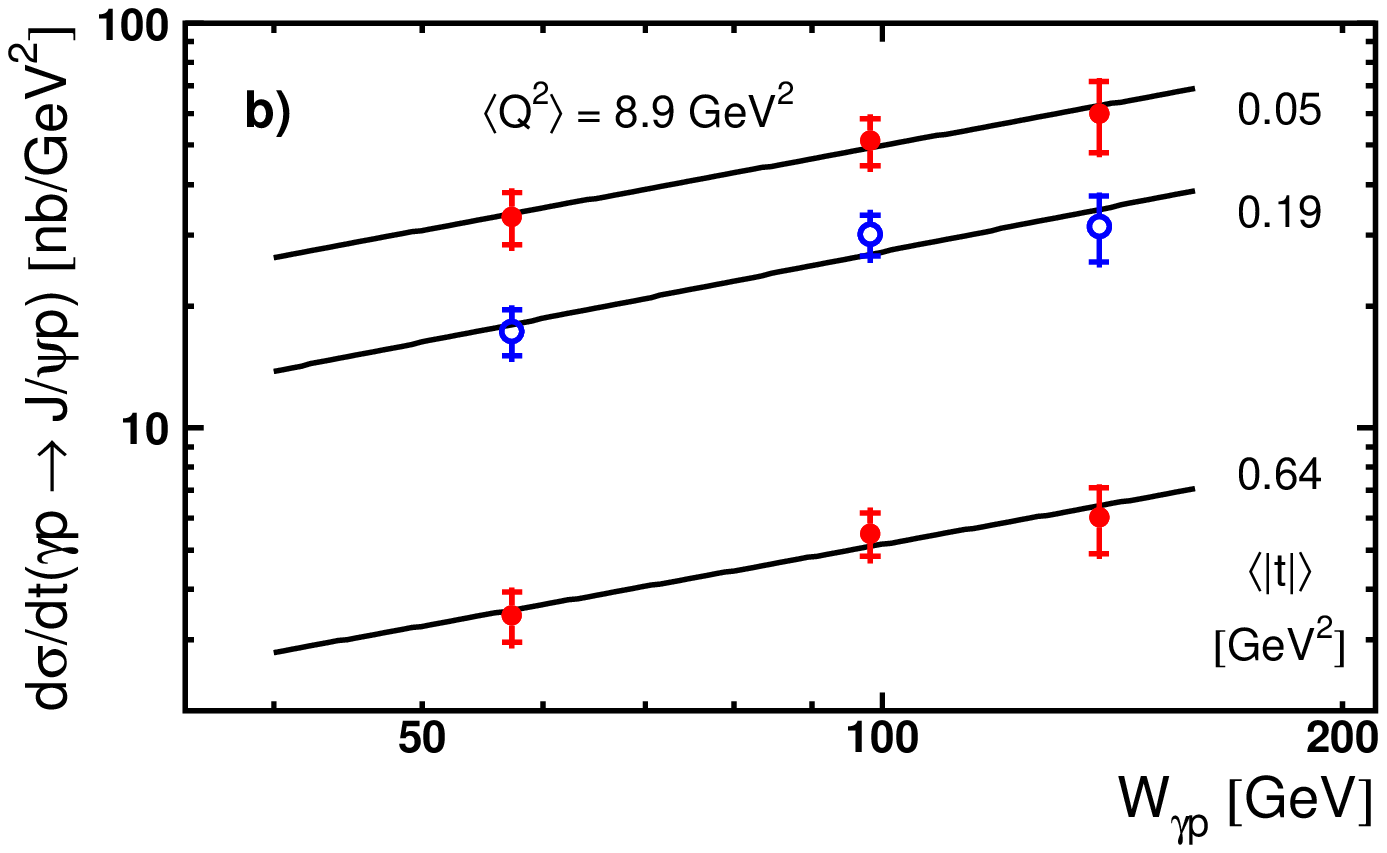}
    \caption{Differential cross section d$\sigma/\mbox{d}t$ for elastic \jpsi
      production as a function of \Wgp in bins of $|t|$ a) in
      photoproduction and b) in electroproduction. $\langle\Qsq\rangle$ and
      $\langle |t|\rangle$ indicate the  bin centre values in the
      ranges considered. The inner error bars show the statistical error,
      while the outer error bars show the statistical and systematic
      uncertainties added in quadrature. The solid lines show the results of
      one-dimensional fits of the form $\sigma
      \propto\Wgsp^{4\left(\alpha(\langle t\rangle)-1\right)}$ in each $t$ bin.}
    \label{fig:wtxsec}
  \end{center}
\end{figure}

\begin{figure}[p]
  \begin{center}
    \includegraphics[scale=0.99,bb=62 461 526 748,keepaspectratio]{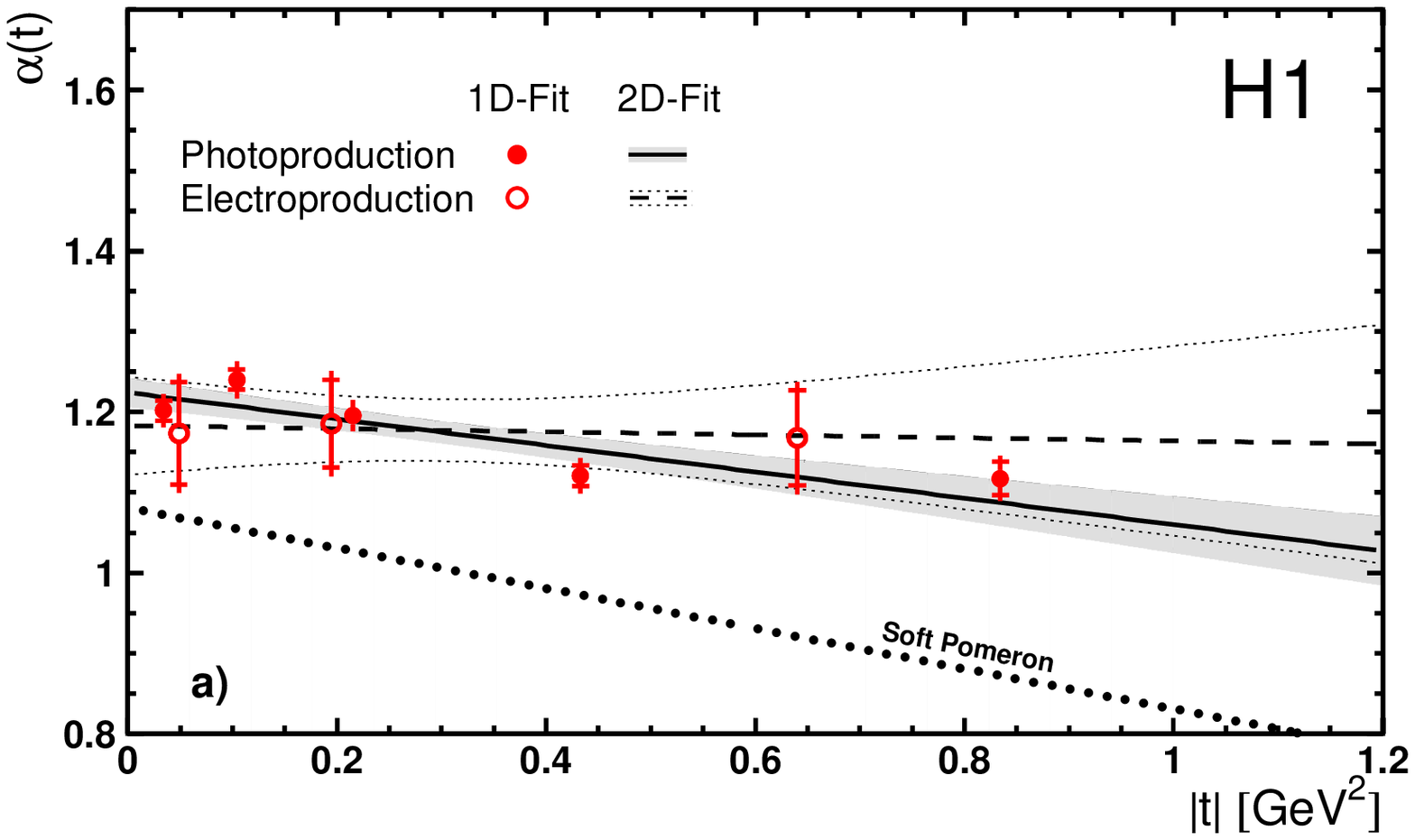}\\[5ex]
    \includegraphics[scale=0.99,bb=62 540 526 733,keepaspectratio]{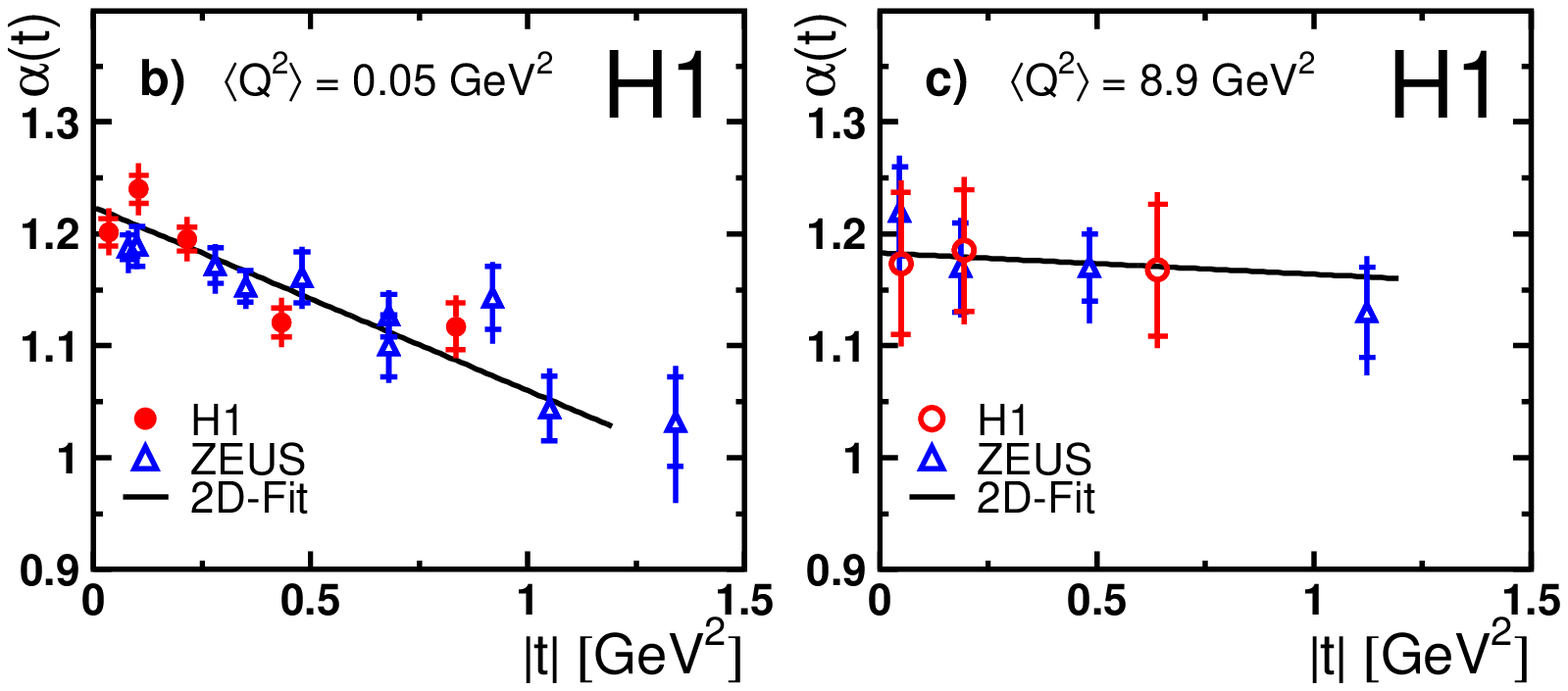}
    \caption{a) The effective trajectory $\alpha(t)$ as a function of $|t|$ in
      the range $40<\Wgsp<305\GeV$ for photoproduction ($\langle\Qsq\rangle=0.05\,\GeVsq$)
      and $40<\Wgsp<160\GeV$
      for electroproduction ($\langle\Qsq\rangle=8.9\,\GeVsq$). The data points
      are the results of the
      one-dimensional fits shown in figure~\ref{fig:wtxsec}.  The inner error
      bars show the statistical error, while the outer error bars show the
      statistical and systematic uncertainties added in quadrature. The solid
      and dashed lines show the results of two-dimensional fits
      (equation~\ref{eq:wtdep}) together with $1\sigma$-error bands, which
      take the correlation between the fit parameters into account.
      A comparison with the results of the ZEUS collaboration
      ~\cite{Chekanov:2002xi,Chekanov:2004mw} is shown in  b) and c) for 
      photoproduction and electroproduction respectively. The data 
      in~\cite{Chekanov:2004mw} are derived at slightly different
      values of $\langle\Qsq\rangle$. The lines are results from the two-dimensional fits.}
    \label{fig:alpha}
  \end{center}
\end{figure}

\vspace*{-1.5ex}
In models based on Regge phenomenology and Pomeron exchange, the energy
dependence of the elastic cross section follows a power law:

\begin{equation}
  \label{eq:wtdep}
  \frac{\mbox{d}\sigma}{\mbox{d}t} = \left.\frac{\mbox{d}\sigma}{\mbox{d}t}\right|_{t=0,\Wgp=W_0}\cdot
  e^{b_0 t}\left(\frac{\Wgp}{W_0}\right)^{4\left(\alpha(t)-1\right)},
\end{equation}
where $\alpha(t)=\alpha_0+\alpha^\prime t$ describes the exchanged 
trajectory and $b_0$ and $W_0$ are constants.
Equation~\ref{eq:wtdep} relates the dependence of the differential
cross section on $t$ to that on \Wgp by

\begin{equation}
  \frac{\mbox{d}\sigma}{\mbox{d}t}\left(t\right) \propto
  e^{\left(b_0+4\alpha^\prime\ln\left(\Wgsp/W_0\right)\right) t}.
\end{equation}
Here only $t$ dependent terms are kept. 
In hard interactions, where Regge phenomenology with a single universal
Pomeron may no longer be applicable, an `effective Pomeron
trajectory'\cite{Bartels:2000ze} is nevertheless  often extracted in order to 
describe the dependence of the
differential cross sections on \Wgp and $t$.  For the determination of
this effective trajectory, a double differential analysis is performed in
which the differential cross section d$\sigma/\mbox{d}t$ is measured in bins of
\Wgp and $t$. The measurements are displayed in figures~\ref{fig:wtxsec}a
and b for photoproduction and electroproduction, respectively
(tables~\ref{tab:wtxsecPhP} and~\ref{tab:wtxsecDIS}). First, one-dimensional 
fits of the form $\Wgsp^{4(\alpha(\langle t\rangle)-1)}$ to the cross
sections in each $|\langle t\rangle|$ bin are performed. The results, which are listed in
table~\ref{tab:alpha} and displayed as solid lines in
figures~\ref{fig:wtxsec}a and b, describe the data well. In
figures~\ref{fig:alpha}b and~c the one-dimensional fit results for $\alpha(t)$
are compared with recent results~\cite{Chekanov:2002xi,Chekanov:2004mw} from
the ZEUS collaboration, which are in good agreement.

A two-dimensional fit of the function given in equation~\ref{eq:wtdep} 
to the data yields values
for $b_0$, $\alpha_0$ and $\alpha^\prime$. The parameter $W_0$ is arbitrarily
chosen to be $90\GeV$; the fit result 
does not depend on this choice. As described before, different
normalisations are allowed for the different data sets in the fit.
Figure~\ref{fig:alpha}a shows the result of the two-dimensional fit for
$\alpha(t)$ as solid and dashed lines for photoproduction and
electroproduction, respectively. Error bands corresponding to one standard
deviation are shown, taking the correlation between $\alpha_0$ and
$\alpha'$ into account. 
The results for $\alpha(\langle t\rangle)$ from the one-dimensional fits are 
shown as points with error bars for comparison.
Good internal consistency is observed. 


The results of the two-dimensional fits are listed in the following table.

  \begin{center}
    \begin{tabular}[c]{|c@{\hspace*{0.99\tabcolsep}}|c@{\hspace*{0.99\tabcolsep}}|c@{\hspace*{1.\tabcolsep}}|c|}
      \hline\\[-2.4ex]
      \Qsq $[\GeVsq]$ & $b_0$ $[\GeV^{-2}]$ & $\alpha_0$ & $\alpha'$ $[\GeV^{-2}]$\\
      \hline\\[-2.4ex]
      $\lesssim 1$ & \bophp & \aophp & \apphp \\
      $2-80$     & \bodis & \aodis & \apdis \\
      \hline
    \end{tabular}
  \end{center}

Here the first errors are statistical and the second reflect the systematic uncertainties.

The \Wgp dependence of the cross section is predominantly determined by
$\alpha_0$ and the fit values lead to a \Wgp dependence very similar to the 
parameterisation with $\delta$ discussed above.  
The parameter $\alpha^\prime$ relates the $t$ and \Wgp dependences and if
non-zero leads to
the `shrinkage' of the diffractive peak. 
For photoproduction $\alpha^\prime$ is larger than zero by four standard deviations 
and is two standard deviations below  the value of $0.25\,\GeV^{-2}$ obtained 
for the soft Pomeron in \cite{Donnachie:1998gm}. For
electroproduction $\alpha^\prime$ is compatible with 0, which matches the 
expectation in~\cite{Bartels:2000ze}, but due to the errors $\alpha^\prime$ 
is also compatible with the value measured for photoproduction. 

\begin{figure}[p]
  \begin{center}
    \includegraphics[scale=0.99,keepaspectratio]{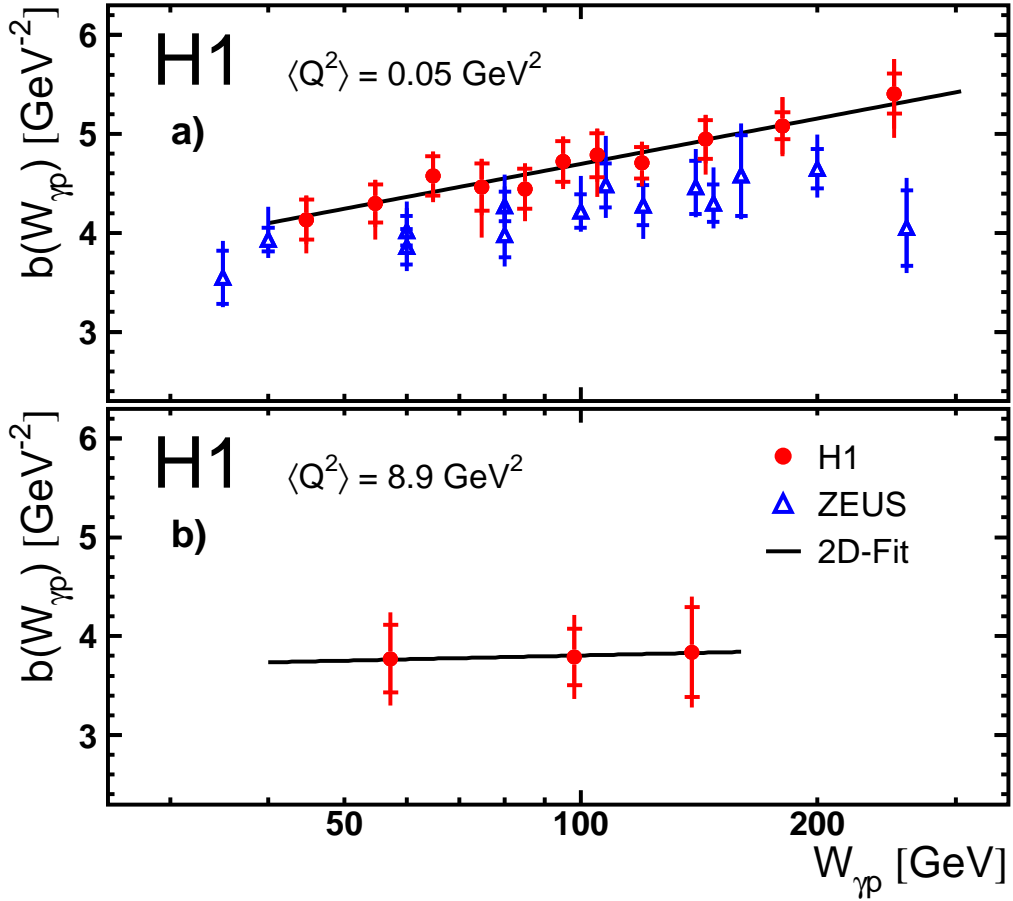}
    \caption{The values of the $t$ slope parameter $b(\Wgsp)$ as a function of \Wgp in
      the range $|t|<1.2\GeVsq$ for a)~photoproduction and
      b)~electroproduction. $\langle\Qsq\rangle$ indicates the  bin
      centre value in the \Qsq range considered. The data points are the
      results of one-dimensional fits of the form d$\sigma/\mbox{d}t\propto e^{bt}$
      in \Wgp bins. The inner error bars show the statistical errors, while
      the outer error bars show the statistical and systematic uncertainties
      added in quadrature. The solid lines show the results of the 
      two-dimensional fits (equation~\ref{eq:wtdep}) as in
      figure~\ref{fig:alpha}. In a) the data are compared with results
      from the ZEUS collaboration~\cite{Chekanov:2002xi}.}
    \label{fig:bslope}
  \end{center}
\end{figure}

Alternatively the value of $\alpha^\prime$ can be measured using the dependence 
of the $t$
slope parameter on \Wgp, using $b(\Wgsp)=b_0+4\alpha'\ln(\Wgsp/W_0)$.  Exponential
fits of the form $e^{bt}$ to the measured differential cross sections
d$\sigma/\mbox{d}t$ in bins of \Wgp are performed and the
resulting values for $b$ are displayed in figure~\ref{fig:bslope}a and b and
listed in table~\ref{tab:bw} for
photoproduction and electroproduction.  For photoproduction the $b$ 
values are seen to increase with $\Wgp$. These $b$ values are
independent of normalisation uncertainties between data sets. 
The curves in figure~\ref{fig:bslope}a and b
show the corresponding result $b(\Wgsp)$ from the two-dimensional fit
described above. 

In figure~\ref{fig:bslope}a photoproduction
results for the slope parameter from the ZEUS experiment~\cite{Chekanov:2002xi} in a similar 
kinematic region are also shown.
They show a similar dependence on \Wgp but are on average $0.5$ GeV$\,^{-2}$
lower. This difference in the absolute size of $b$ may be due to differences 
in the handling of the background from proton dissociative events, which has
a much shallower $b$ slope than for the elastic case ($1.6\,\GeV^{-2}$\cite{Aktas:2003zi}).

\subsection{Helicity Studies}\label{decang}
\vspace*{-1.5ex}
The assumption that the \jpsi meson observed in the
final state keeps the helicity of the photon is referred to as $s$-channel
helicity conservation (SCHC). This assumption can be tested by measurements of the
angles in the production and decay of the \jpsi meson. If SCHC holds, the
angular analysis leads to a separation of the cross sections due to
longitudinally and transversely polarised photons which are both predicted in the MRT 
calculations.  

\begin{figure}[htb]
  \begin{center}
   \includegraphics[width=0.9\textwidth,keepaspectratio]{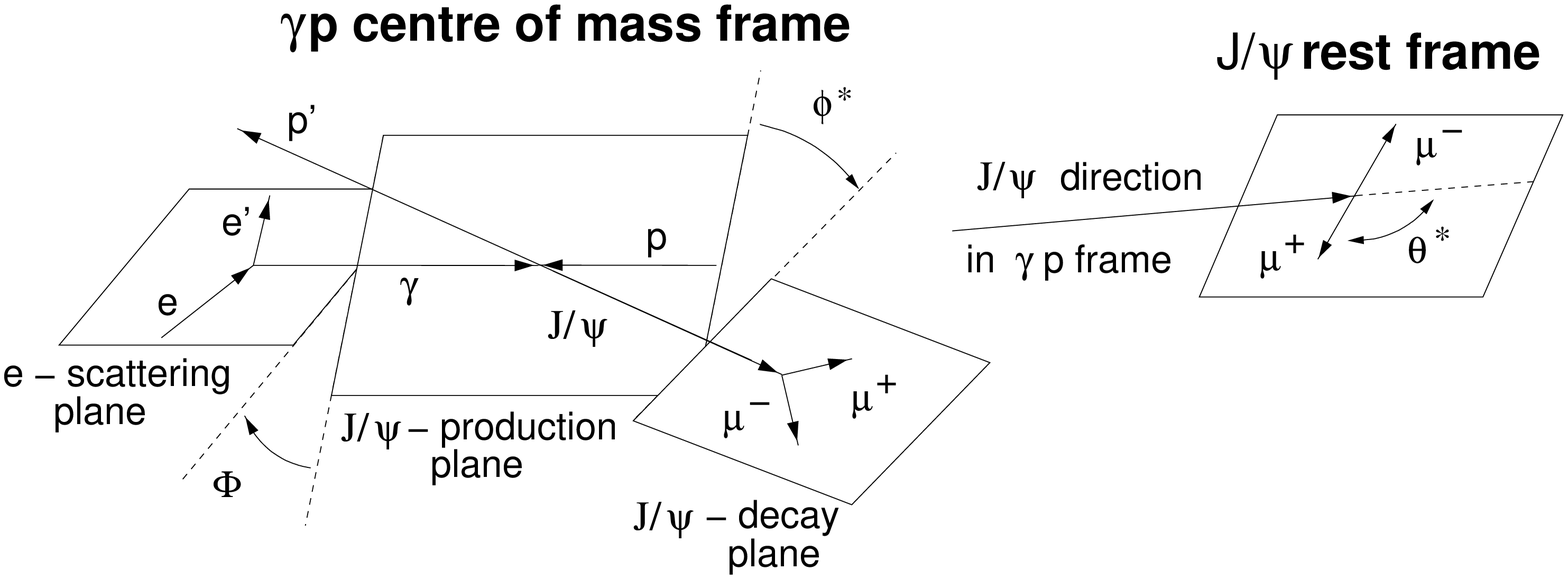}
  \end{center}
\end{figure}

Three angles are defined, which are illustrated in the figure above.  $\theta^*$ is the
polar angle of the decay muon with the charge of the beam lepton in the
\jpsi rest frame. $\theta^*=0\deg$ corresponds to the flight direction of the 
\jpsi in the \gsp 
centre of mass frame. $\phi^*$ is the angle between the \jpsi production
plane, defined by the exchanged photon and the \jpsi meson, and the decay
plane in the \gsp 
centre of mass frame.  $\Phi$ is the angle between the scattering plane of
the beam lepton and the \jpsi production plane. The angle $\Phi$ can only be
measured when the scattered electron is observed, i.e. for electroproduction.
In the case of SCHC and natural parity exchange, the
angular distributions of the \jpsi production and decay are functions of
$\cos\theta^*$ and $\Psi=\phi^*-\Phi$ only~\cite{Schilling:1973ag}. 

The angular distributions are expected to be similar for elastic and
proton dissociative processes. This expectation is
verified within the present statistical accuracy. Therefore, in order to
increase the statistics, the cross sections differential in the angles are 
derived using a data set, which 
includes proton dissociative events in addition to data sets~I and~II.  
The selection of proton dissociative events is similar to that 
for data sets~I and~II (section \ref{secsel}  and table \ref{tab:sel}),
but now a signal in one of the forward detectors is required and one 
additional track is allowed with $\theta<20^\circ$.  Non-diffractive
events are rejected by requiring $z>0.95$,
where $z=E_{\jpsi}/E_\gamma$ in the proton rest frame.
Furthermore, an increased $t$ range, $|t|<5\GeVsq$, is allowed, 
since the 
proton dissociative \jpsi cross section shows a flatter $t$ dependence,
with an exponential slope of about $1.6\,\GeV^{-2}$~\cite{Aktas:2003zi}.

In the present analysis, differential cross sections for the four angles are
used to measure combinations of seven of the 15 spin-density matrix elements,
which describe the spin structure of the interaction completely\footnote{Spin
  density matrix elements $r_{\lambda(\gamma)\lambda(\psi)}^{i}$ or $r_{\lambda(\gamma)\lambda(\psi)}^{ik}$ are linear 
combinations of the transition
  amplitudes $T^{i,k}_{\lambda(\gamma)\lambda(\psi)}$ from a photon of helicity
  $\lambda(\gamma)$ to a \jpsi of helicity $\lambda(\psi)$.}.  The measured
angular distributions and their dependence on the spin-density matrix
elements $r_{\lambda(\gamma)\lambda(\psi)}^{ik}$ are~\cite{Schilling:1973ag}:

\begin{eqnarray}
  \label{eq:costheta}
  \frac{\mbox{d}\sigma}{\mbox{d}\cos\theta^*} & \propto &
  1+r_{00}^{04}+\left(1-3r_{00}^{04}\right)\cos^2\theta^*\\
  \label{eq:phistar}
  \frac{\mbox{d}\sigma}{\mbox{d}\phi^*} &\propto &
  1+r_{1-1}^{04}\cos(2\phi^*)\\
  \label{eq:psi}
  \frac{\mbox{d}\sigma}{\mbox{d}\Psi} & \propto &
  1-\varepsilon r_{1-1}^{1}\cos(2\Psi)\\
  \label{eq:phi}
  \frac{\mbox{d}\sigma}{\mbox{d}\Phi} & \propto &
  1-\varepsilon\left(r_{00}^1+2r_{11}^1\right)\cos2\Phi
  +\sqrt{2\varepsilon\left(1+\varepsilon\right)}
  \left(r_{00}^5+2r_{11}^5\right)\cos\Phi.
\end{eqnarray}
Here $\varepsilon$ is the polarisation parameter of the virtual photon.

\begin{figure}[p]
  \begin{center}
    \includegraphics[scale=0.99,bb=296 513 521 748,keepaspectratio]{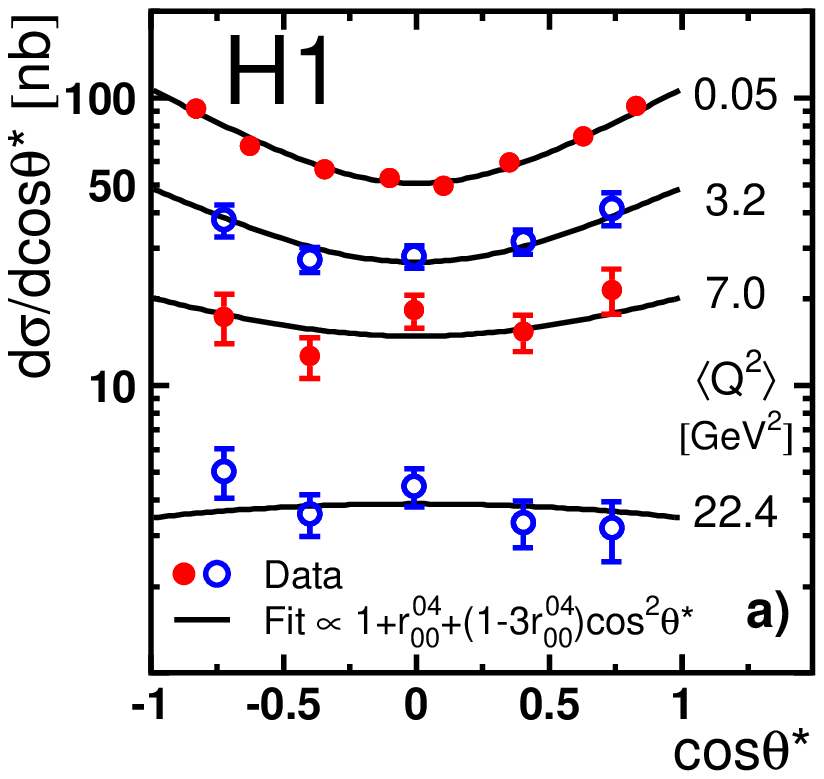}~
    \includegraphics[scale=0.99,bb=296 513 521 748,keepaspectratio]{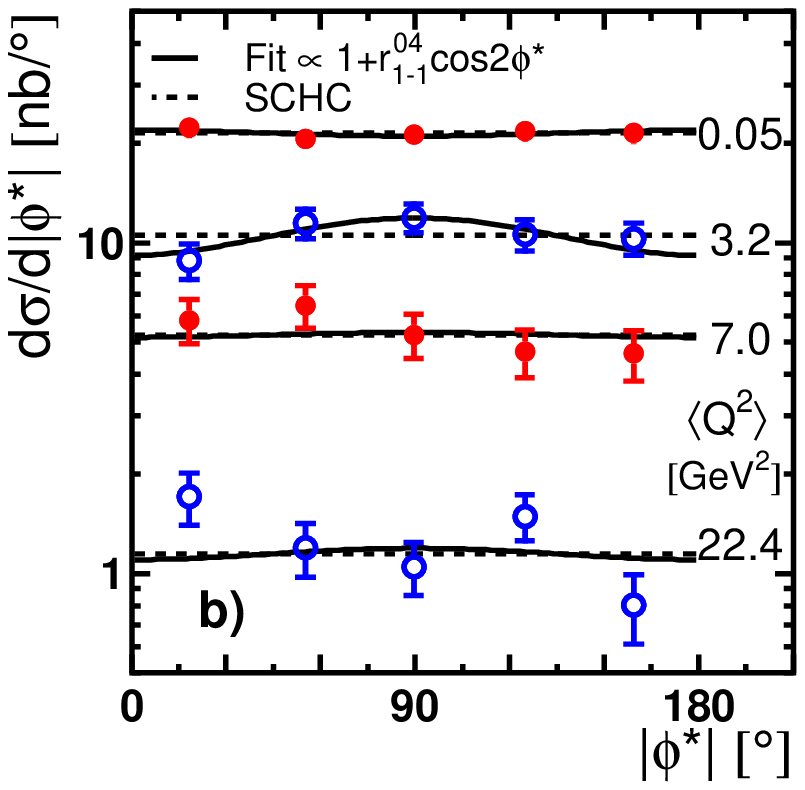}\\[1ex]
    \includegraphics[scale=0.99,bb=296 513 521 748,keepaspectratio]{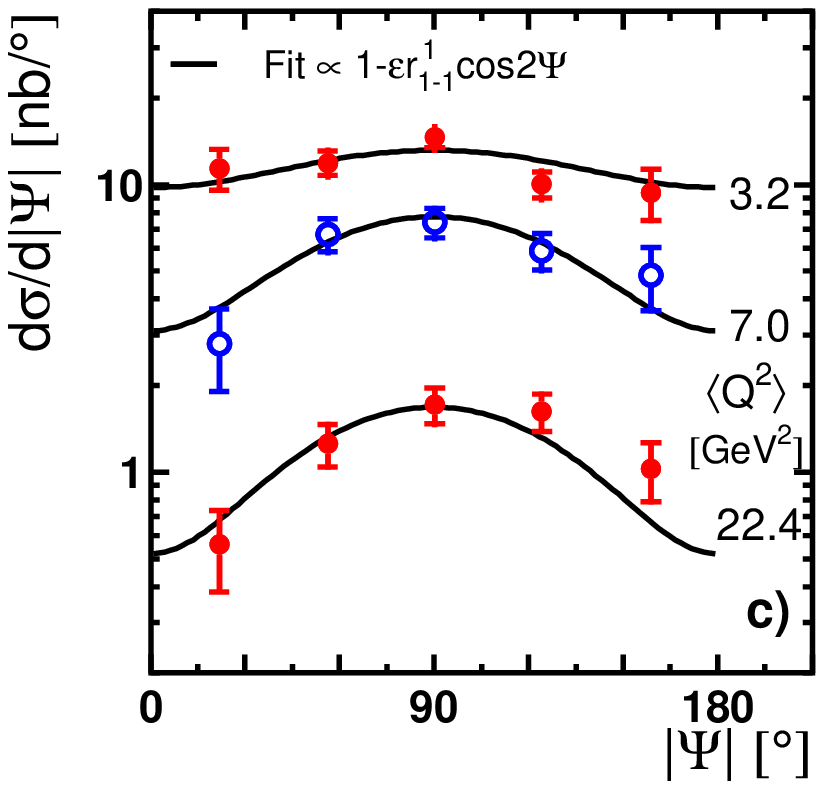}~
    \includegraphics[scale=0.99,bb=296 513 521 748,keepaspectratio]{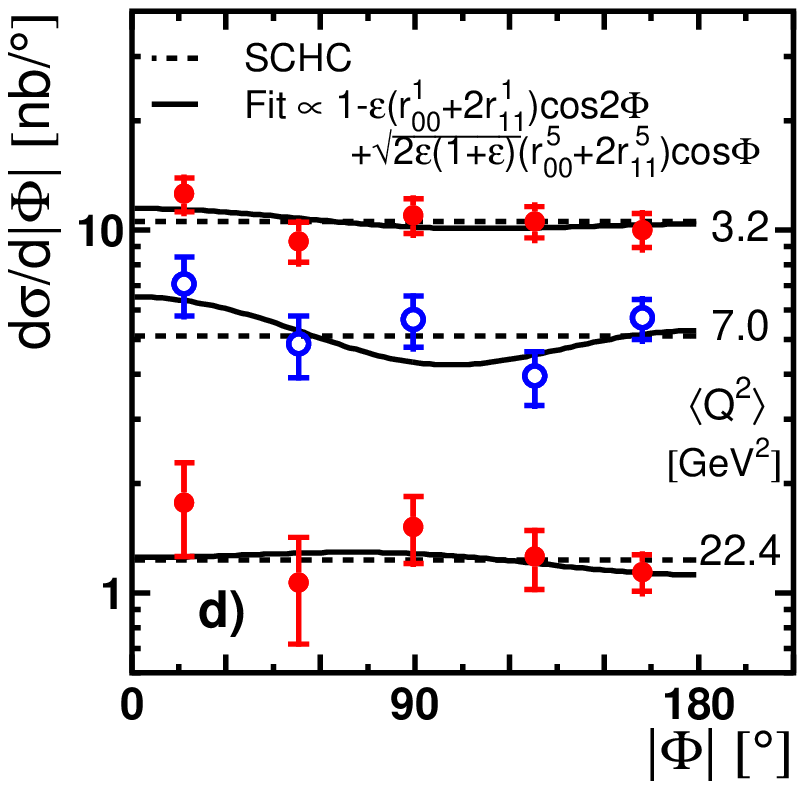}
    \caption{The differential cross sections for diffractive (elastic and
      proton dissociative) \jpsi\ production as functions of the four angles
      $\theta^\ast,\ \phi^\ast,\ \Psi$ and $\Phi$ defined in the text. The
      data are shown in bins of \Qsq for the range $40<\Wgsp<160\GeV$ and
      $|t|<5\GeVsq$. $\langle\Qsq\rangle$ indicates the  bin centre
      value in the \Qsq range considered. The inner error bars show the
      statistical error, while the outer error bars show the statistical and
      systematic uncertainties added in quadrature. The solid lines represent
      the results of fits to the
      data. The dashed lines in b) and d) are fits assuming $s$-channel helicity
      conservation.}
    \label{fig:heli}
  \end{center}
\end{figure}

Figures~\ref{fig:heli}a and b show the differential \gp\ cross sections
d$\sigma/\mbox{d}\cos\theta^*$ and d$\sigma/\mbox{d}\phi^*$ in four
bins of $Q^2$. Figures~\ref{fig:heli}c and d show the differential cross
sections 
d$\sigma/\mbox{d}\Psi$ and d$\sigma/\mbox{d}\Phi$ in three bins of $Q^2$. 
The results of fits of equations~\ref{eq:costheta}, \ref{eq:phistar},
 \ref{eq:psi} and \ref{eq:phi} are shown as full lines. 
In the fits, the spin-density matrix elements or 
the combinations $r_{00}^1+2r_{11}^1$ and
$r_{00}^5+2r_{11}^5$ for equation~\ref{eq:phi} are free parameters. 
In figure~\ref{fig:heli}b and d results from a fit assuming SCHC
are also shown.

\begin{figure}[p]
  \begin{center}
    \includegraphics[scale=0.97,bb=288 294 520 748,keepaspectratio]{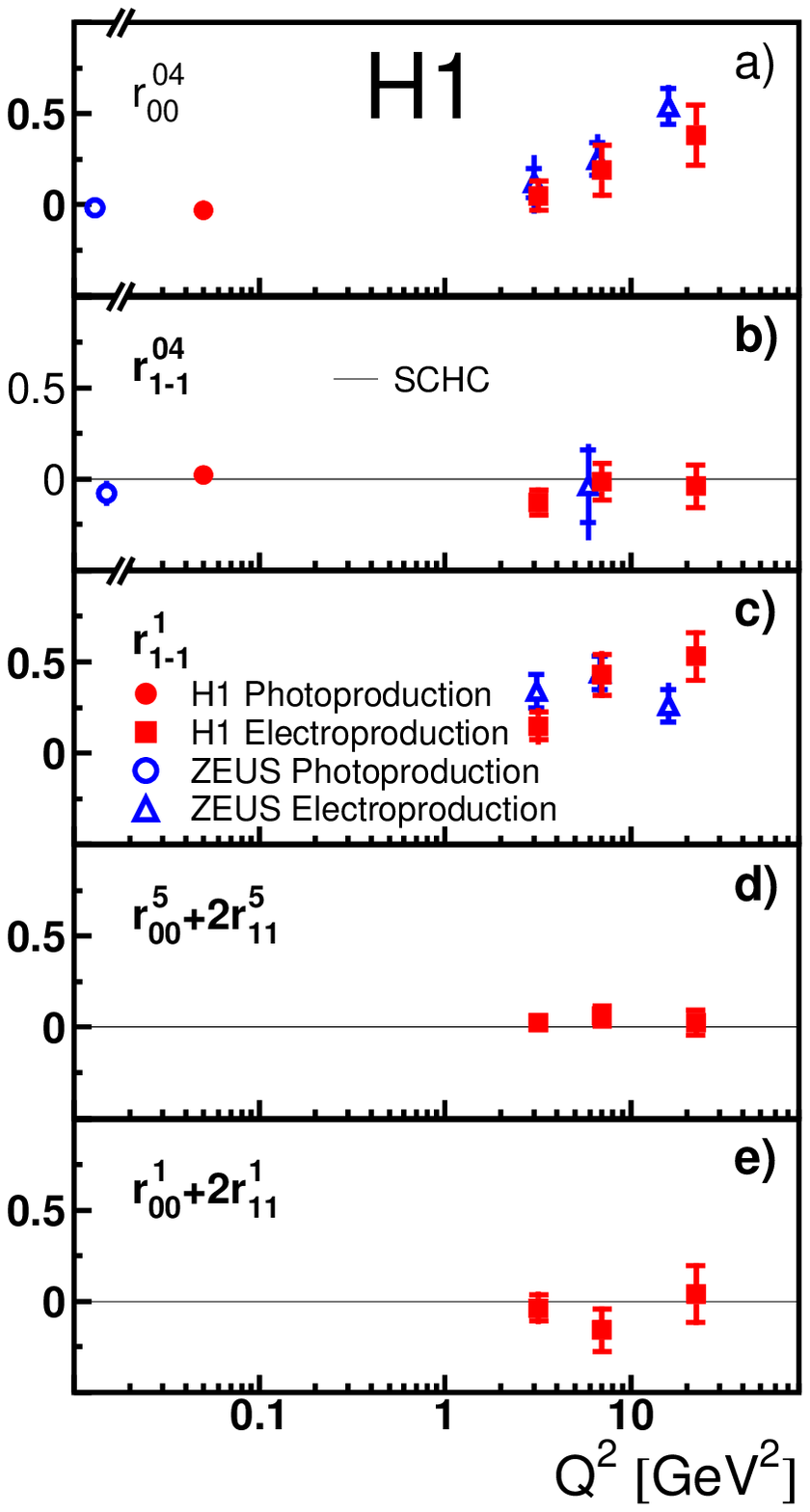}
    \includegraphics[scale=0.97,bb=303 294 535 748,keepaspectratio]{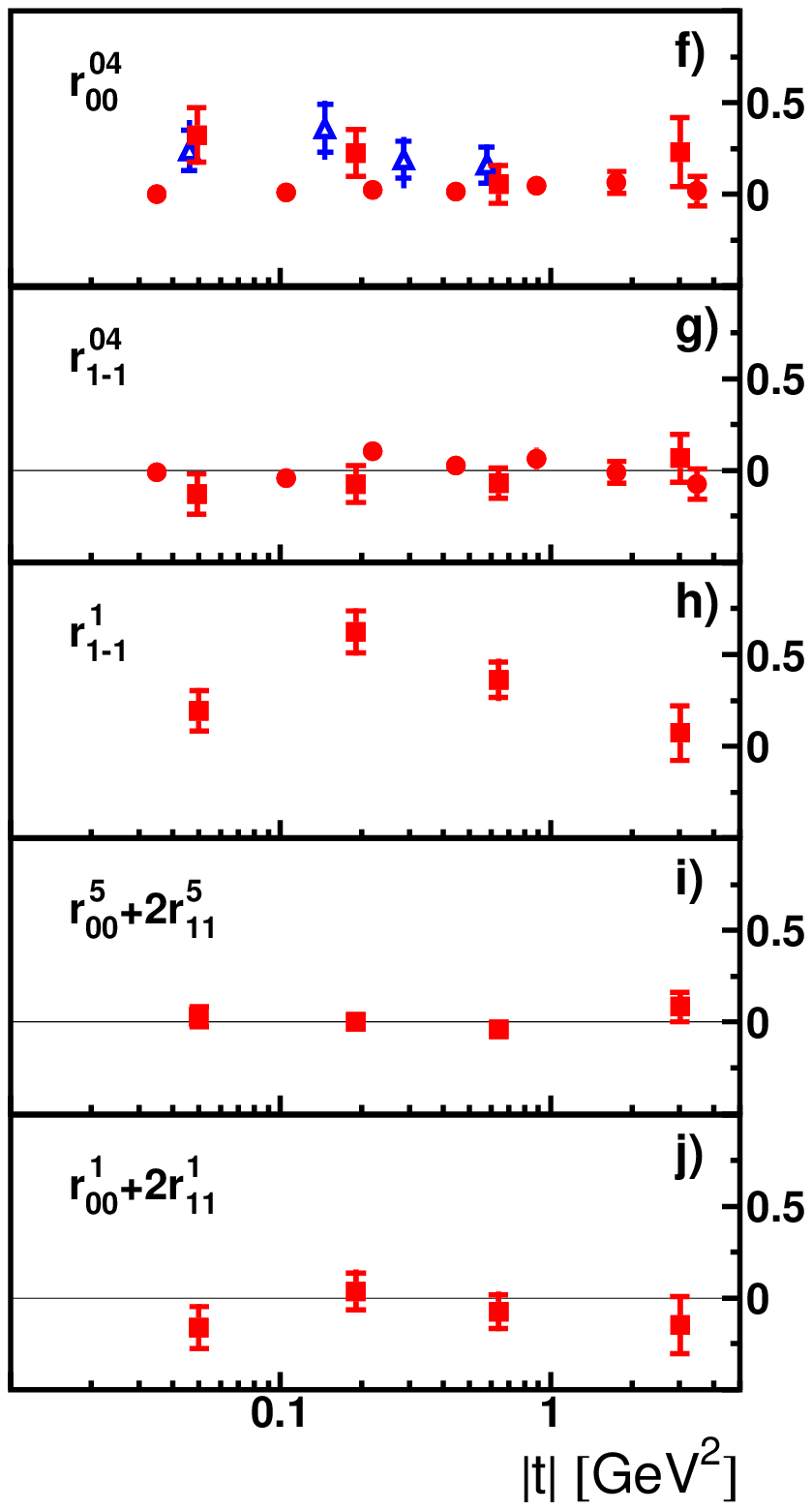}
    \caption{Spin-density matrix elements as functions of \Qsq (a-e) and
      $|t|$ (f-j) for the range $40<\Wgsp<160\GeV$. The data points are the
      results of fits of equations~\ref{eq:costheta}-~\ref{eq:phi} to the
      data shown in figure~\ref{fig:heli}. The inner error bars show the fit
      result including only the statistical error, while the outer error
      bars also include the systematic uncertainties. The expectations
      from SCHC are shown as solid lines. The results
      from the ZEUS collaboration are
      also shown, ( a), c) and f)~\cite{Chekanov:2002xi,Chekanov:2004mw}
      and b)~\cite{Breitweg:1997rg,Breitweg:1998nh}).}
    \label{fig:matrix}
  \end{center}
\end{figure}

The spin density matrix elements,
which are determined by the fits, are shown in figures~\ref{fig:matrix}a--e
as functions of \Qsq (tables \ref{tab:qr} and \ref{tab:qr2}). 
The analysis is also performed in bins of $|t|$ 
and the resulting spin density matrix elements are displayed 
in figures~\ref{fig:matrix}f--j (tables \ref{tab:tr} and \ref{tab:tr2}) 
as functions of $|t|$.

Results from the ZEUS experiment
~\cite{Chekanov:2002xi,Chekanov:2004mw,Breitweg:1997rg,Breitweg:1998nh}  
are also shown in figures
\ref{fig:matrix}a, b, c and f, which are in good agreement with the present
results. 
In figures \ref{fig:matrix}b, d, e, g, i and j the expectation from SCHC,
namely 0, matches the data well.
 SCHC yields a relation
between two spin density matrix elements:
$r_{1-1}^1 = (1 - r_{00}^{04}) / 2$. This is observed to be fulfilled within errors.

In the case of SCHC, the matrix element $r_{00}^{04}$ provides a direct
measurement of $R$, the ratio of the cross sections for longitudinal and
transverse polarised photons, $\sigma^L$ and $\sigma^T$, respectively:

\begin{equation*}
  R = \frac{\sigma^L}{\sigma^T} =
  \frac{1}{\varepsilon}\frac{r_{00}^{04}}{1-r_{00}^{04}}.
\end{equation*}

\begin{figure}[p]
  \begin{center}
 \includegraphics[scale=0.99,bb=141 549 521 735,keepaspectratio]{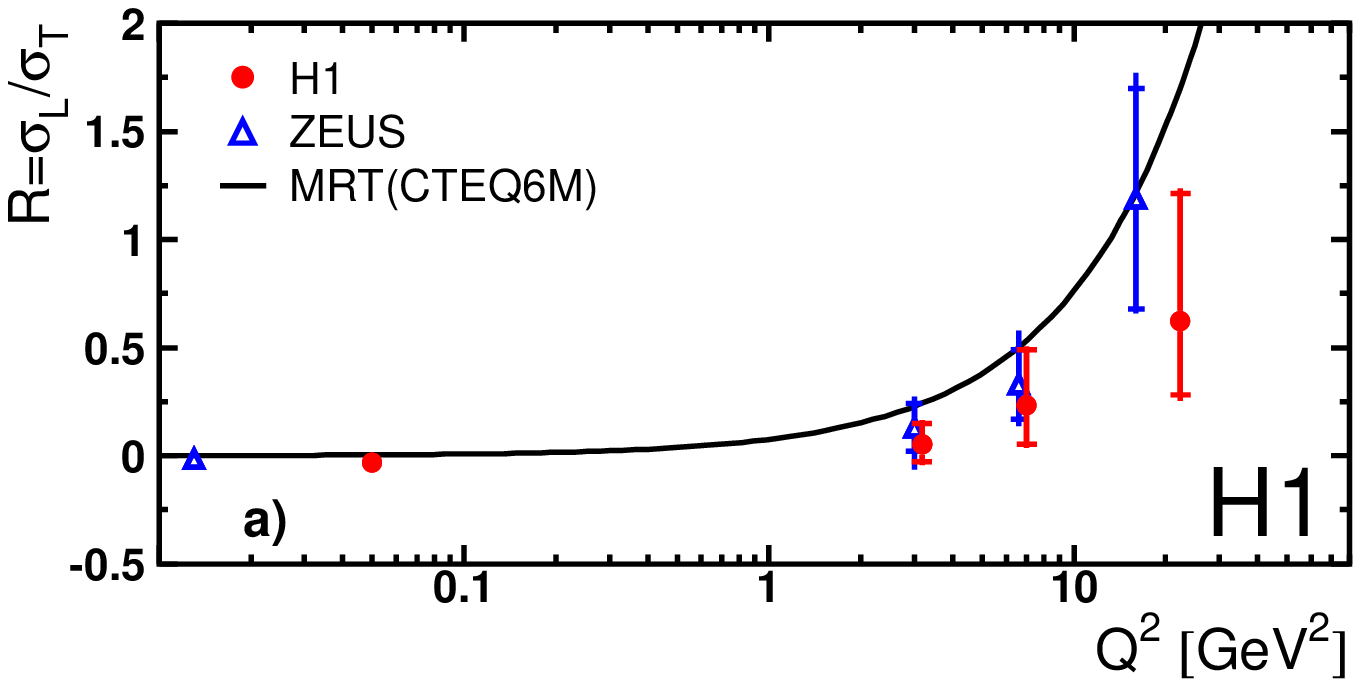}\\[1ex]
    \includegraphics[scale=0.99,bb=141 523 521 733,keepaspectratio]{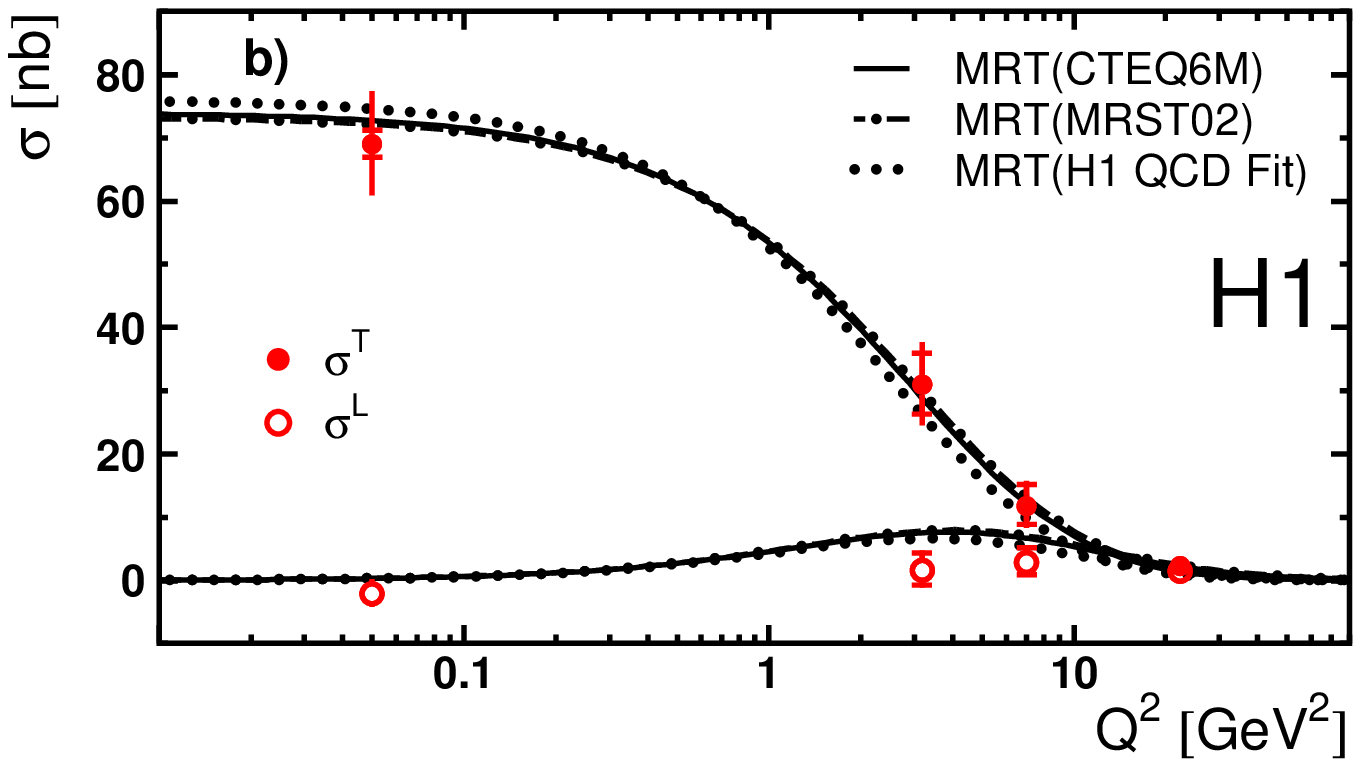}
    \caption{a) Ratio $R=\sigma^L/\sigma^T$  as a function of \Qsq 
      for the
      range $40<\Wgsp<160\GeV$ and $|t|<5\GeVsq$. The data are compared with
      the result of a MRT calculation~\cite{Martin:1999wb} based
      on the CTEQ6M~\cite{Pumplin:2002vw} gluon distribution. Also shown are
      results from the ZEUS collaboration~\cite{Chekanov:2002xi,Chekanov:2004mw}. b)~The
      cross sections for longitudinally and transversely polarised photons
      $\sigma^{L}$ and $\sigma^{T}$ as a function of $\Qsq$. The MRT QCD
      calculations based on different gluon
      distributions~(\cite{Pumplin:2002vw,Martin:2001es,Adloff:2003uh})
      are also shown  with
      the same normalisation factors as derived from figure~\ref{fig:qxsec}.
      The inner error bars show the statistical errors, while the outer error bars
      show the total errors. }
    \label{fig:r}
  \end{center}
\end{figure}

The values of $R$ are presented
in figure~\ref{fig:r}a and in table~\ref{tab:qr}.  For comparison the
prediction from MRT~\cite{Martin:1999wb} is shown, which depends only weakly
on the gluon density. In figure~\ref{fig:r}a the gluon density
from CTEQ6M is chosen with the normalisation as before, 
which gives the best description of the \Qsq and \Wgp
dependences of the cross sections. The prediction is somewhat above the data  
but still describes the \Qsq dependence reasonably well.
Similar results from \cite{Chekanov:2002xi,Chekanov:2004mw} agree also with
the present data.

The values of $R$ can be used to derive the cross sections $\sigma^L$ and
$\sigma^T$ using the relationship $\sigma_{\gsp} = \sigma^T_{\gsp} +
\varepsilon\sigma^L_{\gsp}$. The results are shown in figure~\ref{fig:r}b as
a function of $\Qsq$. $\sigma^T$ dominates at low $Q^2$, while at $\Qsq\sim
M_\psi^2$ both $\sigma^T$ and $\sigma^L$ are of similar magnitude. The MRT
predictions are compared with the data using different gluon density
parameterisations. The differences between the predictions are not very large.
All gluon density parameterisations give a reasonable description of the data,
although $\sigma_L$ is somewhat above the data for \Qsq$\gtrsim 3\,\GeVsq$.

In brief, the helicity studies show consistency with SCHC within
experimental errors. The ratio of cross sections for longitudinally and
transversely polarised photons is extracted and its \Qsq dependence is found to 
be reasonably described by the MRT calculations.

\section{Summary}
\noindent
New measurements are presented of elastic \jpsi photoproduction and electroproduction in the
ranges $40<\Wgsp<305\GeV$ and $40<\Wgsp<160\GeV$, respectively~\footnote{The results 
of the present analysis agree within errors
with our previous results. We consider the new data to supersede them
due to improved statistics and better understanding of the detector
efficiencies.}.

The cross section $\sigma(\gsp\to\jpsi p)$
is measured as a function of \Qsq in the range $0<\Qsq<80\GeVsq$, and a fit
of the form $\sigma_{\gsp}\propto(M_\psi^2+\Qsq)^{-n}$ yields a value of $n=\qexp$. The
shape of the \Qsq distribution is well described by a perturbative QCD
calculation by Martin, Ryskin and Teubner (MRT), almost independently of
the gluon density distribution used. 

The photoproduction cross section is measured as a function of the
photon-proton centre of mass energy \Wgp in the range $40<\Wgp<305\GeV$, and
can be parameterised as $\sigma_{\gsp}\propto\Wgp^\delta$ with
$\delta=0.754\pm0.033\stat\pm0.032\sys$.
The results for $\delta$ in electroproduction, measured in the range
$40<\Wgsp<160\GeV$, are consistent with those in photoproduction and no 
\Qsq dependence is observed within experimental errors.
Predictions of the \Wgp dependence of the cross section in pQCD-based models
 depend strongly on the gluon distribution, as can be seen
explicitly in the MRT model. A good description of
the shape of the data can currently be achieved only with some gluon  
parameterisations. This demonstrates the potential to constrain the
gluon distribution with the elastic \jpsi data in a kinematic region (low $x$,
low \Qsq) where fits from inclusive data yield gluon distributions
with large uncertainties.   

The differential cross section d$\sigma/\mbox{d}t$ for elastic \jpsi photoproduction
for $|t|\leq 1.2\GeVsq$ is measured in the extended range of $40\leq\Wgp\leq305\,\GeV$.
A single exponential function yields a good description of d$\sigma/\mbox{d}t$ in this
range, while a functional form based on a dipole function is strongly disfavoured.
The slope parameter $b$ of the exponential shows a dependence on \Wgp 
which is weaker than expected 
from soft Pomeron phenomenology, but is clearly positive, leading to 
shrinkage of the diffractive peak. The slope parameter $b$ in
electroproduction agrees with the photoproduction values within errors, but has a
tendency to decrease with increasing \Qsq. 

Effective Pomeron trajectories $\alpha_0+\alpha^\prime t$ for elastic \jpsi
photoproduction and 
electroproduction are determined from a simultaneous analysis
of d$\sigma/\mbox{d}t$ as a function of \Wgp and $|t|$. 
The electroproduction and photoproduction results
are consistent with each other within errors. The trajectory for
photoproduction has a $t$ slope which is two standard deviations below the soft Pomeron value
but four standard deviations above zero.

Finally, the helicity structure of diffractive \jpsi production is analysed as
a function of \Qsq and $|t|$. No evidence is found for a violation of
$s$-channel helicity conservation (SCHC). Assuming SCHC, the ratio of the
longitudinal to the transverse polarised photon cross sections is determined as a function of
$Q^2$ and is found to be consistent with QCD calculations.

\section*{Acknowledgments}
\noindent
We are grateful to the HERA machine group whose outstanding efforts have made
this experiment possible.  We thank the engineers and technicians for their
work in constructing and maintaining the H1 detector, our funding agencies
for financial support, the DESY technical staff for continual assistance and
the DESY directorate for support and for the hospitality which they extend to
the non DESY members of the collaboration. We thank M.~Strikman and T.~Teubner 
 for valuable discussions and for making their theoretical
predictions available.


\clearpage
\renewcommand{\captionlabelfont}{\normalsize}

\clearpage
\begin{table}\centering
    \begin{tabular}[c]{r@{\hspace*{0.5\tabcolsep}}c@{\hspace*{0.5\tabcolsep}}l@{\hspace*{0.9\tabcolsep}}c@{\hspace*{0.9\tabcolsep}}c}
      \hline\\[-2.4ex]
      \multicolumn{3}{c}{\Qsq} & $\langle\Qsq\rangle$ & $\sigma$\\
      \multicolumn{3}{c}{$[\GeVsq]$} & $[\GeVsq]$ & $[\nb]$\\
      \hline\hline
      &$\lesssim$&$1$   & $0.05$ & $73.1\pm1.1\pm6.4$\\
      $2$&$-$&$3.2$     & $2.5$  & $37.3\pm3.9\pm3.6$\\
      $3.2$&$-$&$5.0$   & $4.0$  & $31.7\pm2.7\pm3.0$\\
      $5.0$&$-$&$8.0$   & $6.3$  & $21.8\pm2.4\pm2.1$\\
      $8.0$&$-$&$12.7$  & $10.0$ & $13.3\pm1.8\pm1.3$\\
      $12.7$&$-$&$20.1$ & $15.8$ & $7.53\pm1.24\pm0.72$\\
      $20.1$&$-$&$31.8$ & $25.0$ & $3.43\pm0.81\pm0.33$\\
      $31.8$&$-$&$80.0$ & $47.3$ & $0.60\pm0.24\pm0.06$\\
      \hline
    \end{tabular}
    \caption{ Cross section for the elastic process
      $\gsp\to\jpsi p$ measured in bins of \Qsq for $\Wgsp=90\GeV$ and for
      $|t|<1.2\GeVsq$.  $\langle\Qsq\rangle$ indicates the bin
      centre value in the \Qsq range considered. The first error 
      is statistical and the second the total systematic uncertainty.}
    \label{tab:qxsec}
\end{table}

\begin{table}\centering 
    \begin{tabular}[c]{c@{\hspace*{0.9\tabcolsep}}r@{\hspace*{0.5\tabcolsep}}c@{\hspace*{0.5\tabcolsep}}l@{\hspace*{0.9\tabcolsep}}c@{\hspace*{0.9\tabcolsep}}c}
      \hline
      Data set & \multicolumn{3}{c}{\Wgp} & $\langle\Wgp\rangle$ & $\sigma$\\
      & \multicolumn{3}{c}{$[\GeV]$} & $[\GeV]$ & $[\nb]$\\
      \hline\hline
      II  & $40$&$-$&$50$   & $44.8$  & $46.0\pm2.4\pm4.0$\\
          & $50$&$-$&$60$   & $54.8$  & $48.5\pm2.3\pm4.3$\\
          & $60$&$-$&$70$   & $64.8$  & $59.7\pm2.8\pm5.3$\\
          & $70$&$-$&$80$   & $74.8$  & $62.7\pm3.2\pm5.5$\\
          & $80$&$-$&$90$   & $84.9$  & $72.6\pm3.4\pm6.4$\\
          & $90$&$-$&$100$  & $94.9$  & $78.6\pm3.7\pm6.9$\\
          & $100$&$-$&$110$ & $104.9$ & $82.6\pm4.0\pm7.3$\\
          & $110$&$-$&$130$ & $119.5$ & $91.5\pm3.5\pm8.1$\\
          & $130$&$-$&$160$ & $144.1$ & $98.3\pm4.4\pm8.7$\\
      \hline                        
      III & $135$&$-$&$155$ & $144.9$ & $ 98.6\pm6.6\pm9.6$\\ 
          & $155$&$-$&$170$ & $162.5$ & $114\pm8\pm11$\\
          & $170$&$-$&$185$ & $177.3$ & $126\pm8\pm12$\\
          & $185$&$-$&$205$ & $194.8$ & $143\pm10\pm15$\\
          & $205$&$-$&$235$ & $219.6$ & $187\pm14\pm25$\\
      \hline                        
      IV  & $205$&$-$&$235$ & $219.6$ & $133\pm10\pm18$\\ 
          & $235$&$-$&$255$ & $244.8$ & $171\pm13\pm17$\\ 
          & $255$&$-$&$280$ & $267.2$ & $173\pm13\pm18$\\
          & $280$&$-$&$305$ & $292.3$ & $194\pm19\pm23$\\
      \hline
    \end{tabular}
    \caption{ Photoproduction cross section for the elastic
      process $\gp\to\jpsi p$ in bins of \Wgp for $|t|<1.2\GeVsq$ using the
      data sets II-IV (table~\ref{tab:sel}). $\langle\Wgp\rangle$ indicates
      the  bin centre value in the \Wgp range considered. 
      The first error on the cross section
      is statistical and the second the total systematic uncertainty.
       Note that there is an overlapping bin between data sets III
      and IV at $\langle \Wgp\rangle=219.6\,\GeV$, which is averaged for figure \ref{fig:wxsec} to 
      $\sigma=151\pm8\pm20\,\nb$.}
    \label{tab:wxsec_php}
\end{table}

\begin{table}\centering 
    \begin{tabular}[c]{ccr@{\hspace*{0.5\tabcolsep}}c@{\hspace*{0.5\tabcolsep}}lcc}
      \hline\\[-2.4ex]
      \Qsq & $\langle\Qsq\rangle$ & \multicolumn{3}{c}{\Wgp} &
      $\langle\Wgsp\rangle$ & $\sigma$\\
      $[\GeVsq]$ & $[\GeVsq]$ & \multicolumn{3}{c}{$[\GeV]$} & $[\GeV]$ &
      $[\nb]$\\
      \hline\hline
      $2-5$   & $3.2$  & $40$&$-$&$70$   & $53.3$  & $25.1\pm2.9\pm2.4$\\
              &        & $70$&$-$&$100$  & $83.9$  & $30.0\pm3.4\pm2.9$\\
              &        & $100$&$-$&$130$ & $114.1$ & $41.5\pm5.1\pm4.0$\\
              &        & $130$&$-$&$160$ & $144.2$ & $45.0\pm8.8\pm4.5$\\
      \hline                                                      
      $5-10$  & $7.0$  & $40$&$-$&$70$   & $53.3$  & $12.9\pm2.5\pm1.2$\\
              &        & $70$&$-$&$100$  & $83.9$  & $14.5\pm2.5\pm1.4$\\
              &        & $100$&$-$&$130$ & $114.1$ & $24.7\pm4.1\pm2.4$\\
              &        & $130$&$-$&$160$ & $144.2$ & $24.1\pm6.2\pm2.5$\\
      \hline                                                      
      $10-80$ & $22.4$ & $40$&$-$&$70$   & $53.4$  & $3.19\pm0.69\pm0.31$\\
              &        & $70$&$-$&$100$  & $83.9$  & $4.04\pm0.70\pm0.39$\\
              &        & $100$&$-$&$130$ & $114.1$ & $5.29\pm1.0\pm0.5$\\
              &        & $130$&$-$&$160$ & $144.2$ & $6.10\pm1.6\pm0.6$\\
      \hline
    \end{tabular}
    \caption{ Total cross section for the elastic
      process $\gsp\to\jpsi p$ measured in bins of \Qsq and $\Wgp$ for
      $|t|<1.2\GeVsq$. $\langle\Qsq\rangle$ and $\langle\Wgsp\rangle$ are the
       bin centre values in the indicated ranges.  The first error on the 
       cross section
      is statistical and the second the total systematic uncertainty.}
    \label{tab:wxsec_dis}
 \end{table}

\begin{table}\centering 
    \begin{tabular}[c]{r@{\hspace*{0.5\tabcolsep}}c@{\hspace*{0.5\tabcolsep}}lccc}
      \hline\\[-2.4ex]
      \multicolumn{3}{c}{\Qsq $[\GeVsq]$} & $\langle\Qsq\rangle\,[\GeVsq]$ &
      $\delta$ & $b$ $[\GeV^{-2}]$\\[0.5ex]
      \hline\hline\\[-2.4ex]
      &$\lesssim$&$1$& $0.05$ & \wexpphp & $4.57\pm0.06^{+0.11}_{-0.18}$\\
      $2$&$-$&$5$   & $3.2$  & $0.67\pm0.20\pm0.14$ & $4.11\pm0.26\pm0.37$\\
      $5$&$-$&$10$  & $7.0$  & $0.83\pm0.31\pm0.15$ & $3.50\pm0.50\pm0.49$\\
      $10$&$-$&$80$ & $22.4$ & $0.69\pm0.32\pm0.14$ & $3.49\pm0.45\pm0.33$\\
      \hline
    \end{tabular}
    \caption{ The parameters $\delta$ ($\sigma\propto\Wgsp^{\delta}$) and $b$ 
              $(\frac{d\sigma}{dt}\propto e^{bt})$ measured in bins
      of $Q^2$ in the range $40<\Wgsp<160\GeV$ and $|t|<1.2\GeVsq$.
      The values $\langle\Qsq\rangle$ indicate the  bin centre value in the
      \Qsq range considered. The first error is statistical and the
      second systematic.}
    \label{tab:bq}
 \end{table}

\begin{table}\centering 
    \begin{tabular}[c]{c@{\hspace*{0.9\tabcolsep}}c@{\hspace*{0.9\tabcolsep}}r@{\hspace*{0.5\tabcolsep}}c@{\hspace*{0.5\tabcolsep}}l@{\hspace*{0.9\tabcolsep}}c@{\hspace*{0.9\tabcolsep}}c}
      \hline\\[-2.4ex]
      \Qsq & $\langle\Qsq\rangle$ & \multicolumn{3}{c}{$|t|$} & $\langle
      |t|\rangle$ & d$\sigma/\mbox{d}t$\\
      $[\GeVsq]$ & $[\GeVsq]$ & \multicolumn{3}{c}{$[\GeVsq]$} & $[\GeVsq]$ &
      $[\nb/\GeVsq]$\\
      \hline\hline
      $\lesssim1$& $0.05$ & $0$&$-$&$0.07$ & $0.03$ & $285\pm9\pm25$\\
              &        & $0.07$&$-$&$0.14$ & $0.10$ & $180\pm7\pm16$\\
              &        & $0.14$&$-$&$0.21$ & $0.17$ & $130\pm6\pm11$\\
              &        & $0.21$&$-$&$0.30$ & $0.25$ & $92.1\pm4.0\pm8.1$\\
              &        & $0.30$&$-$&$0.40$ & $0.35$ & $61.2\pm3.1\pm5.4$\\
              &        & $0.40$&$-$&$0.60$ & $0.49$ & $32.5\pm1.5\pm2.9$\\
              &        & $0.60$&$-$&$0.90$ & $0.73$ & $10.60\pm0.60\pm0.90$\\
              &        & $0.90$&$-$&$1.20$ & $1.03$ & $2.70\pm0.20\pm0.30$\\
      \hline
      $2-5$   & $3.2$  & $0$&$-$&$0.08$    & $0.04$ & $107\pm14\pm10$\\
              &        & $0.08$&$-$&$0.18$ & $0.13$ & $95.1\pm11.0\pm9.1$\\
              &        & $0.18$&$-$&$0.38$ & $0.27$ & $40.2\pm5.4\pm3.9$\\
              &        & $0.38$&$-$&$1.20$ & $0.68$ & $8.04\pm1.05\pm0.77$\\
      \hline                                                         
      $5-10$  & $7.0$  & $0$&$-$&$0.08$    & $0.04$ & $78.6\pm13.2\pm7.5$\\
              &        & $0.08$&$-$&$0.18$ & $0.13$ & $27.7\pm5.7\pm2.7$\\
              &        & $0.18$&$-$&$0.38$ & $0.27$ & $18.9\pm3.7\pm1.8$\\
              &        & $0.38$&$-$&$1.20$ & $0.68$ & $5.21\pm0.96\pm0.50$\\
      \hline                                                         
      $10-80$ & $22.4$ & $0$&$-$&$0.08$    & $0.04$ & $15.0\pm3.1\pm1.4$\\
              &        & $0.08$&$-$&$0.18$ & $0.13$ & $8.90\pm2.14\pm0.85$\\
              &        & $0.18$&$-$&$0.38$ & $0.27$ & $4.55\pm0.93\pm0.44$\\
              &        & $0.38$&$-$&$1.20$ & $0.68$ & $1.36\pm0.25\pm0.13$\\
      \hline
    \end{tabular}
    \caption{ Differential cross section for the elastic
      process $\gsp\to\jpsi p$ measured in bins of \Qsq and $|t|$ in the
      range $40<\Wgsp<160\GeV$ using the data sets I and II
      (table~\ref{tab:sel}). $\langle\Qsq\rangle$ and $\langle |t|\rangle$
      are the  bin centre values in the indicated ranges. The first
      error on the cross section is statistical and the second
      is the total systematic uncertainty.}
    \label{tab:txsec}
\end{table}

  \begin{table}\centering
\begin{small}
\begin{tabular}[c]{c|lcccc}
      \hline\\[-2.4ex]
      & \multicolumn{5}{c}{d$\sigma/\mbox{d}t$ $[\mbox{nb}/\GeVsq]$}\\\hline
      $\langle\Wgp\rangle$&
            $\ \ \ |t|~~~0-0.07$ & $0.07-0.14$&$0.14-0.30$&$0.30-0.60$&$0.60-1.20$\\
   $[\GeV]$&
            $[\GeVsq]$\hspace{2cm}&\\
    \hline\hline
      $45$ & $182\pm20\pm16$&$115\pm15\pm10$&$64.9\pm6.7\pm5.7$&$35.5\pm3.6\pm3.1$&$5.7\pm0.8\pm0.5$\\
      $55$ &$208\pm20\pm18$&$118\pm14\pm10$&$69.6\pm6.9\pm6.1$&$35.6\pm3.4\pm3.1$&$5.5\pm0.7\pm0.5$\\
      $65$ &$225\pm23\pm20$&$169\pm18\pm15$&$107.1\pm9.4\pm9.4$&$34.3\pm3.7\pm3.0$&$6.2\pm0.9\pm0.5$\\
      $75$ &$321\pm31\pm28$&$151\pm19\pm13$&$93.4\pm9.5\pm8.2$&$38.9\pm4.3\pm3.4$&$7.1\pm1.1\pm0.6$\\
      $85$ &$292\pm29\pm26$&$178\pm20\pm16$&$132\pm11\pm12$&$41.4\pm4.4\pm3.6$&$8.2\pm1.1\pm0.7$\\
      $95$ &$326\pm32\pm29$&$224\pm24\pm20$&$135\pm12\pm12$&$46.3\pm4.8\pm4.1$&$7.5\pm1.1\pm0.7$\\
      $105$ &$392\pm37\pm34$&$224\pm26\pm20$&$125\pm12\pm11$&$48.5\pm5.3\pm4.3$&$7.7\pm1.1\pm0.7$\\
      $119$ &$376\pm31\pm33$&$265\pm24\pm23$&$142\pm11\pm12$&$60.9\pm4.9\pm5.4$&$8.2\pm1.0\pm0.7$\\
      $144$ &$458\pm42\pm40$&$267\pm29\pm23$&$167\pm14\pm15$&$51.5\pm5.3\pm4.5$&$8.4\pm1.1\pm0.7$\\
      $181$&$537\pm28\pm68$&$427\pm18\pm46$&$202\pm9\pm25$&$67.0\pm4.1\pm8.8$&$10.7\pm1.1\pm1.7$\\
      $251$ &$744\pm48\pm88$&$573\pm28\pm75$&$246\pm13\pm32$&$71.6\pm5.5\pm9.8$&$13.1\pm1.8\pm1.9$\\
      \hline
    \end{tabular}
\end{small}
    \caption{ Differential photoproduction cross sections
      d$\sigma/\mbox{d}t$ for the elastic process $\gsp\to\jpsi p$ measured in bins
      of \Wgp and $|t|$ using data sets II-IV (table~\ref{tab:sel}). The
      first error is statistical and the second
      the total systematic uncertainty.}
    \label{tab:wtxsecPhP}
  \end{table}

  \begin{table}\centering
    \begin{tabular}[c]{c|lcc}
      \hline\\[-2.4ex]
      & \multicolumn{3}{c}{d$\sigma/\mbox{d}t$ $[\mbox{nb}/\GeVsq]$}\\\hline
      $\langle\Wgp\rangle\,[\GeV]$&
      $|t|\,[\GeVsq]~~ 0-0.1$&$0.1-0.3$&$0.3-1.2$\\
      \hline\hline
      $57$&$33.3\pm4.9\pm3.2$&$17.3\pm2.2\pm1.7$&$3.4\pm0.5\pm0.3$\\
      $98$&$51.3\pm6.7\pm4.9$&$30.1\pm3.5\pm2.9$&$5.5\pm0.7\pm0.5$\\
      $140$&$~~~~60\pm12\pm6$&$31.5\pm5.8\pm3.0$&$6.0\pm1.1\pm0.6$\\
      \hline
    \end{tabular}
    \caption{ Differential electroproduction ($\langle Q^2\rangle=8.9\,\GeVsq$) 
      cross section
      d$\sigma/\mbox{d}t$ for the elastic process $\gsp\to\jpsi p$ measured in bins
      of \Wgp and $|t|$ using data set I (table~\ref{tab:sel}). The first
      error is statistical and the second
      the total systematic uncertainty.}
    \label{tab:wtxsecDIS}
\end{table}

\begin{table}\centering
    \begin{tabular}[c]{c@{\hspace*{0.9\tabcolsep}}r@{\hspace*{0.5\tabcolsep}}c@{\hspace*{0.5\tabcolsep}}l@{\hspace*{0.9\tabcolsep}}c@{\hspace*{0.9\tabcolsep}}c}
      \hline\\[-2.4ex]
      $\langle\Qsq\rangle$ & \multicolumn{3}{c}{$|t|$} & $\langle |t|\rangle$
      & $\alpha(\langle |t|\rangle)$\\
      $[\GeVsq]$ & \multicolumn{3}{c}{$[\GeVsq]$} & $[\GeVsq]$ & \\
      \hline\hline
       $0.05$ & $0$&$-$&$0.07$    & $0.03$ & $1.202\pm0.012\pm0.017$ \\
       $    $ & $0.07$&$-$&$0.14$ & $0.10$ & $1.240\pm0.012\pm0.019$ \\
       $    $ & $0.14$&$-$&$0.30$ & $0.22$ & $1.195\pm0.011\pm0.016$ \\
       $    $ & $0.30$&$-$&$0.60$ & $0.43$ & $1.121\pm0.013\pm0.018$ \\
       $    $ & $0.60$&$-$&$1.20$ & $0.84$ & $1.117\pm0.021\pm0.019$ \\
      \hline                        
       $8.9 $ & $0$&$-$&$0.1$     & $0.05$ & $ 1.173\pm0.064\pm0.038$ \\
       $    $ & $0.1$&$-$&$0.3$   & $0.19$ & $ 1.185\pm0.054\pm0.037$ \\
       $    $ & $0.3$&$-$&$1.2$   & $0.64$ & $ 1.168\pm0.059\pm0.037$ \\
      \hline
    \end{tabular}
    \caption{ The effective Pomeron trajectories $\alpha (t)$ derived
      from one-dimensional fits of the \Wgp dependence in bins of $t$, in the
      ranges $\Qsq\lesssim1\GeVsq$, $40<\Wgp<305\GeV$ (photoproduction) and
      $2<\Qsq<80\GeVsq$, $40<\Wgp<160\GeV$ (electroproduction). The values 
       $\langle\Qsq\rangle$
      and $\langle |t|\rangle$ are the  bin centre values in the
      indicated ranges. The first error is statistical and the second
      systematic.}
    \label{tab:alpha}
\end{table}

\begin{table}\centering
    \begin{tabular}[c]{c@{\hspace*{0.9\tabcolsep}}c@{\hspace*{0.9\tabcolsep}}r@{\hspace*{0.5\tabcolsep}}c@{\hspace*{0.5\tabcolsep}}l@{\hspace*{0.9\tabcolsep}}c@{\hspace*{0.9\tabcolsep}}c}
      \hline\\[-2.4ex]
      Data set & $\langle\Qsq\rangle$ & \multicolumn{3}{c}{\Wgsp} &
      $\langle\Wgsp\rangle$ & $b(\langle\Wgsp\rangle)$\\
      & $[\GeVsq]$ & \multicolumn{3}{c}{$[\GeV]$} & $[\GeV]$ & $[\GeV^{-2}]$ \\
      \hline\hline
      I   & $8.9 $ & $40$&$-$&$80$   & $57.3$  & $3.77\pm0.34\pm0.33$ \\
          & $    $ & $80$&$-$&$120$  & $98.2$  & $3.79\pm0.29\pm0.32$ \\
          & $    $ & $120$&$-$&$160$ & $139.6$ & $3.84\pm0.45\pm0.33$ \\
      \hline\\[-2.4ex]
      II  & $0.05$ & $40$&$-$&$50$   & $44.8$  & $4.13\pm0.20^{+0.14}_{-0.27}$ \\[0.5ex]
          & $    $ & $50$&$-$&$60$   & $54.8$  & $4.30\pm0.19^{+0.14}_{-0.31}$ \\[0.5ex]
          & $    $ & $60$&$-$&$70$   & $64.8$  & $4.57\pm0.20^{+0.14}_{-0.17}$ \\[0.5ex]
          & $    $ & $70$&$-$&$80$   & $74.8$  & $4.46\pm0.24^{+0.15}_{-0.46}$ \\[0.5ex]
          & $    $ & $80$&$-$&$90$   & $84.9$  & $4.45\pm0.20^{+0.15}_{-0.25}$ \\[0.5ex]
          & $    $ & $90$&$-$&$100$  & $94.9$  & $4.72\pm0.21^{+0.15}_{-0.19}$ \\[0.5ex]
          & $    $ & $100$&$-$&$110$ & $104.9$ & $4.79\pm0.22^{+0.15}_{-0.36}$ \\[0.5ex]
          & $    $ & $110$&$-$&$130$ & $119.5$ & $4.71\pm0.16^{+0.14}_{-0.18}$ \\[0.5ex]
          & $    $ & $130$&$-$&$160$ & $144.1$ & $4.95\pm0.19^{+0.15}_{-0.30}$ \\
      \hline\\[-2.4ex]
      III & $0.05$ & $135$&$-$&$235$ & $180.6$ & $5.08\pm0.14^{+0.25}_{-0.27}$ \\
      \hline\\[-2.4ex]
      IV  & $0.05$ & $205$&$-$&$305$ & $250.7$ & $5.41\pm0.20^{+0.29}_{-0.40}$ \\
      \hline
    \end{tabular}
    \caption{ The slope parameter $b$ derived from
      one-dimensional fits to the $t$ dependence measured in bins of \Wgsp.
      The values $\langle\Qsq\rangle$
      and $\langle\Wgsp\rangle$ are the  bin centre values in the
      indicated ranges. The first error is statistical and the second
      systematic.}
    \label{tab:bw}
  \end{table}

\begin{table}\centering
    \begin{tabular}[c]{crrr}
      \hline\\[-2.4ex]
      $\langle\Qsq\rangle$ & \multicolumn{1}{c}{$r_{00}^{04}$} &
      \multicolumn{1}{c}{$r_{1-1}^{04}$} & \multicolumn{1}{c}{$R$}\\
  $[\GeVsq]$ &\\
    \hline\hline\\[-2.4ex]
      $0.05$ & $-0.030\pm0.016\pm0.027$ & $ 0.020\pm0.016\pm0.042$ & $-0.030^{+0.015+0.026}_{-0.015-0.025}$\\[0.5ex]
      $3.2$  & $ 0.049\pm0.079\pm0.050$ & $-0.129\pm0.070\pm0.039$ & $ 0.052^{+0.096+0.059}_{-0.081-0.053}$\\[0.5ex]
      $7.0$  & $ 0.19\pm0.14\pm0.06$ & $-0.017\pm0.10\pm0.04$ & $ 0.23^{+0.25+0.09}_{-0.18-0.08}$\\[0.5ex]
      $22.4$ & $ 0.38\pm0.16\pm0.06$ & $-0.04\pm0.12\pm0.04$ & $ 0.62^{+0.59+0.17}_{-0.34-0.14}$\\
      \hline
    \end{tabular}
    \caption{ The spin-density matrix elements, $r_{00}^{04}$
      and $r_{1-1}^{04}$, and the ratio of cross sections of longitudinally
      and transversely polarised photons $R$ as a function of \Qsq in the
      range $|t|<5\GeVsq$ and $40<\Wgp<160\GeV$. The values $\langle\Qsq\rangle$
      indicate the  bin centre values in the \Qsq range considered.
      The first error is statistical and the second systematic.}
    \label{tab:qr}
\end{table}

\begin{table}\centering
\begin{tabular}[c]{crrr}
      \hline\\[-2.4ex]
      $\langle\Qsq\rangle$ &
      \multicolumn{1}{c}{$r_{1-1}^{1}$} & \multicolumn{1}{c}{$r_{00}^1+2r_{11}^1$} & \multicolumn{1}{c}{$r_{00}^5+2r_{11}^5$}\\
   $[\GeVsq]$ &\\    \hline\hline
      $3.2$  & $0.149\pm0.077\pm0.064$ & $0.026\pm0.035\pm0.026$ & $-0.035\pm0.072\pm0.055$\\
      $7.0$  & $0.43\pm0.11\pm0.06$ & $0.062\pm0.054\pm0.028$ & $-0.16\pm0.12\pm0.06$\\
      $22.4$ & $0.53\pm0.13\pm0.05$ & $0.026\pm0.069\pm0.031$ & $ 0.04\pm0.16\pm0.06$\\
      \hline
    \end{tabular}
    \caption{ The spin-density matrix element $r_{1-1}^{1}$ and
      the combined elements $r_{00}^1+2r_{11}^1$ and $r_{00}^5+2r_{11}^5$ as
      a function of \Qsq in the range $|t|<5\GeVsq$ and
      $40<\Wgp<160\GeV$.  The values $\langle\Qsq\rangle$ indicate the 
      bin centre value in the \Qsq range considered. The first error is
      statistical and the second systematic.}
    \label{tab:qr2}
  \end{table}


  \begin{table}\centering
    \begin{tabular}[c]{ccrr}
      \hline\\[-2.4ex]
      $\langle\Qsq\rangle$ $[\GeVsq]$ & $\langle |t|\rangle$ $[\GeVsq]$ &
      \multicolumn{1}{c}{$r_{00}^{04}$} & \multicolumn{1}{c}{$r_{1-1}^{04}$} \\
      \hline\hline
       $0.05$ & $0.03$ & $0.003\pm0.039\pm0.028$ & $-0.011\pm0.036\pm0.030$\\
       $    $ & $0.10$ & $0.011\pm0.043\pm0.029$ & $-0.041\pm0.042\pm0.030$\\
       $    $ & $0.22$ & $0.026\pm0.036\pm0.028$ & $ 0.104\pm0.035\pm0.029$\\
       $    $ & $0.43$ & $0.013\pm0.037\pm0.029$ & $ 0.025\pm0.037\pm0.030$\\
       $    $ & $0.84$ & $0.047\pm0.041\pm0.029$ & $ 0.064\pm0.047\pm0.034$\\
       $    $ & $1.8$  & $0.066\pm0.061\pm0.028$ & $-0.010\pm0.060\pm0.030$\\
       $    $ & $3.5$  & $0.018\pm0.081\pm0.028$ & $-0.074\pm0.082\pm0.032$\\
      \hline                     
       $8.9$ & $0.05$  & $0.32\pm0.15\pm0.06$ & $-0.13\pm0.11\pm0.04$\\
       $   $ & $0.19$  & $0.22\pm0.13\pm0.06$ & $-0.07\pm0.10\pm0.04$\\
       $   $ & $0.64$  & $0.05\pm0.10\pm0.05$ & $-0.071\pm0.083\pm0.036$\\
       $   $ & $3.0$   & $0.23\pm0.19\pm0.06$ & $ 0.06\pm0.13\pm0.03$\\
      \hline
    \end{tabular}
    \caption{ The spin-density matrix elements, $r_{00}^{04}$
      and $r_{1-1}^{04}$, as a function of $|t|$ in the
      range $40<\Wgp<160\GeV$ for photoproduction and electroproduction.
      $\langle\Qsq\rangle$ and $\langle |t|\rangle$ are the  bin
      centre values. The first error is
      statistical and the second systematic.}
    \label{tab:tr}
  \end{table}

  \begin{table}\centering
    \begin{tabular}[c]{crrr}
      \hline\\[-2.4ex]
      $\langle |t|\rangle$ $[\GeVsq]$ & \multicolumn{1}{c}{$r_{1-1}^1$} &
      \multicolumn{1}{c}{$r_{00}^1+2r_{11}^1$} & \multicolumn{1}{c}{$r_{00}^5+2r_{11}^5$}\\
      \hline\hline
       $0.05$  & $0.19\pm0.11\pm0.06$ & $-0.16\pm0.12\pm0.06$ & $ 0.031\pm0.055\pm0.029$\\
       $0.19$  & $0.62\pm0.11\pm0.06$ & $ 0.04\pm0.10\pm0.06$ & $ 0.004\pm0.048\pm0.026$\\
       $0.64$  & $0.361\pm0.097\pm0.061$ & $-0.073\pm0.091\pm0.057$ & $-0.039\pm0.043\pm0.027$\\
       $3.0$   & $0.07\pm0.15\pm0.06$ & $-0.15\pm0.15\pm0.06$ & $ 0.083\pm0.079\pm0.030$\\
      \hline
    \end{tabular}
    \caption{ The spin-density matrix element $r_{1-1}^{1}$ and
      the combined elements $r_{00}^1+2r_{11}^1$ and $r_{00}^5+2r_{11}^5$ as
      a function of $|t|$ in the range $40<\Wgp<160\GeV$ and
      $2<\Qsq<80\GeVsq$.  $\langle |t|\rangle$ indicates the 
      bin centre value. The first error
      is statistical and the second systematic.}
    \label{tab:tr2}
 \end{table}
 

\begin{thebibliography}{99}

\bibitem{Aid:1996ee}
S.~Aid {\it et al.}  [H1 Collaboration],
Nucl.\ Phys.\ B {\bf 468} (1996) 3
[hep-ex/9602007].
\bibitem{Derrick:1995vq}
M.~Derrick {\it et al.}  [ZEUS Collaboration],
Z.\ Phys.\ C {\bf 69} (1995) 39
[hep-ex/9507011].
\bibitem{Aid:1996dn}
S.~Aid {\it et al.}  [H1 Collaboration],
Nucl.\ Phys.\ B {\bf 472} (1996) 3
[hep-ex/9603005].
\bibitem{Adloff:2000vm}
C.~Adloff {\it et al.}  [H1 Collaboration],
Phys.\ Lett.\ B {\bf 483} (2000) 23
[hep-ex/0003020].
\bibitem{Breitweg:1997rg}
J.~Breitweg {\it et al.}  [ZEUS Collaboration],
Z.\ Phys.\ C {\bf 75} (1997) 215
[hep-ex/9704013].
\bibitem{Chekanov:2002xi}
S.~Chekanov {\it et al.}  [ZEUS Collaboration],
Eur.\ Phys.\ J.\ C {\bf 24} (2002) 345
[hep-ex/0201043].
\bibitem{Diehl:2003ny}
M.~Diehl,
Phys.\ Rept.\  {\bf 388} (2003) 41
[arXiv:hep-ph/0307382].
\bibitem{Frankfurt:2002ka}
L.~Frankfurt and M.~Strikman,
Phys.\ Rev.\ D {\bf 66} (2002) 031502
[hep-ph/0205223].
%
\bibitem{Frankfurt:2000ez}
L.~Frankfurt, M.~McDermott and M.~Strikman,
JHEP {\bf 0103} (2001) 045
[hep-ph/0009086].
%
\bibitem{Brodsky:1998kn}
S.~J.~Brodsky {\it et al.}, 
JETP Lett.\  {\bf 70} (1999) 155
[hep-ph/9901229];
%
N.~N.~Nikolaev, B.~G.~Zakharov and V.~R.~Zoller,
Phys.\ Lett.\ B {\bf 366} (1996) 337
[hep-ph/9506281].
\bibitem{Adloff:1999zs}
C.~Adloff {\it et al.}  [H1 Collaboration],
Eur.\ Phys.\ J.\ C {\bf 10} (1999) 373
[hep-ex/9903008].
\bibitem{Adloff:2002re}
C.~Adloff {\it et al.}  [H1 Collaboration],
Phys.\ Lett.\ B {\bf 541} (2002) 251
[hep-ex/0205107].
\bibitem{Aktas:2003zi}
A.~Aktas {\it et al.}  [H1 Collaboration],
Phys.\ Lett.\ B {\bf 568} (2003) 205
[hep-ex/0306013].
\bibitem{Chekanov:2002rm}
S.~Chekanov {\it et al.}  [ZEUS Collaboration],
Eur.\ Phys.\ J.\ C {\bf 26} (2003) 389
[hep-ex/0205081].
\bibitem{Breitweg:1998nh}
J.~Breitweg {\it et al.}  [ZEUS Collaboration],
Eur.\ Phys.\ J.\ C {\bf 6} (1999) 603
[hep-ex/9808020].
\bibitem{Chekanov:2004mw}
S.~Chekanov {\it et al.}  [ZEUS Collaboration],
Nucl.\ Phys.\ B {\bf 695} (2004) 3
[hep-ex/0404008].
\bibitem{Collins:1977}
P.~D.~B.~ Collins, {\it An Introduction to Regge Theory and High Energy
  Physics}, Cambridge University Press, 1977.
\bibitem{DoLan:1995}
A.~Donnachie and P.~V.~Landshoff, Phys. Lett. B {\bf 348} (1995) 213. 
\bibitem{DoLan:1998}
A.~Donnachie and P.~V.~Landshoff, Phys. Lett. B {\bf 437} (1998) 408; 
Phys. Lett. B {\bf 470} (1999) 243;
Phys.\ Lett.\ B {\bf 478} (2000) 146 [hep-ph/9912312];
Phys.\ Lett.\ B {\bf 518} (2001) 63 [hep-ph/0105088].
\bibitem{Bartels:2000ze}
J.~Bartels and H.~Kowalski,
Eur.\ Phys.\ J.\ C {\bf 19} (2001) 693
[hep-ph/0010345].
\bibitem{Ryskin:1992ui}
M.~G.~Ryskin,
Z.\ Phys.\ C {\bf 57} (1993) 89.
\bibitem{Martin:1997wy}
A.~D.~Martin and M.~G.~Ryskin,
Phys.\ Rev.\ D {\bf 57} (1998) 6692
[hep-ph/9711371].
\bibitem{Ivanov:2004vd}
D.~Y.~Ivanov, A.~Sch\"afer, L.~Szymanowski and G.~Krasnikov,
Eur.\ Phys.\ J.\ C {\bf 34} (2004) 297
[hep-ph/0401131].
\bibitem{Ivanov:2004ax}
I.~P.~Ivanov, N.~N.~Nikolaev and A.~A.~Savin,
``Diffractive vector meson production at HERA: From soft to hard QCD'',
DESY-04-243 (2004) [hep-ph/0501034].
\bibitem{Teubner:1999pm}
T.~Teubner, Proc. Ringberg Workshop ``New Trends in HERA Physics
        1999'', ed. G.~Grindhammer {\it et al.}, Ringberg Castle, Germany,
        Lecture Notes in Physics, Vol. 546, page 349, Springer-Verlag (2000)
        [hep-ph/9910329].
\bibitem{Martin:1999wb}
A.~D.~Martin, M.~G.~Ryskin and T.~Teubner,
Phys.\ Rev.\ D {\bf 62} (2000) 014022
[hep-ph/9912551].
\bibitem{Shuvaev:1999ce}
A.~G.~Shuvaev, K.~J.~Golec-Biernat, A.~D.~Martin and M.~G.~Ryskin,
Phys.\ Rev.\ D {\bf 60} (1999) 014015
[hep-ph/9902410].
\bibitem{teubnerdis}
T.~Teubner, private communication and contribution to the XIIIth International 
Workshop on Deep Inelastic Scattering (DIS2005), Madison, Wisconsin (in press, 
ed. W.H. Smith, Volume 792 in the AIP Conference Proceedings series).
\bibitem{Abt:1996hi}
I.~Abt {\it et al.}  [H1 Collaboration],
Nucl.\ Instrum.\ Meth.\ A {\bf 386} (1997) 310 and 348.
\bibitem{Eick:1996gv}
W.~Eick {\it et al.},
Nucl.\ Instrum.\ Meth.\ A {\bf 386} (1997) 81.
\bibitem{Nicholls:1995di}
T.~Nicholls {\it et al.}  [H1 SpaCal Group],
Nucl.\ Instrum.\ Meth.\ A {\bf 374} (1996) 149.
\bibitem{Appuhn:1996na}
R.~D.~Appuhn {\it et al.}  [H1 SpaCal Group],
Nucl.\ Instrum.\ Meth.\ A {\bf 386} (1997) 397.
\bibitem{Biddulph:1993bz}
P.~Biddulph {\it et al.},
Nucl.\ Instrum.\ Meth.\ A {\bf 340} (1994) 304.
\bibitem{Kohne:1997ph}
J.~K.~K\"ohne {\it et al.},
Nucl.\ Instrum.\ Meth.\ A {\bf 389} (1997) 128.
\bibitem{Bentvelsen:1992fu}
S.~Bentvelsen, J.~Engelen and P.~Kooijman,
Proceedings of the Workshop ``Physics at HERA",
        vol. 1, eds. W. Buchm\"uller, G. Ingelman, DESY (1992) 23;
        C. Hoeger, ibid., page 43.
\bibitem{Jacquet:1979}
F.~Jacquet and A.~Blondel,
in Proc. of the ``Study for an $ep$ Facility for Europe'',
edited by U.~Amaldi (1979) 391, DESY-79-048.
\bibitem{diffvm} B.~List and 
  A.~Mastroberardino, Proc. of the Workshop on Monte Carlo
  Generators for HERA Physics, eds. A.T. Doyle {\it et al.}, 
DESY-PROC-1999-02 (1999) 396.
\bibitem{Kwiatkowski:1990es}
A.~Kwiatkowski, H.~Spiesberger and H.~J.~M\"ohring,
Comput.\ Phys.\ Commun.\  {\bf 69} (1992) 155.
\bibitem{Baranov:1991yq}
S.~P.~Baranov, O.~D\"unger, H.~Shooshtari and J.~A.~Vermaseren,
Proceedings of the Workshop ``Physics at HERA",
        vol. 3, eds. W. Buchm\"uller, G. Ingelman, DESY (1992) 1478
\bibitem{Carli:1991yn}
T.~Carli, A.~Courau, S.~Kermiche and P.~Kessler,
Proceedings of the Workshop ``Physics at HERA",
        vol. 3, eds. W. Buchm\"uller, G. Ingelman, DESY (1992) 1468
\bibitem{grape}
T.~Abe,
Comput.\ Phys.\ Commun.\  {\bf 136} (2001) 126
[hep-ph/0012029].
\bibitem{Brun:1987ma}
R.~Brun {\it et al.}, 
CERN-DD/EE/84-1 (1987).
\bibitem{Fleischmann:2004uj}
P.~Fleischmann,
PhD thesis, Universit\"at Hamburg,  
DESY-THESIS-2004-013,\\ (available at http://www-h1.desy.de/publications/theses$\_$list.html).
\bibitem{Janauschek:2004}
L.~Janauschek,
PhD thesis, Ludwig-Maximilians-Universit\"at M\"unchen, 2005,\\ 
(available at http://www-h1.desy.de/publications/theses$\_$list.html).
\bibitem{Aktas:2004ek}
A.~Aktas {\it et al.}  [H1 Collaboration],
Phys.\ Lett.\ B {\bf 598} (2004) 159
[hep-ex/0406029].
\bibitem{Frixione:1993yw}
S.~Frixione, M.~L.~Mangano, P.~Nason and G.~Ridolfi,
Phys.\ Lett.\ B {\bf 319} (1993) 339
[hep-ph/9310350].
\bibitem{Eidelman:2004wy}
S.~Eidelman {\it et al.}  [Particle Data Group],
Phys.\ Lett.\ B {\bf 592} (2004) 1.
\bibitem{Pumplin:2002vw}
J.~Pumplin {\it et al.}, 
JHEP {\bf 0207} (2002) 012
[hep-ph/0201195].
\bibitem{Martin:2001es}
A.~D.~Martin, R.~G.~Roberts, W.~J.~Stirling and R.~S.~Thorne,
Eur.\ Phys.\ J.\ C {\bf 23} (2002) 73
[hep-ph/0110215].
\bibitem{Adloff:2003uh}
C.~Adloff {\it et al.}  [H1 Collaboration],
Eur.\ Phys.\ J.\ C {\bf 21} (2001) 33 [hep-ex/0012053].
\bibitem{Chekanov:2002pv}
S.~Chekanov {\it et al.}  [ZEUS Collaboration],
Phys.\ Rev.\ D {\bf 67} (2003) 012007
[hep-ex/0208023].
\bibitem{Lai:1996mg}
H.~L.~Lai {\it et al.},
Phys.\ Rev.\ D {\bf 55} (1997) 1280
[hep-ph/9606399].
\bibitem{nim:DAgostini}
G.~D'Agostini,
Nucl.\ Instrum.\ Meth.\ A {\bf 362} (1995) 487;
preprint DESY-95-242, ROME-1070-1995 [hep-ph/9512295].
\bibitem{Donnachie:1998gm}
A.~Donnachie and P.~V.~Landshoff,
Phys.\ Lett.\ B {\bf 437} (1998) 408
[hep-ph/9806344].
\bibitem{Schilling:1973ag}
K.~Schilling and G.~Wolf,
Nucl.\ Phys.\ B {\bf 61} (1973) 381.
\end{thebibliography}
\end{document}